\documentclass[12pt,a4paper]{report}
\usepackage[T1]{fontenc}
\usepackage{lmodern}
\usepackage[doublespacing]{setspace}
\usepackage{amsmath}
\usepackage{amssymb}
\usepackage{abstract}
\usepackage[linktocpage,pdftitle={Thermal Inflation with a Thermal Waterfall Scalar Field Coupled to a Light Spectator Scalar Field},pdfauthor={Arron Rumsey}]{hyperref}
\usepackage[dvipsnames]{xcolor}
\definecolor{hyperlinks}{HTML}{0000FF}
\hypersetup{colorlinks, urlcolor=hyperlinks, citecolor=hyperlinks, linkcolor=hyperlinks, breaklinks=true}
\usepackage{breakurl}
\usepackage{graphicx}
\graphicspath{ {Images/} }

\usepackage{cite}

\usepackage{enumitem}
\setlist{listparindent=\parindent}
\usepackage{cleveref}

\crefname{chapter}{Chapter}{Chapters}
\Crefname{chapter}{Chapter}{Chapters}
\crefname{section}{Section}{Sections}
\Crefname{section}{Section}{Sections}
\crefname{appendix}{Appendix}{Appendices}
\Crefname{appendix}{Appendix}{Appendices}
\crefname{equation}{Eq.}{Eqs.}
\Crefname{equation}{Eq.}{Eqs.}
\crefname{figure}{Fig.}{Figs.}
\Crefname{figure}{Fig.}{Figs.}
\crefname{table}{Table}{Tables}
\Crefname{table}{Table}{Tables}
\usepackage{tikz}
\newcommand{\Tikzmark}[1]{\tikz[remember picture] \coordinate[shift={(0,.7ex)}](#1);}
\newcommand{\colrow}[3][]{\tikz[overlay,remember picture,line width=12pt] \draw[shorten >=-.1em, shorten <=-.1em, #1] (#2)--(#3);}
\usepackage[title,titletoc]{appendix}
\usepackage{fancyhdr}
\begin{document}
\begin{titlepage}
\vspace*{-37.5mm}
\hspace*{99mm}
\includegraphics[width=55mm]{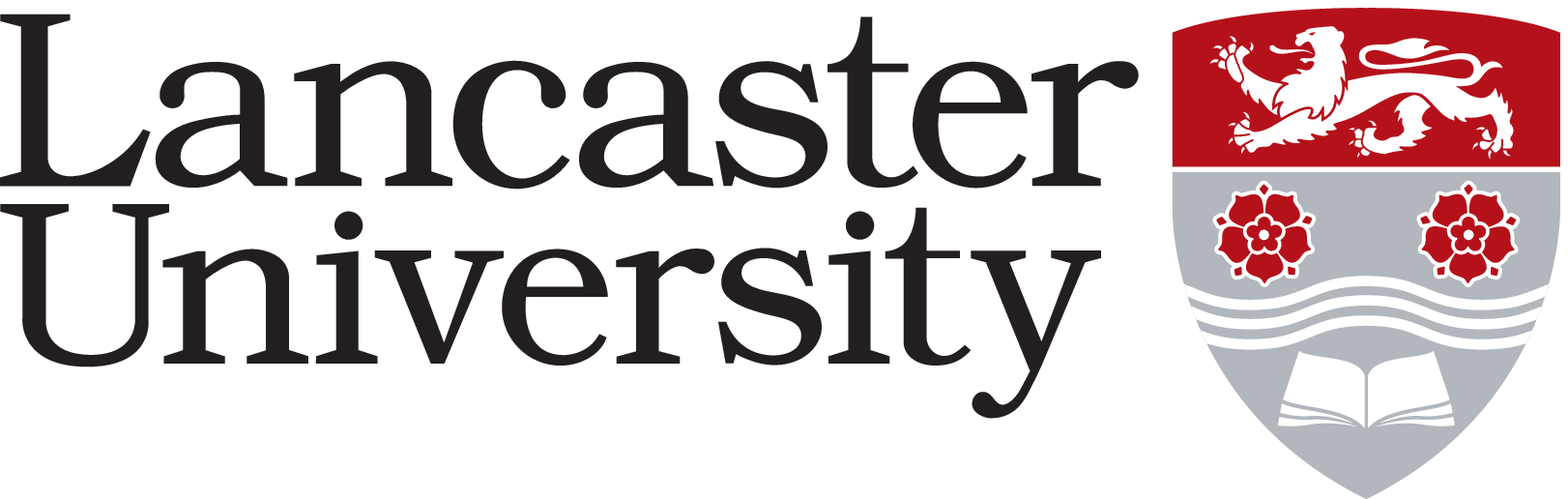}
\vspace*{25mm}
\begin{center}
\huge{\textbf{Thermal Inflation with a Thermal Waterfall Scalar Field Coupled to a Light Spectator Scalar Field}}\\
\vspace*{15mm}
\Large{\textbf{Arron Rumsey}}\\
\vspace*{20mm}
September 2016\\
\vfill
\normalsize{\textit{This thesis is submitted in partial fulfillment of the requirements for the degree of Master of Philosophy (MPhil) at Lancaster University. No part of this thesis has been previously submitted for the award of a higher degree.}}
\end{center}
\end{titlepage}
\begin{abstract}
\noindent
\begin{center}
This thesis begins with an introduction to the state of the art of modern Cosmology. The field of Particle Cosmology is then introduced and explored, in particular with regard to the study of cosmological inflation. We then introduce a new model of Thermal Inflation, in which the mass of the thermal waterfall field responsible for the inflation is dependent on a light spectator scalar field. The model contains a variety of free parameters, two of which control the power of the coupling term and the non-renormalizable term. We use the $\delta N$ formalism to investigate the ``end of inflation'' and modulated decay scenarios in turn to see whether they are able to produce the dominant contribution to the primordial curvature perturbation $\zeta$. We constrain the model and then explore the parameter space. We explore key observational signatures, such as non-Gaussianity, the scalar spectral index and the running of the scalar spectral index. We find that for some regions of the parameter space, the ability of the model to produce the dominant contribution to $\zeta$ is excluded. However, for other regions of the parameter space, we find that the model yields a sharp prediction for a variety of parameters within the model.
\end{center}
\end{abstract}
\chapter*{Acknowledgements}
As is extremely common for large pieces of academic work, this thesis would not exist if it were not for many people other than myself. Also, this thesis was completed in part whilst receiving an STFC PhD studentship.\\
\indent I would like to start by thanking Anupam Mazumdar for several helpful discussions regarding thermalization and thermal interaction rates.\\
\indent I have had many helpful and insightful conversations with my academic peers and friends Phil Stephens, Frankie Doddato, Ernest Pukartas, Lingfei Wang, Mindaugas Karciauskas and Jacques Wagstaff.\\
\indent I would like to thank Jessica Brooks.\\
\indent The majority of the work on the new Thermal Inflation model that is introduced in this thesis was done in collaboration with David Lyth. A special mention needs to be given to David, however, as he was not merely an academic collaborator, but for a large part was also my acting supervisor. I have had many fruitful discussions on Particle Cosmology with David. I feel privileged to have worked with such a pillar of modern Cosmology.\\
\indent I would like to thank my very dear friends and family Linda Rumsey, Rob Bishop, Paul Evans, Matt Eve, Guy Rusha, Kelly Hodder, Tristan Reeves, Chelle Nevill and Alise Kirtley for the support and encouragement that you have provided to me.\\
\indent It would be improper if the final acknowledgment did not go to my supervisor, Kostas Dimopoulos. The work that we have done together has been exciting, interesting and challenging. During some particular bleak times during the course of this work, he has been compassionate and accommodating. I cherish the knowledge, wisdom and conversations that we have shared. I sincerely thank you for everything that you have done for me.
\pagenumbering{roman}
\tableofcontents
\clearpage
\phantomsection
\addcontentsline{toc}{chapter}{\listfigurename}
\listoffigures
\clearpage
\phantomsection
\addcontentsline{toc}{chapter}{\listtablename}
\listoftables
\pagenumbering{arabic}
\pagestyle{fancy}
\renewcommand{\sectionmark}[1]{\markright{\thesection\,\ #1}}
\fancyhf{}
\lhead{\textit{\rightmark}}
\rhead{\thepage}
\chapter{Introduction}
\label{Chapter: Introduction}
Cosmological Inflation is a leading candidate for the solution of the three main problems of the standard Big Bang scenario: the Horizon, Flatness and Relic problems. It also has the ability to seed the initial conditions required to explain the observed large-scale structure of the Universe. (For a textbook on this topic, see \cite{Lyth_&_Liddle:Prim._Den._Pert.}.) In the simplest scenario, quantum fluctuations of a scalar field are converted to classical perturbations around the time of horizon exit, after which they become frozen. This gives rise to the primordial curvature perturbation, $\zeta$, which grows under the influence of gravity to give rise to all of the large-scale structure in the Universe. The observed value of the spectrum of the primordial curvature perturbation is $\mathcal{P}_{\zeta}^{\frac{1}{2}}(k_{0}) \sim 10^{-5}$.\\
\indent Moving away from this simplest scenario, there has been much work done on trying to generate the observed $\zeta$ in other scenarios, such as the curvaton \cite{Lyth_&_Wands:Generating_Curv._Pert._without_Inflaton , Lyth_et_al.:Prim._Den._Pert._in_Curvaton_Scenario , Choi_&_Seto:Modulated_Reheat._Curvaton , Dimopoulos:Can_Vec._Field_be_respons._Cur._Pert._Uni. , Dimopoulos_et_al.:Stat._Ani._Curv._Pert._Vec._Field_Perts. , Dimopoulos_&_Karciauskas:Non-min._Coupled_Vec._Curv. , Karciauskas:Prim._Curv._Pert._Vec._Fields_Gen._Non-Abelian_Groups , Dimopoulos:Stat._Ani._Vec._Curv._Paradigm , Dimopoulos:Stat._Ani._Vec._Curv._Paradigm_(Erratum) , Karciauskas_et_al.:Ani._Non-Gauss._Vec._Field._Perts. , Karciauskas_et_al.:Ani._Non-Gauss._Vec._Field._Perts._(Erratum) , Yokoyama_&_Soda:Prim._Stat._Ani._generated_End_of_Inf. , Dimopoulos_et_al.:Stat._Ani._Vec._Curv._D-Brane_Inf. , Bartolo_et_al.:Ani._Trispectrum_Curv._Perts._Induced_Prim._Non-Abelian_Vec._Fields , Dimopoulos:Supergravity_Inspired_Vec._Curv. , Dimopoulos_&_Karciauskas:Par._Violating_Stat._Ani. , Dimopoulos_et_al.:Elim._eta_Prob._SUGRA_Hyb._Inf._Vec._Backreaction , Assadullahi_et_al.:Mod._Curvaton_Dec. , Langlois_&_Takahashi:Den._Perts._Mod._Dec._Curvaton , Enomoto_et_al.:Mod._Dec._Multi-Comp._Uni. , Kohri_et_al.:Delta-N_Form._Curvaton_Mod._Decay , Dimopoulos_et_al.:Curvaton_Dyn.}, inhomogeneous reheating \cite{Dvali_et_al.:New_Mech._Generating_Den._Perts._from_Inflation , Dvali_et_al.:Cosmo._Perts._Inhomo._Reheat._Freezeout_&_Mass_Dom. , Postma:Inhom._Reheat._Low_Scale_Inflation_and/or_MSSM_Flat_Dirs. , Choi_&_Seto:Modulated_Reheat._Curvaton , Vernizzi:Generating_Cosmo._Perts._Mass_Variations , Vernizzi:Cosmo._Perts._Varying_Masses_&_Couplings , Zaldarriaga:Non-Gauss._in_Models_Varying_Inflaton_Decay_Rate , Kawasaki_et_al.:Den._Fluctuations_in_Thermal_Inflation_&_Non-Gaussianity , Assadullahi_et_al.:Mod._Curvaton_Dec. , Langlois_&_Takahashi:Den._Perts._Mod._Dec._Curvaton , Enomoto_et_al.:Mod._Dec._Multi-Comp._Uni. , Kohri_et_al.:Delta-N_Form._Curvaton_Mod._Decay}, ``end of inflation'' \cite{Lyth:Generating_Curvature_Perturbation_E._of_I. , Yokoyama_&_Soda:Prim._Stat._Ani._generated_End_of_Inf. , Kawasaki_et_al.:Den._Fluctuations_in_Thermal_Inflation_&_Non-Gaussianity , Salem:On_Gen._Den._Perts._End_of_Inf. , Alabidi_&_Lyth:Curv._Pert._Sym._Break._End_of_Inf. , Lyth:Hybrid_Waterfall_and_Curvature_Perturbation , Lyth_&_Riotto:Gen._Curv._Pert._End_of_Inf._Str._Theory , Sasaki:Multi-brid_Inf._Non-Gauss.} (also see \cite{Bernardeau_et_al.:Mod._Flucts._Hyb._Inf.}) and inhomogeneous phase transition \cite{Matsuda:Cosmo._Perts._from_Inhomogeneous_Phase_Transition}. (Also see \cite{Alabidi_et_al.:How_Curvaton_Mod._Reheat._and_Inhomo._End_of_Inf._are_related}.)\\
\indent This thesis is structured as follows. In \cref{Chapter: Concordance Model of Cosmology - LambdaCDM} we talk about the standard model of Cosmology. We then go on to talk about the field of Particle Cosmology in \cref{Chapter: Particle Cosmology}. In \cref{Chapter: A New Thermal Inflation Model} we give a detailed account of a new Thermal Inflation model that we have created, where we give expressions for key observational quantities that are predicted by the model. We finish with \cref{Chapter: Summary and Conclusions}, in which we conclude.
\chapter{Concordance Model of Cosmology {---} $\Lambda$CDM}
\label{Chapter: Concordance Model of Cosmology - LambdaCDM}
At the beginning of the 20\textsuperscript{th} century, the field of Cosmology was devoid of General Relativity, proof of the existence of other galaxies and that the Universe was expanding, evidence of the Cosmic Microwave Background, hereafter referred to as the CMB, as well as the Big Bang Theory, or indeed any theory of the genesis of the Universe that was scientifically based. By the \textit{end} of the 20\textsuperscript{th} century however, a consistent and hugely successful model of the entire history of the Universe (save for the very first moments after the creation of spacetime, if indeed even such a creation occurred) was firmly established.\\
\indent Several hundred years ago, the Polish astronomer Nicolaus Copernicus proposed an alternative to the Ptolemaic view, which stated that the Earth was at the centre of the Universe.\footnote{The Heliocentric system was originally proposed by Aristarchus of Samos in the 3\textsuperscript{rd} century BC, whom Copernicus was aware of and cited.} Copernicus stated that it was the Sun that is at the centre of our planetary system. Moreover, the Earth did not reside in any special place in the Universe and as such, all physical laws that apply on Earth should apply in the same way in other parts of the Universe. This is known as the \textbf{Copernican Principle}.\\
\indent Analysis of observations of the large-scale structure of the Universe and generalizing the Copernican Principle allows us to state that, on large scales, the Universe is statistically both \textbf{homogeneous} and \textbf{isotropic}. Homogeneity states that observations made at any point in the Universe will be statistically representative of those made at any other point and hence there is no preferred place in the Universe. Isotropy states that the Universe looks statistically the same in all directions. These two aspects of our Universe, when taken together, define what is termed the \textbf{Cosmological Principle}, which is a foundational pillar in the current standard model of Cosmology.
\section{General Relativity}
\label{Section: General Relativity}
2015 was a milestone year for Albert Einstein's \textbf{General Theory of Relativity}, being the 100\textsuperscript{th} anniversary of its first presentation to the world by Einstein. After 100 years, it has survived the test of time and scientific rigor to remain the principle theory that humankind possesses regarding the behavior of gravity on large (cosmological) scales.\\
\indent Let us start with a reminder of the basic mathematics of Special Relativity. We define a 4-vector
\[
x^{\mu} \equiv
\begin{bmatrix}
x^{0}\\
x^{1}\\
x^{2}\\
x^{3}
\end{bmatrix}
\equiv
\begin{bmatrix}
ct\\
x\\
y\\
z
\end{bmatrix}
=
\begin{bmatrix}
ct\\
\textbf{x}
\end{bmatrix}
\]
in \textbf{Minkowski} coordinates (in a Minkowski space). This defines a \textbf{threading} of spacetime, which can be represented by a series of lines, corresponding to a fixed $x^{i}$ ($i\!=\!1,2,3$), as well as a \textbf{slicing} of spacetime into hypersurfaces, corresponding to a fixed $x^{0}$.\\
\indent The line element between two spacetime points is given by
\begin{equation}
ds^{2} = -dt^{2} + dx^{2} + dy^{2} + dz^{2}
\label{Eq.: Line Element (Coordinates)}
\end{equation}
where we use natural units for which $c\!=\!\hbar\!=\!k_{B}\!=\!1$. By defining a symmetric \textbf{Minkowski Metric Tensor}
\[
\eta_{\mu\nu} \equiv
\begin{bmatrix}
-1 & 0 & 0 & 0\\
0 & 1 & 0 & 0\\
0 & 0 & 1 & 0\\
0 & 0 & 0 & 1
\end{bmatrix}
\]
we can denote the line element as
\begin{equation}
ds^{2} = \eta_{\mu\nu}dx^{\mu}dx^{\nu}
\label{Eq.: Line Element (Minkowski Metric Tensor)}
\end{equation}
where we are using \textbf{Einstein summation} convention.\\
\indent Going from a flat (Minkowski) manifold to the generically curved one of spacetime, which is an example of a \textbf{pseudo-Riemannian} manifold, we have the general symmetric metric tensor $g_{\mu\nu}$ (as opposed to the flat $\eta_{\mu\nu}$). The line element now becomes
\begin{equation}
ds^{2} = g_{\mu\nu}dx^{\mu}dx^{\nu}
\label{Eq.: Line Element (General Metric Tensor)}
\end{equation}
The essence of General Relativity, hereafter referred to as GR, is reflected in the \textbf{Einstein Field Equation}, which is
\begin{equation}
R_{\mu\nu} - \frac{1}{2}g_{\mu\nu}R = 8\pi GT_{\mu\nu}
\label{Eq.: Einstein Field Equation}
\end{equation}
where $R_{\mu\nu}$ is the \textbf{Ricci Tensor}, given by
\begin{equation}
R_{\mu\nu} \equiv R^{\lambda}_{\ \mu\lambda\nu}
\label{Eq.: Ricci Tensor Definition}
\end{equation}
where $R^{\mu}_{\ \nu\beta\alpha}$ is the \textbf{Riemann Tensor}, given by
\begin{equation}
R^{\mu}_{\ \nu\beta\alpha} \equiv \partial_{\alpha}\Gamma^{\mu}_{\nu\beta} - \partial_{\beta}\Gamma^{\mu}_{\nu\alpha} + \Gamma^{\mu}_{\sigma\alpha}\Gamma^{\sigma}_{\nu\beta} - \Gamma^{\mu}_{\sigma\beta}\Gamma^{\sigma}_{\nu\alpha}
\label{Eq.: Riemann Tensor Definition}
\end{equation}
where $\Gamma^{\mu}_{\nu\beta}$ is the \textbf{Christoffel Symbol}, given by
\begin{equation}
\Gamma^{\mu}_{\nu\beta} = \frac{1}{2}g^{\alpha\mu}(\partial_{\beta}g_{\alpha\nu} + \partial_{\nu}g_{\alpha\beta} - \partial_{\alpha}g_{\nu\beta})
\label{Eq.: Christoffel Symbol}
\end{equation}
$R$ is the \textbf{Ricci Scalar}, given by
\begin{equation}
R \equiv g^{\mu\nu}R_{\mu\nu}
\label{Eq.: Ricci Scalar Definition}
\end{equation}
with $g^{\mu\nu}$ being the inverse metric. In \cref{Eq.: Einstein Field Equation}, $G$ is \textbf{Newton's Gravitational Constant} and $T_{\mu\nu}$ is the \textbf{Energy-Momentum Tensor}, also known as the Stress-Energy Tensor, which is defined as
\[
T_{\mu\nu} \equiv
\begin{bmatrix}
\Tikzmark{00}\boldsymbol{T_{00}}\Tikzmark{00end} & \Tikzmark{01}\boldsymbol{T_{01}} & \boldsymbol{T_{02}} & \boldsymbol{T_{03}}\Tikzmark{03}\\
\Tikzmark{10}\boldsymbol{T_{10}}\Tikzmark{10end} & \Tikzmark{11}\boldsymbol{T_{11}}\Tikzmark{11end} & \Tikzmark{12}\boldsymbol{T_{12}} & \boldsymbol{T_{13}}\Tikzmark{13}\\
\Tikzmark{20}\boldsymbol{T_{20}}\Tikzmark{20end} & \Tikzmark{21}\boldsymbol{T_{21}}\Tikzmark{21end} & \Tikzmark{22}\boldsymbol{T_{22}}\Tikzmark{22end} & \Tikzmark{23}\boldsymbol{T_{23}}\Tikzmark{23end}\\
\Tikzmark{30}\boldsymbol{T_{30}}\Tikzmark{30end} & \Tikzmark{31}\boldsymbol{T_{31}} & \boldsymbol{T_{32}}\Tikzmark{32} & \Tikzmark{33}\boldsymbol{T_{33}}\Tikzmark{33end}
\end{bmatrix}
\colrow[red,opacity=0.5]{00}{00end}
\colrow[YellowOrange,opacity=0.3]{01}{03}
\colrow[YellowOrange,opacity=0.3]{10}{10end}
\colrow[YellowOrange,opacity=0.3]{20}{20end}
\colrow[YellowOrange,opacity=0.3]{30}{30end}
\colrow[green,opacity=0.3]{11}{11end}
\colrow[blue,opacity=0.3]{12}{13}
\colrow[blue,opacity=0.3]{21}{21end}
\colrow[green,opacity=0.3]{22}{22end}
\colrow[blue,opacity=0.3]{23}{23end}
\colrow[blue,opacity=0.3]{31}{32}
\colrow[green,opacity=0.3]{33}{33end}
\]
where the $T_{00}$ component in \textcolor{red}{\textbf{red}} is the energy density, the $T_{i0}$ components in \textcolor{YellowOrange}{\textbf{orange}} are the momentum density, the $T_{0i}$ components in \textcolor{YellowOrange}{\textbf{orange}} are the energy flux, the components in \textcolor{blue}{\textbf{blue}} are the shear stress and the $T_{ii}$ components in \textcolor{green}{\textbf{green}} are the pressure. The $T_{ij}$ components are the momentum flux. Sometimes the LHS of \cref{Eq.: Einstein Field Equation} is combined into a single tensor, known as the \textbf{Einstein Tensor}
\begin{equation}
G_{\mu\nu} \equiv R_{\mu\nu} - \frac{1}{2}g_{\mu\nu}R
\label{Eq.: Einstein Tensor}
\end{equation}
The Einstein Field Equation relates the curvature of a region of spacetime, i.e. the strength of gravity, to the amount of energy, momentum and stress that is present within that spacetime region.
\subsection{FLRW Universe}
\label{Subsection: FLRW Universe}
During the 1920s and 1930s, four scientists worked independently on problems concerning the geometry and evolution of a homogeneous and isotropic Universe. These were Alexander Friedmann, Georges Lemaître, Howard P. Robertson and Arthur Geoffrey Walker. The most general form of the line element in polar coordinates that satisfies homogeneity and isotropy, as well as allowing for uniform expansion is
\begin{equation}
ds^{2} = -dt^{2} +a^{2}(t)\left[\frac{dr^{2}}{1-Kr^{2}} + r^{2}\left(d\theta^{2} + \sin^{2}(\theta)d\phi^{2}\right)\right]
\label{Eq.: Line Element (Polar Coordinates)}
\end{equation}
where $a(t)$ is the scale factor and $K$ is the spatial intrinsic curvature. The value $K\!=\!0$ corresponds to a spatially flat (Euclidean) Universe. A value of $K\!<\!0$ and $K\!>\!0$ correspond to a spatially open hyperbolic and spatially closed elliptical Universe respectively. This is known as the \textbf{FLRW metric line element}, after the authors mentioned above. Analysis of the CMB shows that our Universe is spatially flat to a very high degree of precision. For a $K\!=\!0$ spatially flat Universe, the line element in Cartesian coordinates is
\begin{equation}
ds^{2} = -dt^{2} +a^{2}(t)\left(dx^{2} + dy^{2} + dz^{2}\right)
\label{Eq.: Line Element (Cartesian Coordinates - Proper Time)}
\end{equation}
where the metric tensor is
\[
g_{\mu\nu} =
\begin{bmatrix}
-1 & 0 & 0 & 0\\
0 & a^{2}(t) & 0 & 0\\
0 & 0 & a^{2}(t) & 0\\
0 & 0 & 0 & a^{2}(t)
\end{bmatrix}
\]
We can rearrange \cref{Eq.: Line Element (Cartesian Coordinates - Proper Time)} slightly so that all four spacetime coordinates have the same scale factor, by defining a \textbf{Conformal Time} variable
\begin{equation}
d\eta = \frac{dt}{a}
\label{Eq.: Conformal Time Definition}
\end{equation}
The line element now becomes
\begin{equation}
ds^{2} = a^{2}(\eta)\left(-d\eta^{2} + dx^{2} + dy^{2} + dz^{2}\right)
\label{Eq.: Line Element (Cartesian Coordinates - Conformal Time)}
\end{equation}
\indent Let us now assume that the content of the Universe is analogous to a \textbf{perfect fluid}, which is a fluid that has no shear stress, no viscosity and which does not conduct heat. It can be described entirely by its energy density $\boldsymbol{\rho}$ and its isotropic pressure \textbf{P}. In the local rest frame, the Energy-Momentum tensor becomes simply
\[
T_{\mu\nu} =
\begin{bmatrix}
\rho & 0 & 0 & 0\\
0 & P & 0 & 0\\
0 & 0 & P & 0\\
0 & 0 & 0 & P
\end{bmatrix}
\]
\indent We have not yet employed GR in the discussion regarding our FLRW Universe. All we have assumed so far is a uniformly expanding/contracting homogeneous and isotropic Universe, which has a content that can be described as a perfect fluid. We now take the ``$00$'' (``$tt$'') component of the Einstein field equation
\begin{align}
R_{00} - \frac{1}{2}g_{00}R &= 8\pi GT_{00}
\\
\nonumber
\\
R_{00} + \frac{1}{2}R &= 8\pi G\rho
\label{Eq.: Einstein Field Equation - 00}
\end{align}
where the Ricci tensor is
\begin{equation}
R_{00} \equiv R^{\lambda}_{\ 0\lambda0} \equiv \partial_{0}\Gamma^{\lambda}_{0\lambda} - \partial_{\lambda}\Gamma^{\lambda}_{00} + \Gamma^{\lambda}_{\alpha0}\Gamma^{\alpha}_{0\lambda} - \Gamma^{\lambda}_{\alpha\lambda}\Gamma^{\alpha}_{00}
\label{Eq.: Ricci Tensor - 00}
\end{equation}
We calculate the Christoffel symbols from \cref{Eq.: Christoffel Symbol}. We also calculate the Ricci scalar in \cref{Eq.: Einstein Field Equation - 00} from \cref{Eq.: Ricci Scalar Definition}. After this work, we obtain what is known as the \textbf{Friedmann Equation}
\begin{equation}
H^{2} = \frac{\rho}{3M_{P}^{2}} - \frac{K}{a^{2}}
\label{Eq.: Friedmann Equation}
\end{equation}
where $M_{P}\!=\!2.436\times10^{18}\ \mbox{GeV}$ is the \textbf{Reduced Planck Mass}, for which $8\pi G\!\equiv\!M_{P}^{-2}$ in natural units and $H\!\equiv\!\frac{\dot{a}}{a}$ is the \textbf{Hubble Parameter}, with the dot denoting derivative with respect to the cosmic time $t$. The Hubble parameter is the rate of expansion/contraction of the Universe. For a spatially flat Universe, $K\!=\!0$ and \cref{Eq.: Friedmann Equation} is simply
\begin{equation}
\rho = 3M_{P}^{2}H^{2}
\label{Eq.: Friedmann Equation (K=0)}
\end{equation}
From energy conservation ($\nabla_\mu T^{\mu\nu}\!=\!0$) we obtain the \textbf{Continuity Equation}
\begin{equation}
\dot{\rho} = -3H(\rho + P)
\label{Eq.: Continuity Equation}
\end{equation}
If we differentiate \cref{Eq.: Friedmann Equation (K=0)} with respect to $t$ and employ \cref{Eq.: Continuity Equation} we obtain the \textbf{(Friedmann) Acceleration Equation}
\begin{equation}
\frac{\ddot{a}}{a} = -\frac{\rho + 3P}{6M_{P}^{2}}
\label{Eq.: Acceleration Equation}
\end{equation}
\indent In order to solve the Friedmann equation for the time-evolution of the scale factor, we first need an equation of state giving the relationship between the energy density and the pressure of the perfect cosmic fluid. The equation of state is barotropic ($P\!=\!P(\rho)$) and is parameterized as
\begin{equation}
w = \frac{P}{\rho}
\label{Eq.: Equation of State Parametrization}
\end{equation}
With $c\!=\!1$, if the cosmic fluid has velocity $v_{rms}\!\ll\!1$, it is called matter (or non-relativistic matter), whereas if it has $v_{rms}\!\approx\!1$, then it is called radiation (or relativistic matter). For the case of matter, we have $w\!=\!0$, as $\rho\!\gg\!P$. The continuity equation then gives
\begin{equation}
\rho_{\textrm{mat}} \propto a^{-3}
\label{Eq.: Matter Energy Density "a" Proportionality}
\end{equation}
For the case of radiation, we have $w\!=\!\frac{1}{3}$, as $\rho\!=\!3P$. The continuity equation then gives
\begin{equation}
\rho_{\textrm{rad}} \propto a^{-4}
\label{Eq.: Radiation Energy Density "a" Proportionality}
\end{equation}
The physical interpretation of the extra ``$a$'' factor compared to the case of matter is that the frequency of the radiation is red-shifted as the Universe expands, hence it loses energy.\\
\indent In the case of $K\!=\!0$ (i.e. our Universe), the Friedmann equation yields the solution for the evolution of the scale factor as
\begin{equation}
a(t) \propto t^{\frac{2}{3(w+1)}}
\label{Eq.: a(t) Friedmann Equation Solution}
\end{equation}
for $w\!>\!-1$ and constant. Therefore, for matter domination we have
\begin{equation}
a(t) \propto t^{\frac{2}{3}}
\label{Eq.: a(t) Evolution for Matter Domination}
\end{equation}
and for radiation domination we have
\begin{equation}
a(t) \propto t^{\frac{1}{2}}
\label{Eq.: a(t) Evolution for Radiation Domination}
\end{equation}
\indent In the case of a cosmological constant (see \cref{Section: LambdaCDM}) (and also for inflation (see \cref{Subsection: Scalar Fields})), which is causing the expansion of the Universe to accelerate, beginning from around the time of the current epoch, the equation of state is $w\!=\!-1$. For this situation, \cref{Eq.: a(t) Friedmann Equation Solution} is not valid. Instead, \cref{Eq.: Continuity Equation,Eq.: Equation of State Parametrization} give that $\dot{\rho}\!=\!0$ (c.f. \cref{Eq.: Friedmann Equation (K=0)}). $H\!\equiv\!\frac{\dot{a}}{a}$ is constant and the evolution of the scale factor is
\begin{equation}
a(t) \propto e^{Ht}
\label{Eq.: a(t) Evolution for Dark Energy}
\end{equation}
\indent Another important quantity is the \textbf{Density Parameter}
\begin{equation}
\Omega(t) \equiv \frac{\rho(t)}{\rho_{\textrm{crit}}(t)}
\label{Eq.: Density Parameter Definition}
\end{equation}
with the \textbf{Critical Energy Density} being defined as $\rho_{\textrm{crit}}(t)\!\equiv\!3M_{P}^{2}H^{2}(t)$. The critical energy density is the total density that a spatially flat Universe would have for a given value of the Hubble parameter. We also define
\begin{equation}
\Omega-1 \equiv \frac{K}{(aH)^{2}} \equiv \frac{K}{\dot{a}^{2}}
\label{Eq.: Omega-1 Definition}
\end{equation}
Precise measurements of the geometry and energy density of the Universe made by the Planck spacecraft, when combined with other data \cite{Planck_2015_Cosmo._Results}, yield the current value
\begin{equation}
\Omega_{0} = 1.000 \pm 0.005
\label{Eq.: Omega_{0}}
\end{equation}
which is consistent with the time-independent value of $\Omega\!=\!1$. We therefore have that the energy density of the Universe is very close to the critical density ($\rho\!=\!\rho_{\textrm{crit}}$) and that the geometry of the Universe is very close to spatially flat ($K\!=\!0$). The observed energy density of the Universe is made-up of three components: Baryonic Matter (5\%), Dark Matter (26\%) and Dark Energy (69\%), the percentages indicating the approximate relative amount of each component.
\section{The Big Bang}
\label{Section: The Big Bang}
\begin{figure}[h!]
\begin{center}
\includegraphics[scale=0.13]{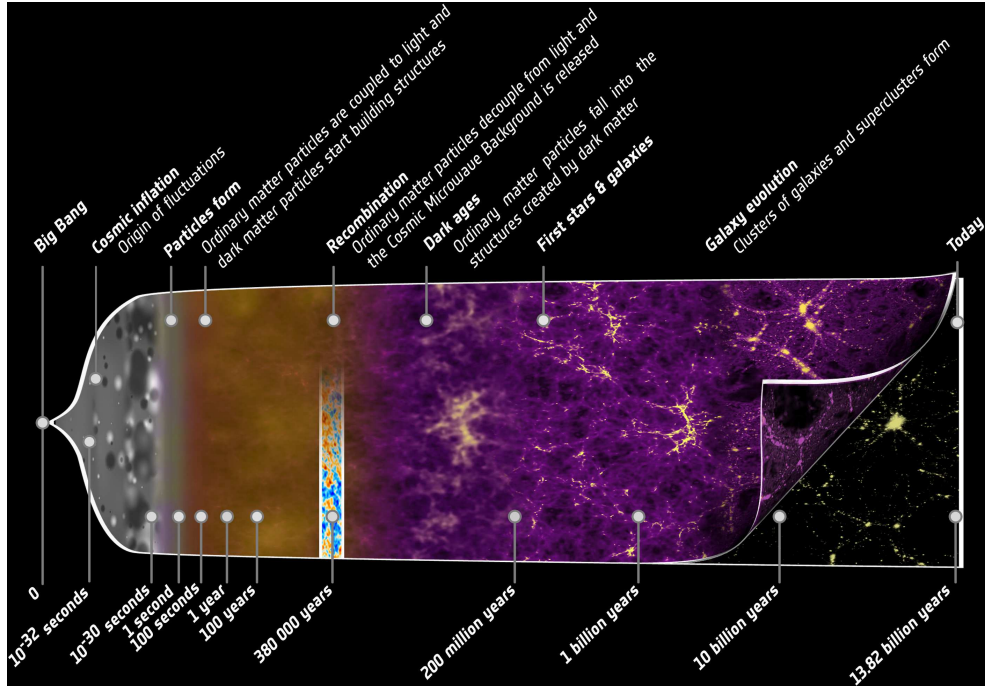}
\caption[Schematic diagram of history of universe]{A schematic diagram of the history of the Universe, taken from \cite{IMAGE:History_of_Universe}. The accelerated expansion that the Universe is currently experiencing, due to Dark Energy, is not depicted in this diagram.}
\label{Figure: History of Universe}
\end{center}
\end{figure}

In 1929, Edwin Hubble discovered what is known as \textbf{Hubble's Law}, which states that there exists a linear relationship between the distance of a galaxy from us and its recession velocity, with the constant of proportionality being \textbf{Hubble's Constant}, $H_{0}$, which has been observed using a variety of sources, with the combined data value \cite{Planck_2015_Cosmo._Results} being
\begin{equation}
H_{0} = 67.74 \pm 0.46\ \mbox{km s$^{-1}$\ Mpc$^{-1}$}
\label{Eq.: H_{0}}
\end{equation}
Therefore, the further away a galaxy is from our Milky Way, the faster it is traveling away from us. As a result of the Cosmological Principle, Hubble's Law implies that, over large scales, each galaxy is moving away from every other galaxy. If we consider this fact in reverse time, it is clear that galaxies will get ever closer to each other. At some point in the past, the Universe will be sufficiently dense, hot and energetic that the equations of GR will break down and leave us with a spacetime singularity. It is this point that we label as the beginning of our Universe, which occurred around $13.8\mbox{Gyr}$ ago. (It is not clear whether such a singularity actually existed. Given the extremely high energies, small scales and small time intervals that existed in the very early Universe, we are required to obtain a theory of Quantum Gravity in order to fully explain the physics of this time, which is not yet fully available to us.)\\
\indent We define the \textbf{Hot Big Bang}, hereafter referred to as HBB, to begin at the time when the reheating process from inflation (discussed in \cref{Subsection: Particle Production}) is complete. This process must leave us with a radiation-dominated Universe at the time of neutrino decoupling, when the temperature $T\!\sim\!1\mbox{MeV}$, with all the standard model particles present at that time being in thermal equilibrium. For a particular particle process to be in thermal equilibrium, the interaction rate between the particles has to be much larger than the Hubble parameter (expansion rate), $\Gamma\!\gg\!H$, so that the process has the ``time'' to occur, before the expansion of the Universe stifles the process.\\
\indent For a collection of particles in thermal equilibrium, the distribution function is
\begin{equation}
f(E) = \frac{1}{e^{\left(\frac{E-\mu}{T}\right)}\pm1}
\label{Eq.: Distribution Function}
\end{equation}
with $+$ for bosons and $-$ for fermions and where $E$ is the particle energy and $\mu$ is the \textbf{Chemical Potential}. For the case where the temperature $T$ is much larger than the mass and chemical potential of the particle species, which was applicable in the very early Universe, the distribution function simplifies to
\begin{equation}
f(p) \approx \frac{1}{e^{\frac{p}{T}}\pm1}
\label{Eq.: Distribution Function (T>>m,mu)}
\end{equation}
where $E\!\sim\!T\!\simeq p$, with $p$ being the momentum of the particle. This distribution function yields the blackbody distribution of photons. The number density is given by
\begin{align}
n &= \frac{g}{2\pi^{2}}\int_{0}^{\infty}f(p)p^{2}\,\mathrm{d}p
\\
\nonumber
\\
&= A\frac{\zeta(3)g}{\pi^{2}}T^{3}
\label{Eq.: Number Density of Blackbody Distribution}
\end{align}
where $A\!=\!1$ for bosons and $A\!=\!\frac{3}{4}$ for fermions and $g$ is the number of spin states of the particle (relativistic degrees of freedom). The energy density is given by
\begin{align}
\rho &= \frac{g}{2\pi^{2}}\int_{0}^{\infty}f(p)p^{3}\,\mathrm{d}p
\\
\nonumber
\\
&= B\frac{\pi^{2}g}{30}T^{4}
\label{Eq.: Energy Density of Blackbody Distribution}
\end{align}
where $B\!=\!1$ for bosons and $B\!=\!\frac{7}{8}$ for fermions. For the Universe, the energy density is given by the weighted sum of all the particles as
\begin{equation}
\rho = \frac{\pi^2}{30}g_{\ast}T^4
\label{Eq.: Universe Energy Density}
\end{equation}
with
\begin{equation}
g_{\ast} \equiv \sum_{i}^{\textrm{bosons}}g_{i} + \frac{7}{8}\sum_{j}^{\textrm{fermions}}g_{j}
\label{Eq.: g_{*} Definition}
\end{equation}
being the total number of spin states of all of the constituent particles (the effective number of relativistic degrees of freedom).\\
\indent As the temperature of the Universe falls due to the expansion, $\Gamma$ for a particular particle species will fall below $H$ at some time. Different particle species will start to fall out of thermal equilibrium at different times and thus \textbf{decouple} from each other.\\
\indent One of the big successes of the HBB is the agreement between analytical/numerical calculations and observation of \textbf{Big Bang Nucleosynthesis}, hereafter referred to as BBN. After an excess of ${}^1H$ (over anti-${}^1H$) had been created via Baryogenesis, BBN then produced the lightest nuclei that existed in the early Universe. By far the most abundant was ${}^4He$. ${}^2H$ and ${}^3He$ were produced in smaller quantities and a very small amount of ${}^7Li$ was also produced. The abundance of these nuclei depend strongly on the baryon number to photon number ratio
\begin{equation}
\eta \equiv \frac{n_{B}}{n_{\gamma}}
\label{Eq.: Baryon Number to Photon Number Ratio}
\end{equation}
which has a value of $\eta\!\sim\!10^{-10}$.\\
\indent Another great success of the HBB concerns the observed large-scale structure, \textbf{LSS}, of the Universe. The theory predicts an expanding Universe, which we observe as the \textbf{Hubble Flow}, with structure forming under the influence of gravity according to the laws of GR. On cosmologically small scales, matter (baryonic and dark matter) is bound gravitationally into galaxies, with there existing a hierarchy of structure, consisting of galaxies, galaxy groups, galaxy clusters, galaxy superclusters and galaxy filaments. The last of these structures form the boundaries with the voids of the Universe. The entire collection of components of the LSS is referred to as the \textbf{Cosmic Web}.\\
\indent Further evidence for a HBB model concerns the CMB. Due to the hot, dense early stages of the Universe, the HBB predicts that there should be a remnant blackbody radiation that was emitted when the Universe was young that should still be observable today. For a long time after BBN, photons were being continuously and rapidly scattered off of free electrons, due to Thomson scattering. As the temperature of the Universe continued to fall due to the expansion, there came a time when the free electrons became bound with the nuclei that was present, in a process known as \textbf{Recombination}. At this point, the photons fell out of thermal equilibrium with the electrons (the latter are now bound inside neutral atoms) and thus became decoupled, which allowed them to travel freely, following a geodesic. Therefore, there should exist a \textbf{last scattering surface}, corresponding to the time of recombination. (In reality, recombination did not occur instantaneously and so the last scattering surface actually has a thickness to it.)\\
\indent The CMB radiation was discovered in 1964 by Arno Penzias and Robert Wilson, a discovery which earned them the 1978 Nobel Prize for Physics. When combined with other data, the data obtained from the Planck spacecraft \cite{Planck_2015_Cosmo._Results} yields values of the corresponding blackbody temperature and redshift of the last scattering surface of
\begin{equation}
T_{0} = 2.718 \pm 0.021\ \mbox{K}
\label{Eq.: T_{0}}
\end{equation}
\begin{equation}
z_{\textrm{ls}} = 1089.90 \pm 0.23
\label{Eq.: z_{ls}}
\end{equation}
The redshift of the last scattering surface corresponds to a time of $\approx\!378000$ years after the birth of the Universe.\\
\indent A diagram depicting the main topics of our discussion so far in the history of the Universe is shown in \cref{Figure: History of Universe}.
\section{$\Lambda$CDM}
\label{Section: LambdaCDM}
\begin{figure}[h!]
\begin{center}
\includegraphics[scale=0.4]{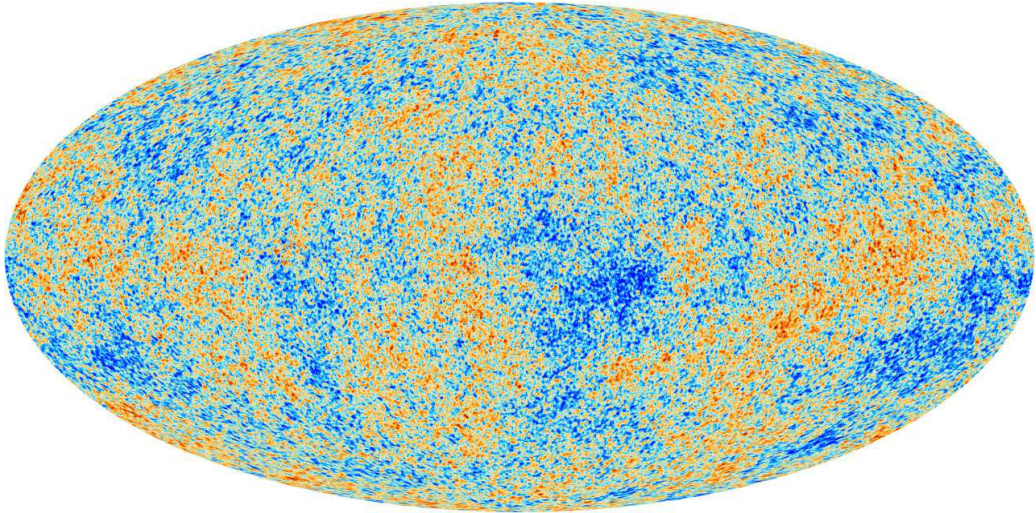}
\caption[Mollweide projection of CMB, as observed by Planck spacecraft]{A Mollweide projection of the CMB in galactic coordinates, as observed by the Planck spacecraft. Temperature perturbations are depicted, with $\frac{\Delta T}{T}\!\sim\!10^{-5}$. Taken from \cite{IMAGE:CMB_Mollweide_Planck}.}
\label{Figure: CMB Mollweide Planck}
\end{center}
\end{figure}

\begin{figure}[h!]
\begin{center}
\includegraphics[scale=0.6]{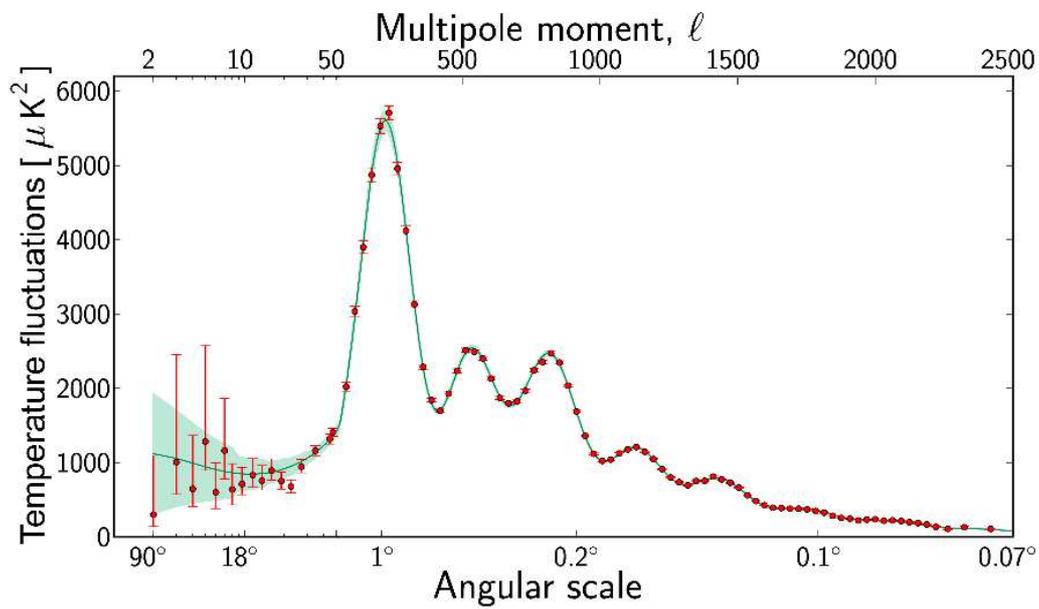}
\caption[Analysis of $\frac{\Delta T}{T}$ of CMB in spherical harmonics, as observed by Planck spacecraft]{Analysis of $\frac{\Delta T}{T}$ of the CMB in spherical harmonics, as observed by the Planck spacecraft. \textcolor{red}{Red Points}: Planck Data. \ \textcolor{Green}{Green Curve}: Best Fit of $\Lambda$CDM Model to Planck Data. \ \textcolor{CornflowerBlue}{Light Blue Shading}: Predictions of all Variations of $\Lambda$CDM Model that Best Agree with Planck Data. Taken from \cite{IMAGE:CMB_Power_Spectrum_Planck}.}
\label{Figure: CMB Power Spectrum Planck}
\end{center}
\end{figure}

All of what has been talked about so far forms part of the current standard model of Cosmology. However, to complete the model, we need to consider the effects of reionization. \textbf{Reionization} amounts to the partial ionization of the primordial gas from starlight produced by the first stars (the so-called Population III stars). The affect that reionization has on the matter in the Universe at early times is captured by using the \textbf{optical depth}
\begin{equation}
\tau(t) \equiv \sigma_{T}\int_{t}^{t_{0}}n_{e}(t)\,\mathrm{d}t
\label{Eq.: Optical Depth Definition}
\end{equation}
where $n_{e}$ is the number density of free electrons and $\sigma_{T}$ is the Thomson scattering cross-section. With this definition, the probability that a photon that is observed now that was emitted between the time of recombination and reionization, at a time $t$, has traveled freely is $e^{-\tau(t)}$, with the value being practically constant for emission times between recombination and reionization. When combined with other data, the data obtained from the Planck spacecraft \cite{Planck_2015_Cosmo._Results} yields values for the practically constant optical depth and the redshift at which reionization occurred (assuming it was sudden) of
\begin{equation}
\tau = 0.066 \pm 0.012
\label{Eq.: Observed Optical Depth Value}
\end{equation}
\begin{equation}
z_{\textrm{reion}} = 8.8^{+1.2}_{-1.1}
\label{Eq.: z_{reion}}
\end{equation}
\indent The final piece of the standard model concerns the small perturbations in the cosmic fluid density that existed in the early Universe. On large scales, the Universe is homogeneous and isotropic. However, on smaller scales, it is clear that this is violated, as the Universe contains planets, galaxies, empty space etc. Therefore, there must have existed some small differences in the density of the cosmic fluid at very early times, which then grew under the influence of gravity and the Hubble flow. These tiny perturbations present themselves in the CMB, as anisotropies in the average temperature of the microwave radiation. The Planck spacecraft made precise all-sky measurements of the CMB, which is displayed in \cref{Figure: CMB Mollweide Planck}.
\\
\indent A crucial concept in early Universe Cosmology is that of the \textbf{Primordial Curvature Perturbation}, labeled as $\boldsymbol{\zeta}$. This will be discussed more fully in \cref{Subsection: deltaN Formalism}. However, we will briefly discuss two aspects of it here. The \textbf{spectrum}, $\mathcal{P}_{\zeta}(k)$, of the curvature perturbation conveys how much power is in the perturbation as a function of scale $k$. We also have the \textbf{spectral index}, $n_{s}$, which tells us how the spectrum varies with scale $k$. (The subscript $s$ denotes that this is for a scalar perturbation.) The spectral index is defined as
\begin{equation}
n_{s}-1 \equiv \frac{\mathrm{d}\ln{(\mathcal{P}_{\zeta}(k))}}{\mathrm{d}\ln{(k)}}
\label{Eq.: Spectral Index Definition}
\end{equation}
with $n_{s}-1$ being referred to as the \textbf{tilt} of the spectrum. For constant spectral index, $\mathcal{P}_{\zeta}(k)\!\propto\!k^{n_{s}-1}$. $\mathcal{P}_{\zeta}$ is scale invariant for $n_{s}\!=\!1$. In general, $n_{s}\!=\!n_{s}(k)$, with cosmic inflation predicting $n_{s}$ close but not exactly equal to unity, so that the spectrum is approximately (but not quite) scale invariant. The Planck spacecraft made measurements of the spectrum at what is known as the \textbf{pivot scale}, which is $k_{0}\!\equiv\!0.002\ \mbox{Mpc$^{-1}$}$. When combined with other data, the data obtained from the Planck spacecraft \cite{Planck_2015_Cosmo._Results} yields values for the spectrum and spectral index of
\begin{equation}
\mathcal{P}_{\zeta}(k_{0}) = (2.142 \pm 0.049) \times 10^{-9}
\label{Eq.: Observed Spectrum Value}
\end{equation}
\begin{equation}
n_{s} = 0.9667 \pm 0.0040
\label{Eq.: Observed Spectral Index Value}
\end{equation}
\indent Lastly, we say a few words about the nature of the dark matter and the dark energy. From calculations and observations, the dark matter must be in the form of \textbf{CDM}, \textbf{C}old \textbf{D}ark \textbf{M}atter. A CDM particle is cold in that it has negligible (meaning non-relativistic) random motion. It has negligible interaction with other particles and also negligible self-interaction, hence must be non-baryonic, with its only real presence being observed via its gravitational effect on galactic dynamics. Regarding dark energy, the simplest realization is a \textbf{Cosmological Constant}, denoted by $\Lambda$. This is a spatially and temporally constant term that is added to Einstein's Field Equation, having the effect of negative pressure.\\
\indent Taking all of our discussion so far into account, we are presented with the $\boldsymbol{\Lambda}$\textbf{CDM} model as our \textbf{Concordance Model} of Cosmology, which contains just six independent parameters: $\boldsymbol{H_{0}}$, $\boldsymbol{\Omega_{\textrm{\textbf{B}}}}$, $\boldsymbol{\Omega_{\textrm{\textbf{CDM}}}}$, $\boldsymbol{\tau}$, $\boldsymbol{\mathcal{P}_{\zeta}(k_{0})}$ and $\boldsymbol{n_{s}}$. \cref{Figure: CMB Power Spectrum Planck} shows a plot of the agreement between the $\Lambda$CDM model and the data obtained from the Planck spacecraft.
\chapter{Particle Cosmology}
\label{Chapter: Particle Cosmology}
So far, we have been concentrating on the $\Lambda$CDM model of the Universe. We are now going to discuss \textbf{Particle Cosmology}, which is the field concerned with a particle physics description of the (early) Universe. The temperature and energy scales that were dominant during the very early Universe were such that Quantum Field Theory is required for a complete description of that period.
\section{Problems of Big Bang Cosmology}
\label{Section: Problems of Big Bang Cosmology}
As already discussed, the Big Bang Theory is hugely successful in explaining many of the properties and features that we observe in our Universe. However, it will become clear that it is not a sufficient theory on its own to explain everything that we observe. There are five main problems with the Big Bang and each of these will be discussed now. We will then see in \cref{Section: Cosmic Inflation} how the theory of Inflation can solve these problems.
\subsection{Horizon Problem}
\label{Subsection: Horizon Problem}
With the definition of the Hubble parameter being $H\!\equiv\!\frac{\dot{a}}{a}$, we define the \textbf{horizon} as the distance that light (information) can travel within one \textbf{Hubble Time} $H^{-1}$.\\
\indent We define the comoving \textbf{Particle Horizon} as
\begin{align}
x_{\textrm{PH}}(t) &\equiv \int_{0}^{t}\frac{1}{a(t)}\,\mathrm{d}t
\\
\nonumber
\\
&= \eta(t) - \eta(0)
\\
\nonumber
\\
x_{\textrm{PH}}(a) &= \int_{0}^{a}\frac{1}{a^{2}H}\,\mathrm{d}a
\label{Eq.: Particle Horizon Definition - Comoving Coordinates}
\end{align}
with the third line coming from the fact that we take $\dot{a}(t)\!>\!0$ (i.e. the Universe is expanding). This is the maximum distance that light (information) could have traveled since the birth of the Universe at $t,a\!=\!0$. Any two events that are separated by a distance of more than twice the particle horizon are out of casual contact. The Particle Horizon in physical coordinates is given by
\begin{align}
a(t)x_{\textrm{PH}}(t) &\equiv a(t)\int_{0}^{t}\frac{1}{a(t^{\prime})}\,\mathrm{d}t^{\prime}
\\
\nonumber
\\
a(t)x_{\textrm{PH}}(a) &= a(t)\int_{0}^{a}\frac{1}{a^{2}H}\,\mathrm{d}a
\label{Eq.: Particle Horizon Definition - Physical Coordinates}
\end{align}
\indent We make another definition, the comoving \textbf{Event Horizon}, as
\begin{align}
x_{\textrm{EH}}(t) &\equiv \int_{t}^{\infty}\frac{1}{a(t)}\,\mathrm{d}t
\\
\nonumber
\\
&= \eta(\infty) - \eta(t)
\\
\nonumber
\\
x_{\textrm{EH}}(a) &= \int_{a}^{\infty}\frac{1}{a^{2}H}\,\mathrm{d}a
\label{Eq.: Event Horizon Definition - Comoving Coordinates}
\end{align}
again with the third line coming from the fact that we take $\dot{a}(t)\!>\!0$. This is the maximum distance that light (information) can travel in the (infinite) future. An event that occurs at a spacetime point cannot influence an event at a future spacetime point if the latter is outside of the former's event horizon. The Event Horizon in physical coordinates is given by
\begin{align}
a(t)x_{\textrm{EH}}(t) &\equiv a(t)\int_{t}^{\infty}\frac{1}{a(t^{\prime})}\,\mathrm{d}t^{\prime}
\\
\nonumber
\\
a(t)x_{\textrm{EH}}(a) &= a(t)\int_{a}^{\infty}\frac{1}{a^{2}H}\,\mathrm{d}a
\label{Eq.: Event Horizon Definition - Physical Coordinates}
\end{align}
\indent In reality, \cref{Eq.: Particle Horizon Definition - Physical Coordinates,Eq.: Event Horizon Definition - Physical Coordinates} are definitions that use unrealistic boundary conditions, as $a\!=\!0,\infty$ is unphysical. Therefore, we usually calculate particle and event horizons between two well-defined limits
\begin{equation}
a(t)x_{\textrm{PH}}(a_{1},a) = a(t)\int_{a_{1}}^{a}\frac{1}{a^{2}H}\,\mathrm{d}a \hspace{5em} a(t)x_{\textrm{EH}}(a,a_{2}) = a(t)\int_{a}^{a_{2}}\frac{1}{a^{2}H}\,\mathrm{d}a
\label{Eq.: Particle & Event Horizon - a_{1},a_{2},a}
\end{equation}
with $a\!\gg\!a_{1}$ or $a\!\ll\!a_{2}$.\\
\indent The Horizon Problem arises when we consider how the horizon of various places in the observable Universe has varied over the lifetime of the Universe and then compare this with observation. Our observable Universe is statistically homogeneous and isotropic on cosmological scales. One manifestation of this is that of the large isotropy that exists in the CMB. Over the entire sky, the temperature of the CMB is the same to within 1 part in $\sim\!10^{5}$. Therefore, it is very safe to assume that all parts of the CMB were in thermal equilibrium with each other at the time of last scattering and thus were in casual contact with each other. However, when we consider two parts of the CMB that are separated by more than a few degrees across the sky, we find that the particle horizon's of these two parts do not overlap. Therefore, they have never been in casual contact with each other, as there has not been enough time for light (information) to travel between the two parts given the age of the Universe. The Horizon Problem is thus to explain how each part of the observable Universe is so extremely statistically similar to every other part, given that the vast majority of parts have never been in casual contact with each other.
\subsection{Flatness Problem}
\label{Subsection: Flatness Problem}
From looking at \cref{Eq.: Omega-1 Definition} regarding the Density Parameter, we see that as time evolves from the birth of the Universe, i.e. as $\dot{a}^{2}$ decreases rapidly, $\Omega$ grows rapidly, away from a value of 1. We observe for the Density Parameter the current value $\Omega_{0}\!=\!1.000 \pm 0.005$ (\cref{Eq.: Omega_{0}}), which is very close to 1, i.e. a spatially flat Universe. Therefore, in the past, the value must have been even more precisely close to 1, with it having had a value of 1 to a precision of at least $\sim\!10^{-62}$ around the Planck time. This Flatness Problem is thus a fine-tuning problem for the initial conditions of the Universe.\\
\indent The Flatness Problem, when looked at from a slightly different viewpoint, can also be regarded as an Age Problem. If the value of $\Omega$ had been only slightly larger than it appears to have been, then the Universe would have been of a sufficient density that it would have collapsed at a time much sooner than the age of our observable Universe. If, on the other hand, the value of $\Omega$ had been only slightly smaller than it appears to have been, then the Universe would have expanded too quickly for stars and galaxies to have formed the LSS that we observe today. Therefore, the fine-tuning problem can be regarded as an Age Problem, in that how has our observable Universe become as old as it is?
\subsection{Relic Problem}
\label{Subsection: Relic Problem}
In contrast to the two problems already mentioned, the Relic Problem is not specific to a Hot Big Bang model of the Universe, but rather something that has to be considered within the total picture of Particle Cosmology. There exist a large variety of particle physics models and applications of these to Cosmology that have the ability to produce many types of relic; particles and other components that contradict established theory and/or observation, such as gravitinos coming from SUGRA and moduli coming from string theory. Another example, which was the subject of the original work of Alan Guth on inflation \cite{Guth:Original_Inflation_Paper}, considered the effect of magnetic monopole creation during a GUT phase transition in the very early Universe. Any monopoles that are produced in the very early Universe must neither spoil the success of BBN nor be in direct tension with observations of such particles. An abundant creation of monopoles in the early Universe would have the affect of overclosing the Universe. Therefore, we must have the situation where either any such model of the very early Universe does not produce too many relics or there must exist some mechanism such that these are not in conflict with our observations, such as having them be diluted away somehow.
\subsection{Baryon Asymmetry Problem}
\label{Subsection: Baryon Asymmetry Problem}
It is clear that there exist many more particles than anti-particles in the Universe. However, there is no explicit mechanism in the Hot Big Bang model itself that can account for this \textbf{Baryon Asymmetry} and according to the model, identically equal amounts of matter and anti-matter should have been created in the early Universe.
\subsection{Initial Perturbation/Structure Problem}
\label{Subsection: Initial Perturbation/Structure Problem}
Lastly, there is the issue of the creation of the LSS of the observable Universe. Assuming just an expanding FLRW Universe, there arises the question of how the observable structure in the Universe came into existence, as this expanding spacetime is exactly homogeneous and isotropic. Although our observable Universe is highly homogeneous and isotropic on cosmological scales, it is clear that this must break down at some level, in order to facilitate the existence of stars, galaxies and humans! In addition, when we look at the CMB, we see tiny perturbations in the temperature across the sky. What mechanism was responsible for producing the associated density perturbations in the very early Universe?
\section{Cosmic Inflation}
\label{Section: Cosmic Inflation}
We now discuss one of the key areas of Particle Cosmology in the 21\textsuperscript{st} century and indeed the main topic of work in this thesis: Cosmic Inflation. Firstly, let us define inflation. With regard to the scale factor $a(t)$, inflation is defined as any period of spacetime expansion in which we have
\begin{equation}
\ddot{a} > 0
\label{Eq.: Inflation Definition - 1}
\end{equation}
We therefore have a period of ``repulsive gravity''. Another, extremely important, definition of inflation is
\begin{equation}
\frac{\mathrm{d}}{\mathrm{d}t}\frac{H^{-1}}{a} < 0
\label{Eq.: Inflation Definition - 2}
\end{equation}
This defines inflation as a period of expansion in which the comoving Hubble length is decreasing. Another definition of inflation, in terms of the time evolution of the Hubble parameter, is
\begin{equation}
-\frac{\dot{H}}{H^{2}} < 1
\label{Eq.: Inflation Definition - 3}
\end{equation}
As this inequality becomes stronger, $H$ becomes more and more constant during the period of inflation, with the expansion becoming more and more exponential, i.e. $a\!\propto\!e^{Ht}$. It should be noted that \cref{Eq.: Inflation Definition - 1,Eq.: Inflation Definition - 2,Eq.: Inflation Definition - 3} are all equivalent.\\
\indent If we now include GR in our discussion, we have one final definition of inflation. Using the (Friedmann) Acceleration Equation, \cref{Eq.: Acceleration Equation}, we have
\begin{equation}
\rho + 3P < 0
\label{Eq.: Inflation Definition - 4}
\end{equation}
This implies that we require negative enough pressure (``repulsive gravity''), $P\!<\!-\frac{\rho}{3}$, in order to achieve a period of inflation.\\
\indent Now we will discuss how a period of inflation can solve the five problems of an isolated Big Bang Cosmology mentioned above. Firstly, the Horizon Problem. The solution to this problem is to say that the entire observable Universe \textit{used to be} inside the horizon, prior to the end of inflation. During inflation however, different scales of relevance to our observable Universe were ``stretched'' to \textit{outside} of the horizon at different times, remembering that one of the definitions of inflation is that of a decreasing comoving Hubble length, \cref{Eq.: Inflation Definition - 2}. At later times, after inflation, these scales then started to re-enter the horizon, again at different times for different scales.\\
\indent Now, the Flatness Problem. From looking at \cref{Eq.: Omega-1 Definition} regarding the Density Parameter, we can see that the RHS will tend towards 0 during inflation. The reason for this, is that during inflation $\ddot{a}\!>\!0$ by definition. Therefore, $\dot{a}^{2}$ will grow rapidly, which will have the affect of rapidly driving the value of $\Omega$ towards 1 throughout the entire period of inflation. However, for the problem to be solved sufficiently, we require that the entire observable Universe is well within the horizon, $aH\!\ll\!H_{0}$, a long time before cosmological scales exit the horizon during inflation.\footnote{We assume the current value of $a$ being $a_{0}\!=\!1$.}\\
\indent We now discuss the solution to the Relic Problem. Any relics that exist before or during inflation will be rapidly diluted away by the inflation. They will quickly become non-interacting as their number density decreases, due to the rapid expansion, as the interaction rates quickly fall below $H$. However, we may still have relics that are produced thermally after the end of inflation. This will depend mainly on the reheat temperature of the Universe after the end of inflation. Therefore, to avoid specific types of relics that may be an issue, for example in how they spoil BBN, any particular model of inflation must have a reheat temperature low enough so as to not produce an unacceptable amount of such relics. Alternatively, a second period of inflation, known as Thermal Inflation (see \cref{Chapter: A New Thermal Inflation Model}), can affectively eradicate them.\\
\indent Let us now discuss the Baryon Asymmetry Problem. Inflation can accommodate Baryogenesis mechanisms that produce the asymmetry between matter and anti-matter. In order to achieve Baryogenesis, we require the following conditions, known as the \textbf{Sakharov Conditions}
\begin{enumerate}[label=\Alph*)]
\item Violation of \textbf{B} (Baryon Number) conservation
\item Violation of \textbf{CP} symmetry
\item Absence of thermal equilibrium
\end{enumerate}
It is the last of these conditions that can be achieved during a period of inflation.\\
\indent Lastly, regarding the problem of how initial perturbations were seeded that gave rise to the LSS that we observe, inflation naturally produces such perturbations that can indeed give rise to the structure that we see in our observable Universe. This is because inflation generates the required primordial curvature perturbation, which we will now discuss. (A much more detailed discussion is found in \cref{Subsection: Particle Production}).
\subsection{$\delta N$ Formalism}
\label{Subsection: deltaN Formalism}
We now go on to define a crucial quantity known as the Primordial Curvature Perturbation, $\zeta$. We will then see how we can use the so-called $\delta N$ formalism to calculate this. Let us consider a coordinate gauge in which the spatial threads are comoving and the temporal slices are such that each one has a uniform energy density. The spatial part of the spacetime metric is
\begin{equation}
g_{ij} = a^{2}(\textbf{x},t)\gamma_{ij}(\textbf{x})
\label{Eq.: g_{ij}}
\end{equation}
where we have
\begin{equation}
a(\textbf{x},t) \equiv a(t)e^{\zeta(\textbf{x},t)}
\label{Eq.: a(x,t) Definition}
\end{equation}
and
\begin{equation}
\gamma_{ij}(\textbf{x}) \equiv \left(\textbf{I}e^{2h(\textbf{x})}\right)_{ij}
\label{Eq.: gamma_{ij}(x) Definition}
\end{equation}
where $h(\textbf{x})$ is the primordial tensor perturbation. We therefore have as our definition of $\zeta$
\begin{equation}
\zeta(\textbf{x},t) \equiv \ln{\left(\frac{a(\textbf{x},t)}{a(t)}\right)}
\label{Eq.: zeta Definition}
\end{equation}
\indent Now let us consider a generic coordinate gauge, in which we still have a comoving threading but a generic slicing, as opposed to one in which each slice has a uniform energy density. We have
\begin{equation}
\tilde{g}_{ij} = \tilde{a}^{2}(\textbf{x},t)\tilde{\gamma}_{ij}(\textbf{x})
\label{Eq.: tilde{g}_{ij}}
\end{equation}
where
\begin{equation}
\tilde{a}(\textbf{x},t) \equiv a(t)e^{\psi(\textbf{x},t)}
\label{Eq.: tilde{a}(x,t) Definition}
\end{equation}
and
\begin{equation}
\tilde{\gamma}_{ij}(\textbf{x}) \equiv \left(\textbf{I}e^{2\tilde{h}(\textbf{x})}\right)_{ij}
\label{Eq.: tilde{gamma}_{ij}(x) Definition}
\end{equation}
At any given value of $t$, the scale factors $a(\textbf{x},t)$ and $\tilde{a}(\textbf{x},t)$ differ only because of the difference in the time coordinate of the two spacetime slices. Therefore, in order to maintain generality, we will drop the \textasciitilde\ on the scale factor. We can define a quantity called the e-folding number as
\begin{equation}
N \equiv \ln{\left(\frac{a_{2}}{a_{1}}\right)}
\label{Eq.: N Definition}
\end{equation}
which is the number of exponential expansions of the Universe between when the scale factor is $a_{1}$ and when it is $a_{2}$. As $H\!\equiv\!\frac{\dot{a}}{a}$, this can also be expressed as
\begin{equation}
N = \int_{t_{1}}^{t_{2}}H\,\mathrm{d}t
\label{Eq.: N - Hubble Times}
\end{equation}
and is thus sometimes called the number of Hubble times. The difference in the number of e-foldings between any two generic spacetime slices is given by
\begin{align}
\delta N(\textbf{x}) &= \delta\int_{t_{1}}^{t_{2}}\frac{1}{a(\textbf{x},t)}\frac{\mathrm{d}a(\textbf{x},t)}{\mathrm{d}t}\,\mathrm{d}t
\\
\nonumber
\\
&= \psi(\textbf{x},t_{2}) - \psi(\textbf{x},t_{1})
\label{Eq.: N - 2 Generic Spacetime Slices}
\end{align}
We will define what we will call a flat spacetime slice as the one where
\begin{equation}
\psi(\textbf{x},t) = 0
\label{Eq.: Flat Slice}
\end{equation}
We define the term $\delta N(\textbf{x},t)$ to denote the number of e-foldings between the flat slice and a slice of uniform energy density at time $t$. Therefore, we reach what is called the $\delta N$ formalism
\begin{equation}
\zeta(\textbf{x},t) = \delta N(\textbf{x},t)
\label{Eq.: deltaN Formalism}
\end{equation}
which thus allows us to calculate the primordial curvature perturbation from calculating the difference in the number of e-foldings of expansion between a flat slice and a latter uniform energy density slice, which is extremely useful when considering the perturbation that is produced from a particular inflation model.
\subsection{Scalar Fields}
\label{Subsection: Scalar Fields}
Inflation models most often assume that the content of the Universe during and immediately after inflation is dominated by the presence of one or more scalar fields, $\phi_{n}$. A scalar field is homogeneous (being as it is homogenized by inflation) and will behave like a perfect fluid, as its stress is isotropic.\\
\indent Let us consider the contents of the Universe to simply be a single scalar field, $\phi$. The action that governs this scenario is
\begin{equation}
S = \int\sqrt{-g}\left(\frac{1}{2}M_{P}^{2}R + \mathcal{L}\right)\,\mathrm{d^{4}}\!x
\label{Eq.: GR Action with Scalar Field}
\end{equation}
where $g$ is the determinant of the metric tensor $g_{\mu\nu}$, $R$ is the Ricci scalar and $\mathcal{L}$ is the Lagrangian density of the scalar field, which is
\begin{equation}
\mathcal{L} = -\frac{1}{2}\partial^{\mu}\phi\partial_{\mu}\phi - V(\phi)
\label{Eq.: Lagrangian Density of Scalar Field}
\end{equation}
where the first term is called the kinetic term and the second term is called the scalar field potential. By using the action principle, $\delta S\!=\!0$, we can obtain the Energy-Momentum tensor for the scalar field, which is
\begin{equation}
T_{\mu\nu} = -2\frac{\partial\mathcal{L}}{\partial g^{\mu\nu}} + g_{\mu\nu}\mathcal{L}
\label{Eq.: Energy-Momentum Tensor of Scalar Field - Lagrangian Density}
\end{equation}
Substituting \cref{Eq.: Lagrangian Density of Scalar Field} into here gives
\begin{equation}
T^{\mu}_{\nu} = \partial^{\mu}\phi\partial_{\nu}\phi - \delta^{\mu}_{\nu}\left(\frac{1}{2}\partial^{\alpha}\phi\partial_{\alpha}\phi + V(\phi)\right)
\label{Eq.: Energy-Momentum Tensor of Scalar Field}
\end{equation}
The ``$00$'' (``$tt$'') component of this Energy-Momentum tensor gives the energy density for a homogeneous scalar field, which is
\begin{equation}
T^{0}_{0} = \rho = \frac{1}{2}\dot{\phi}^{2} + V(\phi)
\label{Eq.: Scalar Field Energy Density}
\end{equation}
and the $T^{1}_{1}\!=\!T^{2}_{2}\!=\!T^{3}_{3}$ components give the pressure for a homogeneous scalar field, which is
\begin{equation}
P = \frac{1}{2}\dot{\phi}^{2} - V(\phi)
\label{Eq.: Scalar Field Pressure}
\end{equation}
\indent For simplicity, let us continue to assume that the cosmological fluid during and immediately after inflation consists principally of just the one scalar field, $\phi$. If the kinetic energy density term of this field, $\frac{1}{2}\dot{\phi}^{2}$, is small, i.e. if the field varies slowly, or not at all, then we will have a situation in which
\begin{equation}
P \approx -\rho
\label{Eq.: P approx -rho}
\end{equation}
Therefore, the equation of state will be
\begin{equation}
w = \frac{P}{\rho} \approx -1
\label{Eq.: w approx -1}
\end{equation}
and we will therefore have a period of (quasi-de Sitter) inflation, in which the scale factor goes (nearly) like $a(t)\!\propto\!e^{Ht}$ (see \cref{Subsection: FLRW Universe}). As $\phi$ is the component in the cosmological fluid that is responsible for driving inflation, we call the field, as well as its associated particle within the context of QFT, the \textbf{Inflaton}.
\subsection{Particle Production}
\label{Subsection: Particle Production}
We now briefly discuss the method by which we actually obtain an energy density perturbation from a scalar field. During inflation, the inflaton field $\phi$ will naturally acquire quantum fluctuations, $\delta\phi$, as a direct result of the uncertainty principle. We assume that $\phi$ is in a vacuum state and so we have 0 particles as the eigenvalue of the number operator. The field equation of the first-order perturbation $\delta\phi$ is
\begin{equation}
\ddot{\delta\phi}_{\textbf{k}} + 3H\dot{\delta\phi}_{\textbf{k}} + \left(\frac{k}{a}\right)^{2}\delta\phi_{\textbf{k}} + V^{\prime\prime}\delta\phi_{\textbf{k}} = 0
\label{Eq.: Perturbation Field Equation}
\end{equation}
where $\phi_{\textbf{k}}$ is the Fourier transform of $\phi(\textbf{x})$
\begin{equation}
\phi_{\textbf{k}} = \frac{1}{(2\pi)^{\frac{2}{3}}}\int\phi(\textbf{x})e^{-i\textbf{k}\cdot\textbf{x}}\,\mathrm{d^{3}}\!x
\label{Eq.: phi_{k} as Fourier Transform of phi(x)}
\end{equation}
We concern ourselves only with a light field. Given this, we have
\begin{equation}
V^{\prime\prime} \ll \left(\frac{k}{a}\right)^{2}
\label{Eq.: Light Field for Perturbation Equation}
\end{equation}
Therefore, we have the field equation
\begin{equation}
\ddot{\delta\phi}_{\textbf{k}} + 3H\dot{\delta\phi}_{\textbf{k}} + \left(\frac{k}{a}\right)^{2}\delta\phi_{\textbf{k}} \simeq 0
\label{Eq.: Perturbation Field Equation - Light Field}
\end{equation}
As the key interest in this discussion is the time around horizon exit, let us concentrate on this and so let us set
\begin{equation}
H = H_{\ast}
\label{Eq.: H=H*}
\end{equation}
where $H_{\ast}$ is the scale-independent constant value of $H$ at around the time of horizon exit. The use of the constant value $H_{\ast}$ as opposed to the scale-dependent value $H_{k}$, for when the scale $k$ exits the horizon, is an approximation, used to simplify the derivation here. The approximation is valid, as in quasi-de Sitter inflation we have $H\!\approx\!constant$, i.e. $H$ varies extremely slowly with scale. We now transform from cosmic time $t$ to conformal time
\begin{equation}
\eta = -\frac{1}{aH_{\ast}}
\label{Eq.: Conformal Time - Perturbation Spectrum Work}
\end{equation}
and also consider instead the comoving field perturbation
\begin{equation}
\varphi \equiv a\delta\phi
\label{Eq.: Comoving Perturbation Definition}
\end{equation}
\Cref{Eq.: Perturbation Field Equation - Light Field} now becomes
\begin{equation}
\frac{\mathrm{d^{2}}\varphi_{\textbf{k}}(\eta)}{\mathrm{d}\eta^{2}} + \omega_{k}^{2}(\eta)\varphi_{\textbf{k}}(\eta) = 0
\label{Eq.: Perturbation Field Equation - Comoving Perturbation}
\end{equation}
where
\begin{equation}
\omega_{k}^{2}(\eta) = k^{2} - \frac{2}{\eta^{2}}
\label{Eq.: Wavenumber - Perturbation Spectrum Work}
\end{equation}
where we have assumed $a(t)\!\propto\!e^{Ht}$. We have thus obtained a harmonic oscillator scenario. Following the usual procedure of QFT, we will now promote variables to operators and quantize this harmonic oscillator. We express the comoving perturbation in terms of Fourier components as
\begin{equation}
\hat{\varphi}_{\textbf{k}}(\eta) = \varphi_{k}(\eta)\hat{a}(\textbf{k}) + \varphi_{k}^{\ast}(\eta)\hat{a}^{\dagger}(-\textbf{k})
\label{Eq.: Comoving Perturbation in Fourier Components}
\end{equation}
where $\hat{a}^{\dagger}$ and $\hat{a}$ are creation and annihilation operators respectively, that satisfy
\begin{equation}
\left[\hat{a}(\textbf{k}),\hat{a}^{\dagger}(\textbf{k}^{\prime})\right] = (2\pi)^{3}\delta^{3}\!\left(\textbf{k}-\textbf{k}^{\prime}\right)
\label{Eq.: Annih. Creat. Comm. - Perturbation Spectrum Work}
\end{equation}
and
\begin{equation}
\left[\hat{a}(\textbf{k}),\hat{a}(\textbf{k}^{\prime})\right] = 0
\label{Eq.: Annih. Comm. - Perturbation Spectrum Work}
\end{equation}
We assume that the vacuum state is that of the Bunch-Davies vacuum. This vacuum state is the ground state of the system within a curved spacetime background. For very early times, as $\eta\!\to\!-\infty$, the Bunch-Davies vacuum gives the initial condition
\begin{equation}
\varphi_{k}(\eta) = \frac{1}{\sqrt{2k}}e^{-ik\eta}
\label{Eq.: varphi_{k}(eta) Initial Condition (Bunch-Davies vacuum)}
\end{equation}
As this is for very early times, it also corresponds to very small wavelengths and so corresponds to the Minkowski vacuum, i.e. the vacuum of the system within a flat spacetime background. The solution for $\varphi_{k}(\eta)$ is
\begin{equation}
\varphi_{k}(\eta) = \frac{(k\eta-i)}{k\eta}\frac{e^{-ik\eta}}{\sqrt{2k}}
\label{Eq.: varphi_{k}(eta) Solution}
\end{equation}
which is the mode function for the Bunch-Davies vacuum. The spectrum is given by
\begin{equation}
\left\langle\hat{\varphi}_{\textbf{k}}\hat{\varphi}_{\textbf{k}^{\prime}}\right\rangle = \frac{1}{4\pi k^{3}}\mathcal{P}_{\varphi}(k)\delta^{3}\!\left(\textbf{k}+\textbf{k}^{\prime}\right)
\label{Eq.: Spectrum - Comoving Perturbation}
\end{equation}
Substituting \cref{Eq.: Comoving Perturbation in Fourier Components} and the commutation relations \cref{Eq.: Annih. Creat. Comm. - Perturbation Spectrum Work,Eq.: Annih. Comm. - Perturbation Spectrum Work} into \cref{Eq.: Spectrum - Comoving Perturbation} yields
\begin{equation}
\mathcal{P}_{\varphi}(k,\eta) = \frac{k^{3}}{2\pi^{2}}|\varphi_{k}(\eta)|^{2}
\label{Eq.: Comoving Perturbation Spectrum}
\end{equation}
Substituting \cref{Eq.: varphi_{k}(eta) Solution} into this, dividing by $a^{2}$ (to return back to the $\delta\phi$ perturbations) and evaluating it a few Hubble times after horizon exit gives the time-independent result
\begin{equation}
\mathcal{P}_{\delta\phi}(k) = \left(\frac{H_{\ast}}{2\pi}\right)^{2}
\label{Eq.: Perturbation Spectrum}
\end{equation}
which is the Hawking Temperature for de Sitter spacetime. See \cite{Bunch_&_Davies:Quan._Field_Theory_de_Sit._Space:Renorm._Point_Splitting} for the original derivation of this result.\\
\indent Now let us consider a time well after horizon exit. The solution for $\varphi_{k}(\eta)$, \cref{Eq.: varphi_{k}(eta) Solution}, tends to
\begin{equation}
\varphi_{k}(\eta) \to \frac{-i}{k\eta\sqrt{2k}}
\label{Eq.: varphi_{k}(eta) Solution well after Horizon Exit}
\end{equation}
i.e. a purely imaginary solution. \Cref{Eq.: Comoving Perturbation in Fourier Components} now becomes
\begin{equation}
\hat{\varphi}_{\textbf{k}}(t) = \varphi_{k}(t)\left(\hat{a}(\textbf{k}) - \hat{a}^{\dagger}(-\textbf{k})\right)
\label{Eq.: Comoving Perturbation in Fourier Components well after Horizon Exit}
\end{equation}
Therefore, we can see that, before horizon exit, the perturbation $\varphi_{\textbf{k}}$ of the scalar field $\phi$ is a quantum object. However, well after horizon exit, the perturbation has become an almost scale-invariant classical perturbation. The classical perturbation is conserved whilst outside the horizon \cite{Lyth_et_al.:Gen._Proof_Cons._Cur._Pert.}.\\
\indent After the end of inflation, there must exist a mechanism for transferring the energy density of $\phi$ into the components that will initiate the Hot Big Bang. This mechanism is called \textbf{Reheating}. At around the time of horizon entry, the classical perturbation starts to oscillate and thus we have a particle interpretation for $\phi$ within the context of QFT. Reheating can be sudden or take some cosmic time to complete and is complete when we have a cosmic fluid whose components are radiation (i.e. relativistic), which are all in thermal equilibrium with each other and that this fluid is the initiation of the Hot Big Bang. Reheating is typically complete when the Hubble parameter $H$ has fallen to the same order as the decay rate of the $\phi$ field
\begin{equation}
H \sim \Gamma_{\phi}
\label{Eq.: H order Gamma_{phi}}
\end{equation}
The temperature at the point where reheating is complete is known as the \textbf{Reheat Temperature} and is given by
\begin{equation}
T_{\textrm{reh}} \sim \sqrt{M_{P}\Gamma_{\phi}}
\label{Eq.: Reheat Temperature}
\end{equation}
\indent There also exists the possibility of having a period of \textbf{Preheating}. This is where most of the energy density of the inflaton decays immediately (explosively) into radiation, due to non-perturbative effects. However, preheating is typically incomplete. Therefore, the final stages of inflaton decay are perturbative. If $\Gamma_{\phi}\!\ll\!H_{\ast}$, then preheating products are irrelevant, as the energy density of the Universe becomes dominated by the oscillating inflaton again. However, if $\Gamma_{\phi}\!\lesssim\!H_{\ast}$, then the Hot Big Bang will begin after preheating.
\chapter{A New Thermal Inflation Model}
\label{Chapter: A New Thermal Inflation Model}
\section{Thermal Inflation}
\label{Section: Thermal Inflation}
\begin{figure}[h!]
\begin{center}
\includegraphics[scale=1.3]{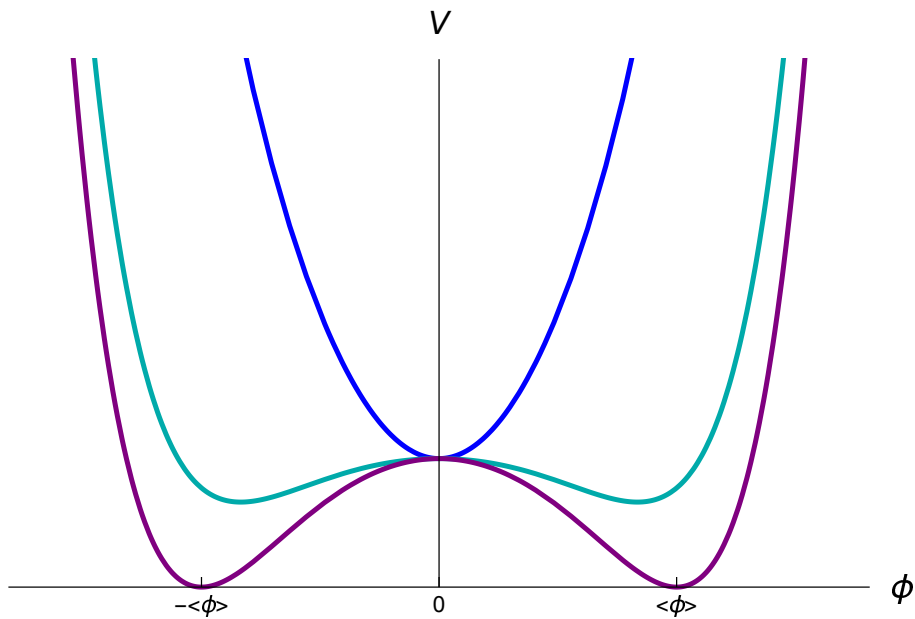}
\caption[Potentials in Thermal Inflation scenario]{\textcolor{blue}{Blue}: $T\!>\!\frac{m_{0}}{\sqrt{2}g}$ \ \ \textcolor{Emerald}{Cyan}: $T\!<\!\frac{m_{0}}{\sqrt{2}g}$ \ \ \textcolor{violet}{Purple}: $T\!=\!0$}
\label{Figure: Thermal Inflation Potential}
\end{center}
\end{figure}

Thermal Inflation \cite{Lyth_&_Stewart:Cos._TeV_Mass_Higgs_Field_break._GUT_Gauge._Sym. , Lyth_&_Stewart:Therm._Inf._Moduli_Prob. , Barreiro_et_al.:Aspects_Therm._Inf.:_Finite_Temp._Pot._Top._Defects , Asaka_&_Kawasaki:Cos._Moduli_Problem_Therm._Inf._Models} is a brief period of inflation (lasting about 10 e-folds) that could have occurred after a period of prior primordial inflation. It occurs due to finite-temperature effects arising from a coupling between a thermal waterfall field and the thermal bath created from the partial or complete reheating from the prior inflation. If we start with a zero-temperature scalar field theory, we can calculate the affect that placing the system in a thermal bath at temperature $T$ has on the theory by introducing an interaction term in the form of a 1-loop correction. After calculating the appropriate variables within the context of thermal field theory, we can take the high-$T$ approximation of the correction, which gives a thermal contribution $V_{\textrm{T}}\!\simeq\!g^{2}T^{2}\phi^{2}$ to the effective potential $V_{\textrm{eff}}$, where $g$ is the coupling constant of the interaction between $\phi$ and the thermal bath. This results in a thermal correction to the effective mass of $m_{\textrm{T}}^{2}\!\simeq\!g^{2}T^{2}$. Within the context of statistical mechanics, the interpretation of $V_{\textrm{eff}}$ is that of the free energy of $\phi$ when the field is in thermal equilibrium with the thermal bath at temperature $T$, with the minima in $V_{\textrm{eff}}$ defining the equilibrium states, with $\left\langle\phi\right\rangle$ representing the thermal average, as opposed to the vacuum expectation value (VEV).\\
\indent Let us take the following potential
\begin{equation}
V(\phi,T) = V_{0} + \left(g^{2}T^{2}-\frac{1}{2}m_{0}^{2}\right)\phi^{2} + \lambda\frac{\phi^{6}}{M_{P}^{2}}
\label{Eq.: Example Thermal Inflation Potential}
\end{equation}
Initially, the temperature of the Universe will be sufficiently high that the temperature term in the brackets will be greater than the mass term in the brackets. This will have the affect of holding the $\phi$ field at $\phi\!=0$. When the energy density of the Universe falls below the value of $V_{0}$ in the thermal inflation potential, the $V_{0}$ term will come to dominate the energy density of the Universe and thermal inflation will begin. It will continue until very shortly after the point when the temperature term has become smaller than the mass term, at which point spontaneous symmetry breaking will occur and so $\phi$ will start to roll down the potential towards either the positive or negative VEV.\footnote{For ease of visualization and calculation, it is usually assumed that a scalar field rolls down to the positive VEV.} The shape of the potential in this scenario is displayed in \cref{Figure: Thermal Inflation Potential}.
\\
\indent This scenario is quite general and would not be particularly unexpected in the early Universe. However, Thermal Inflation was originally proposed as a solution to the moduli problem \cite{Lyth_&_Stewart:Cos._TeV_Mass_Higgs_Field_break._GUT_Gauge._Sym. , Lyth_&_Stewart:Therm._Inf._Moduli_Prob.}. Moduli are scalar fields that arise in string theory. They are flaton fields, which are flat directions in SUSY. These have no tree-level terms in the potential from SUSY and they get a mass term from SUSY breaking (they do not have a quartic (self-interaction) term). Flaton fields have nearly flat potentials (with $\sqrt{V^{\prime\prime}}$ being the relevant quantity, with a prime indicating the (partial) derivative of $V$ with respect to the flaton field) and large VEVs, $\sim\!M_{P}$. The problem is that when inflation ends and a modulus starts to oscillate around its large VEV, the oscillations will also be very large and the energy density of the moduli will start to dominate the energy density of the Universe. This has the affect of creating an abundance of moduli particles that are long-lived and do not decay prior to BBN, thus creating unwanted relics. Thermal Inflation alleviates this problem by diluting away the moduli during the period of inflation. They are not re-created in abundance after thermal inflation, as the typical energy scales involved after thermal inflation are much lower than those typical of a prior period of inflation.
\section{The Model}
\label{Section: The Model}
\begin{figure}[h!]
\begin{center}
\includegraphics[scale=1.3]{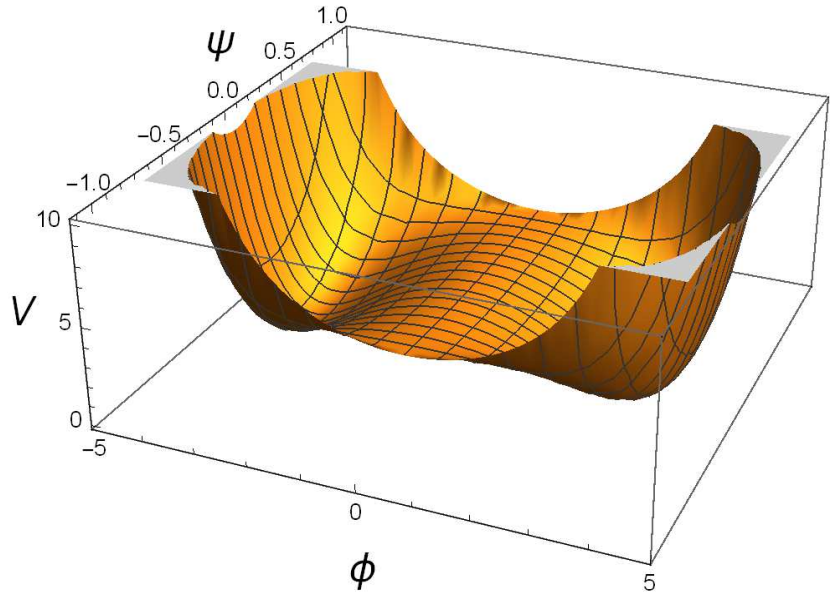}
\\
\textcolor{red}{Arbitrary Units}
\caption[Potential given by \cref{Eq.: Potential}]{The potential given by \cref{Eq.: Potential}.}
\label{Figure: Our Potential}
\end{center}
\end{figure}

It is possible for the mass of a certain scalar field to be dependent on another scalar field \cite{Vernizzi:Generating_Cosmo._Perts._Mass_Variations , Vernizzi:Cosmo._Perts._Varying_Masses_&_Couplings , Lyth:Generating_Curvature_Perturbation_E._of_I. , Matsuda:Cosmo._Perts._from_Inhomogeneous_Phase_Transition , Yokoyama_&_Soda:Prim._Stat._Ani._generated_End_of_Inf. , Dvali_et_al.:Cosmo._Perts._Inhomo._Reheat._Freezeout_&_Mass_Dom. , Postma:Inhom._Reheat._Low_Scale_Inflation_and/or_MSSM_Flat_Dirs. , Salem:On_Gen._Den._Perts._End_of_Inf. , Alabidi_&_Lyth:Curv._Pert._Sym._Break._End_of_Inf. , Lyth:Hybrid_Waterfall_and_Curvature_Perturbation , Alabidi_et_al.:How_Curvaton_Mod._Reheat._and_Inhomo._End_of_Inf._are_related , Kohri_et_al.:Delta-N_Form._Curvaton_Mod._Decay}. More specifically, the mass of a thermal waterfall field that is responsible for a bout of thermal inflation could be dependent on another scalar field. If the latter is light during primordial inflation, quantum fluctuations of the field are converted to almost scale-invariant classical field perturbations at around the time of horizon exit. If the scalar field remains light all the way up to the end of thermal inflation, then thermal inflation will end at different times in different parts of the Universe, because the value of the spectator field determines the mass of the thermal waterfall field, which in turn determines the end of thermal inflation. This is the ``end of inflation'' mechanism \cite{Lyth:Generating_Curvature_Perturbation_E._of_I.} and it will generate a contribution to the primordial curvature perturbation $\zeta$. In addition to this, if the scalar field remains light up until the decay of the thermal waterfall field, the decay rate of the thermal waterfall field will be modulated, due to the mass of the thermal waterfall field (which controls the decay rate) being dependent on the light scalar field. The decay of the thermal waterfall field will generate a second contribution to $\zeta$. The motivation of this work is to explore these two scenarios to see if either of them can produce the dominant contribution to the primordial curvature perturbation with characteristic observational signatures. We consider that these are the dominant contributions to the curvature perturbation, so that the inflaton's contribution can be ignored.\\
\indent It should be noted that the first scenario is very similar to that in Ref. \cite{Kawasaki_et_al.:Den._Fluctuations_in_Thermal_Inflation_&_Non-Gaussianity}. However, in that paper the authors use a modulated coupling constant rather than a modulated mass. Also, the treatment that has been given to the work in this thesis is much more comprehensive. One example of this is in the consideration of the effect that the thermal fluctuation of the thermal waterfall field has on the model (see \cref{Subsubsection: Thermal Fluctuation of phi}). Another example is the requirement that the thermal waterfall field is thermalized (see \cref{Subsubsection: Thermalization of phi}). Also, there is no consideration given in Ref. \cite{Kawasaki_et_al.:Den._Fluctuations_in_Thermal_Inflation_&_Non-Gaussianity} to requiring a fast transition from thermal inflation to thermal waterfall field oscillation (see \cref{Subsubsection: Time of Transition from Thermal Inflation to Thermal Waterfall Field Oscillation}), as detailed in Ref. \cite{Lyth:Hybrid_Waterfall_and_Curvature_Perturbation}, as this paper appeared after Ref. \cite{Kawasaki_et_al.:Den._Fluctuations_in_Thermal_Inflation_&_Non-Gaussianity}.\\
\indent Throughout this work, units are used where $c\!=\!\hbar\!=\!k_{B}\!=\!1$ and the reduced Planck Mass is $M_{P}\!=\!2.436\times10^{18}\ \mbox{GeV}$.\\
\indent The potential that is considered in this model is
\begin{equation}
V(\phi,T,\psi) = V_{0} + \left(g^{2}T^{2}-\frac{1}{2}m_{0}^{2}+h^{2}\frac{\psi^{2\alpha}}{M_{P}^{2\alpha-2}}\right)\phi^{2} + \lambda\frac{\phi^{2n+4}}{M_{P}^{2n}} + \frac{1}{2}m_{\psi}^{2}\psi^{2}
\label{Eq.: Full Potential}
\end{equation}
where $\phi$ is the thermal waterfall scalar field, $\psi$ is a light spectator scalar field, $T$ is the temperature of the thermal bath, $g$, $h$ and $\lambda$ are dimensionless coupling constants, $\alpha\ge1$ and $n\ge1$ are integers and the $-\frac{1}{2}m_{0}^{2}$ and $\frac{1}{2}m_{\psi}^{2}$ terms come from soft SUSY breaking.\footnote{$h$ has a factorial term absorbed into it. Also, we are absorbing the $\frac{1}{\left(2n+4\right)!}$ factor into $\lambda$.} We do not include a $\phi^4$ term, because the thermal waterfall field is a flaton, whose potential is stabilised by the higher-order non-renormalizable term.\\
\indent We make the following definition
\begin{equation}
m^{2} \equiv m_{0}^{2} - 2h^{2}\frac{\psi^{2\alpha}}{M_{P}^{2\alpha-2}}
\label{Eq.: Mass Definition}
\end{equation}
i.e. we combine the bare mass and coupling term into a new mass quantity. The variation of $m$, which is due only to the variation of $\psi$, is
\begin{equation}
\delta m=-\frac{2\alpha h^{2}\psi^{2\alpha-1}}{mM_{P}^{2\alpha-2}}\delta\psi
\label{Eq.: Mass Variation}
\end{equation}
\indent We only consider the case where the mass of $\phi$ is coupled to one field. If the mass were coupled to several similar fields, the results are just multiplied by the number of fields. If the multiple fields are different, then there will be only a small number that dominate the contribution to the mass perturbation. Therefore we consider only one for simplicity.\\
\indent Using \cref{Eq.: Mass Definition}, our redefined mass quantity, the potential becomes
\begin{equation}
V(\phi,T,\psi)=V_{0} + \left(g^{2}T^{2}-\frac{1}{2}m^{2}\right)\phi^{2} + \lambda\frac{\phi^{2n+4}}{M_{P}^{2n}} + \frac{1}{2}m_{\psi}^{2}\psi^{2}
\label{Eq.: Potential}
\end{equation}
This potential is shown in \cref{Figure: Our Potential}.
It would appear from the potential that domain walls will be produced, due to the fact that in some parts of the Universe $\phi$ will roll down to $+\left\langle\phi\right\rangle$ while in others parts it will roll down to $-\left\langle\phi\right\rangle$. However, this does not occur, as we can interpret $\phi$ as being the real part of a complex field whose potential contains only one continuous VEV.\footnote{A complex $\phi$ may result in the copious appearance of cosmic strings after the end of thermal inflation. However, we assume that their energy scale is very low and so they will not have any serious affect on the CMB observables. Moreover, depending on the overall background theory, such cosmic strings may well be unstable. Thus, we ignore them.}\\
\indent The zero temperature potential is
\begin{equation}
V(\phi,0,\psi)=V_{0} -\frac{1}{2}m^{2}\phi^{2} + \lambda\frac{\phi^{2n+4}}{M_{P}^{2n}} + \frac{1}{2}m_{\psi}^{2}\psi^{2}
\label{Eq.: 0 T Potential}
\end{equation}
To obtain the VEV of $\phi$, we find the minimum of the zero temperature potential. The VEV is
\begin{equation}
\left\langle\phi\right\rangle = \left(\frac{mM_{P}^{n}}{\sqrt{\left(2n+4\right)\lambda}}\right)^{\frac{1}{n+1}}
\label{Eq.: Phi VEV}
\end{equation}
$V\!=\!0$ at the VEV.\footnote{We are ignoring a cosmological constant as it is negligible. Considering it would give $V\!\neq\!0$ at the VEV.} $V_{0}$ is obtained by inserting the VEV into the zero temperature potential and then looking along the $\psi\!=\!0$ direction. We obtain
\begin{equation}
V_{0} \sim \left(\frac{m_{0}^{2n+4}M_{P}^{2n}}{\lambda}\right)^{\frac{1}{n+1}}
\label{Eq.: V_{0}}
\end{equation}
We use the Friedmann equation
\begin{equation}
M_{P}^{2}H_{\textrm{TI}}^{2} \sim V_{0}
\label{Eq.: V_{0} Energy Density Thermal Inflation}
\end{equation}
giving the energy density of the Universe during thermal inflation to obtain the Hubble parameter during thermal inflation as
\begin{equation}
H_{\textrm{TI}} \sim \left(\frac{m_{0}^{n+2}}{\sqrt{\lambda}\,M_{P}}\right)^{\frac{1}{n+1}}
\label{Eq.: H_{TI}}
\end{equation}
\indent Within this model, we will consider two cases regarding the decay rate of the inflaton, $\Gamma_{\varphi}$, with $\varphi$, the inflaton, being the field driving the period of primordial inflation prior to thermal inflation. Firstly, the case that $\Gamma_{\varphi}\!\gtrsim\!H_{\textrm{TI}}$, i.e. that reheating from primordial inflation occurs before or around the time of the start of thermal inflation. Secondly, we will consider the case that $\Gamma_{\varphi}\!\ll\!H_{\textrm{TI}}$, i.e. that reheating from primordial inflation occurs at some time after the end of thermal inflation. In the case of $\Gamma_{\varphi}\!\gtrsim\!H_{\textrm{TI}}$, thermal inflation will begin at a temperature
\begin{equation}
T_{1} \sim V_{0}^{\frac{1}{4}}
\label{Eq.: T_{1} (Gamma_{varphi} gtrsim H_{TI})}
\end{equation}
$T_{1}$ corresponds to the temperature when the potential energy density becomes comparable with the energy density of the thermal bath, for which the density is $\rho_{\gamma}\!\sim\!T^4$. In the case of $\Gamma_{\varphi}\!\ll\!H_{\textrm{TI}}$, thermal inflation will begin at a temperature
\begin{equation}
T_{1} \sim \left(M_{P}^{2}\,H_{\textrm{TI}}\Gamma_{\varphi}\right)^{\frac{1}{4}}
\label{Eq.: T_{1} (Gamma_{varphi} << H_{TI})}
\end{equation}
In both cases, thermal inflation ends at a temperature
\begin{equation}
T_{2} = \frac{m}{\sqrt{2}\,g}
\label{Eq.: T_{2}}
\end{equation}
$T_{2}$ corresponds to the temperature when the tachyonic mass term of the thermal waterfall field becomes equal to the thermally-induced mass term
in \cref{Eq.: Potential}.
\section{$\phi$ Decay Rate, Spectral Index and Tensor Fraction}
\label{Section: phi Decay Rate Spectral Index and Tensor Fraction}
\subsection{$\phi$ Decay Rate}
\label{Subsection: phi Decay Rate}
The decay rate of $\phi$ is given by
\begin{equation}
\Gamma \sim \mathrm{max}\left\{g^{2}m_{\phi,\textrm{osc}} \hspace{0.5em},\hspace{0.5em} \frac{m_{\phi,\textrm{osc}}^{3}}{M_{P}^{2}}\right\}
\label{Eq.: Phi Decay Rate -- m_{phi,osc}}
\end{equation}
where $m_{\phi,\textrm{osc}}$ is the effective mass of $\phi$ during the time of $\phi$'s oscillations around its VEV after the end of thermal inflation. This is calculated as
\begin{equation}
m_{\phi,\textrm{osc}} \sim m
\label{Eq.: Phi Effective Mass Oscillating}
\end{equation}
Therefore we obtain
\begin{equation}
\Gamma \sim \mathrm{max}\left\{g^{2}m \hspace{0.5em},\hspace{0.5em} \frac{m^{3}}{M_{P}^{2}}\right\}
\label{Eq.: Phi Decay Rate (Both Decay Channels)}
\end{equation}
The first expression is for decay into the thermal bath via direct interactions and the second is for gravitational decay. We will only consider the case in which the direct decay is the dominant channel ($g$ is not taken to be very small). This is the case when
\begin{equation}
m \ll gM_{P}
\label{Eq.: m << gM_{P}}
\end{equation}
Therefore we have just
\begin{equation}
\Gamma \sim g^{2}m
\label{Eq.: phi Decay Rate - Decay Rate Subsection}
\end{equation}
\subsection{Spectral Index {---} $n_{s}$ and $n_{s}^{\prime}$}
\label{Subsection: Spectral Index --- n_{s} and n_{s}^{'}}
Thermal Inflation has the effect of changing the number of e-folds before the end of primordial inflation at which cosmological scales exit the horizon. This affects the value of the spectral index $n_{s}$ of the curvature perturbation $\zeta$, assuming $\zeta$ is generated due to the perturbations of the spectator scalar field. The spectral index is given by \cite{Lyth_&_Liddle:Prim._Den._Pert.}
\begin{equation}
n_{s} \simeq 1 - 2\epsilon_{H} + 2\eta_{\psi\psi}
\label{Eq.: Spectral Index Definition}
\end{equation}
where $\epsilon_{H}$ and $\eta_{\psi\psi}$ are slow-roll parameters, defined as
\begin{equation}
\epsilon_{H} \equiv 2M_{P}^{2}\left(\frac{H^{\prime}(\varphi)}{H(\varphi)}\right)^{2}
\label{Eq.: epsilon_{H} Definition}
\end{equation}
and
\begin{equation}
\eta_{\psi\psi} \equiv \frac{V_{\psi\psi}}{3H^{2}}
\label{Eq.: eta_{psi psi} Definition}
\end{equation}
where $H^{\prime}(\varphi)$ is the derivative of the Hubble parameter with respect to the inflaton field $\varphi$ and $V_{\psi\psi} \equiv \frac{\partial^{2}V}{\partial \psi^{2}}$. $\epsilon_{H}$ and $\eta_{\psi\psi}$ are to be evaluated at the point where cosmological scales exit the horizon during primordial inflation. In the limit of slow-roll inflation, which we consider to be the case for our primordial inflation period, $\epsilon_{H} \rightarrow \epsilon$, which is defined as
\begin{equation}
\epsilon \equiv \frac{M_{P}^{2}}{2}\left(\frac{V^{\prime}(\varphi)}{V(\varphi)}\right)^{2}
\label{Eq.: epsilon Definition}
\end{equation}
where $V(\varphi)$ is the inflaton potential and $V^{\prime}(\varphi)$ is the derivative of that potential with respect to the inflaton field $\varphi$. $\epsilon$ is to be evaluated at the point where cosmological scales exit the horizon during primordial inflation. The spectral index now becomes
\begin{equation}
n_{s} \simeq 1 - 2\epsilon + 2\eta_{\psi\psi}
\label{Eq.: Spectral Index - Spectral Index Section}
\end{equation}
Regarding the various scalar fields involved in this model, the reason why $\epsilon$ depends only on $\varphi$ is because this slow-roll parameter captures the inflationary dynamics of primordial inflation, which is governed only by $\varphi$ in our model (we are assuming that both $\psi$ and $\phi$ have settled to a constant value (\cref{Subsubsection: The Field Value phi_{*},Subsubsection: The Field Value psi_{*}} respectively) by the time cosmological scales exit the horizon during primordial inflation). In a similar fashion, the reason why the slow-roll parameter $\eta$ depends only on $\psi$ is because this parameter captures the dependance on the spectral index of the field(s) whose perturbations contribute to the observed primordial curvature perturbation $\zeta$. In our case, this is only the $\psi$ field.\\
\indent The definition of the running of the spectral index is \cite{Lyth_&_Liddle:Prim._Den._Pert.}
\begin{align}
n_{s}^{\prime} &\equiv \frac{\mathrm{d}n_{s}}{\mathrm{d}\ln{(k)}}
\\
\nonumber
\\
&\simeq -\frac{\mathrm{d}n_{s}}{\mathrm{d}N}
\label{Eq.: Running of Spectral Index Definition}
\end{align}
the second line coming from $\mathrm{d}\ln{(k)}\!=\!\mathrm{d}\ln{(aH)}\!\simeq\!H\mathrm{d}t\!\equiv\!-\mathrm{d}N$, where $k\!=\!aH$. From \cref{Eq.: Spectral Index - Spectral Index Section} we have
\begin{align}
n_{s}^{\prime} &\simeq 2\frac{\mathrm{d}\epsilon}{\mathrm{d}N} - 2\frac{\mathrm{d}\eta_{\psi\psi}}{\mathrm{d}N}
\\
\nonumber
\\
&\simeq 2\epsilon\frac{\mathrm{d}\ln{(\epsilon)}}{\mathrm{d}N} - 2\frac{\mathrm{d}\eta_{\psi\psi}}{\mathrm{d}N}
\label{Eq.: Running of Spectral Index Working}
\end{align}
By differentiating the natural log of $\epsilon$ with respect to $N$, we obtain \cite{Lyth_&_Liddle:Prim._Den._Pert.}
\begin{equation}
\frac{\mathrm{d}\ln{(\epsilon)}}{\mathrm{d}N} \simeq -4\epsilon + 2\eta
\label{Eq.: d ln(epsilon)/dN}
\end{equation}
where $\eta$ is a slow-roll parameter given by
\begin{equation}
\eta \equiv M_{P}^{2}\frac{V^{\prime\prime}(\varphi)}{V(\varphi)}
\label{Eq.: eta Definition}
\end{equation}
where $V^{\prime\prime}(\varphi)$ is the second derivative of the inflaton potential with respect to the inflaton field $\varphi$. $\eta$ is to be evaluated at the point where cosmological scales exit the horizon during primordial inflation. By differentiating $\eta_{\psi\psi}$ with respect to $N$ using the quotient rule, we obtain
\begin{align}
\frac{\mathrm{d}\eta_{\psi\psi}}{\mathrm{d}N} &= \frac{3H^{2}\frac{\mathrm{d}V_{\psi\psi}}{\mathrm{d}N} - 3V_{\psi\psi}\frac{\mathrm{d}H^{2}}{\mathrm{d}N}}{9H^{4}}
\\
\nonumber
\\
&= -\frac{V_{\psi\psi}\frac{\mathrm{d}H^{2}}{\mathrm{d}N}}{3H^{4}}
\\
\nonumber
\\
&= -\eta_{\psi\psi}\frac{1}{H^{2}}\frac{\mathrm{d}H^{2}}{\mathrm{d}N}
\\
\nonumber
\\
&= -2\eta_{\psi\psi}\frac{\mathrm{d}\ln{(H)}}{\mathrm{d}N}
\label{Eq.: d eta_{psi psi}/dN Working}
\end{align}
with the second line coming from the fact that we have $V_{\psi\psi}$ not depending on $N$, as we are assuming that both $\psi$ and $\phi$ have settled to a constant value (\cref{Subsubsection: The Field Value phi_{*},Subsubsection: The Field Value psi_{*}} respectively) by the time cosmological scales exit the horizon during primordial inflation. By differentiating the natural log of $H$ with respect to $N$, we obtain \cite{Lyth_&_Liddle:Prim._Den._Pert.}
\begin{equation}
\frac{\mathrm{d}\ln{(H)}}{\mathrm{d}N} \simeq \epsilon
\label{Eq.: d ln(H)/dN}
\end{equation}
Therefore we have
\begin{equation}
\frac{\mathrm{d}\eta_{\psi\psi}}{\mathrm{d}N} \simeq -2\epsilon\eta_{\psi\psi}
\label{Eq.: d eta_{psi psi}/dN}
\end{equation}
Therefore, the final result for the running of the spectral index is
\begin{equation}
n_{s}^{\prime} \simeq -8\epsilon^{2} + 4\epsilon\eta + 4\epsilon\eta_{\psi\psi}
\label{Eq.: Running of Spectral Index - Spectral Index Section}
\end{equation}
\indent From now on we assume that $H$ has the constant value $H_{\ast}$ by the time cosmological scales exit the horizon up until the end of primordial inflation. In order to obtain $\epsilon$ and $\eta$, we require $N_{\ast}$, the number of e-folds before the end of primordial inflation at which cosmological scales exit the horizon. We consider the period between when the pivot scale, $k_{0}\!\equiv\!0.002\ \mbox{Mpc$^{-1}$}$, exits the horizon during primordial inflation and when it reenters the horizon long after the end of thermal inflation. We have
\begin{equation}
R_{\ast} = H_{\ast}^{-1} \hspace{5em} k^{-1} = H_{\textrm{piv}}^{-1}
\label{Eq.: R{*} & R_{pivot}}
\end{equation}
where $R_{\ast}$ is a length scale when the pivot scale exits the horizon during primordial inflation. Therefore
\begin{equation}
H_{\ast}^{-1} = \frac{a_{\ast}}{a_{\textrm{piv}}}H_{\textrm{piv}}^{-1}
\label{Eq.: H_{*}^{-1}}
\end{equation}
where $a_{\ast}$ is the scale factor at the time when the pivot scale exits the horizon during primordial inflation and $a_{\textrm{piv}}$ is the scale factor at the time when the pivot scale reenters the horizon.
\subsubsection{The Case $\Gamma_{\varphi}\!\gtrsim\!H_{\textrm{TI}}$}
\label{Subsubsection: The Case Gamma_{varphi} gtrsim H_{TI} (n_{s} and n_{s}^{'})}
We have
\begin{align}
H_{\ast}^{-1} &= \frac{a_{\ast}}{a_{\textrm{end,inf}}}\frac{a_{\textrm{end,inf}}}{a_{\textrm{reh,inf}}}\frac{a_{\textrm{reh,inf}}}{a_{\textrm{start,TI}}}\frac{a_{\textrm{start,TI}}}{a_{\textrm{end,TI}}}\frac{a_{\textrm{end,TI}}}{a_{\textrm{reh,TI}}}\frac{a_{\textrm{reh,TI}}}{a_{\textrm{eq}}}\frac{a_{\textrm{eq}}}{a_{\textrm{piv}}}H_{\textrm{piv}}^{-1}
\\
\nonumber
\\
&= e^{-N_{\ast}}\frac{a_{\textrm{end,inf}}}{a_{\textrm{reh,inf}}}\frac{a_{\textrm{reh,inf}}}{a_{\textrm{start,TI}}}e^{-N_{\textrm{TI}}}\frac{a_{\textrm{end,TI}}}{a_{\textrm{reh,TI}}}\frac{a_{\textrm{reh,TI}}}{a_{\textrm{eq}}}\frac{a_{\textrm{eq}}}{a_{\textrm{piv}}}H_{\textrm{piv}}^{-1}
\label{Eq.: H_{*}^{-1} (Gamma_{varphi} gtrsim H_{TI})}
\end{align}
where $N_{\textrm{TI}}$ is the number of e-folds of thermal inflation and the scale factors are the following: $a_{\textrm{end,inf}}$ is at the end of primordial inflation, $a_{\textrm{reh,inf}}$ is at primordial inflation reheating, $a_{\textrm{start,TI}}$ is at the start of thermal inflation, $a_{\textrm{end,TI}}$ is at the end of thermal inflation, $a_{\textrm{reh,TI}}$ is at thermal inflation reheating and $a_{\textrm{eq}}$ is at the time of matter-radiation equality. For the period between the end of primordial/thermal inflation and primordial/thermal inflation reheating, $a$ goes as ${T}^{-\frac{8}{3}}$. The proof is as follows. During this time, $T\!\sim\!\left(M_{P}^{2}H\Gamma\right)^{\frac{1}{4}}$. As $H$ goes as $t^{-1}$ we have $T\!\propto\!t^{-\frac{1}{4}}$. During the field oscillations, the Universe is matter dominated and so we have $a\!\propto\!t^{\frac{2}{3}}$. Therefore $t\!\propto\!a^{\frac{3}{2}}$. Putting this all together we find $T\!\propto\!t^{-\frac{1}{4}}\!\propto\!a^{-\frac{3}{8}}$ and therefore $a\!\propto\!T^{-\frac{8}{3}}$. For all other times, $a$ goes as $T^{-1}$ and so we have
\begin{equation}
e^{N_{\ast}} = \frac{H_{\ast}}{k}\left(\frac{T_{\textrm{reh,inf}}}{T_{\textrm{end,inf}}}\right)^{\frac{8}{3}}\frac{T_{\textrm{start,TI}}}{T_{\textrm{reh,inf}}}\left(\frac{T_{\textrm{reh,TI}}}{T_{\textrm{end,TI}}}\right)^{\frac{8}{3}}\frac{T_{\textrm{piv}}}{T_{\textrm{reh,TI}}}e^{-N_{\textrm{TI}}}
\label{Eq.: e^{N_{*}} (Gamma_{varphi} gtrsim H_{TI})}
\end{equation}
We need to calculate $T_{\textrm{piv}}$. We consider the period between when the pivot scale reenters the horizon and the present. Throughout this period the Universe is matter-dominated (ignoring dark energy). Therefore we have
\begin{align}
\rho &\propto a^{-3}
\\
&\propto T^{3}
\label{Eq.: Energy Density propto T^3 a}
\end{align}
Therefore, from the Friedmann equation, we have
\begin{equation}
3M_{P}^{2}H^{2} \propto T^{3}
\label{Eq.: Energy Density propto T^3 b}
\end{equation}
This gives
\begin{equation}
\frac{H_{\textrm{piv}}^{2}}{H_{0}^{2}} = \frac{T_{\textrm{piv}}^{3}}{T_{0}^{3}}
\label{Eq.: T pivot Equation a}
\end{equation}
\begin{align}
T_{\textrm{piv}} &= \left(\frac{0.002^{2}\ \mbox{Mpc$^{-2}$}\ T_{0}^{3}}{H_{0}^{2}}\right)^{\frac{1}{3}}
\\
\nonumber
\\
&= 9.830\times10^{-13}\ \mbox{GeV (4 s.f.)}
\label{Eq.: T pivot Value}
\end{align}
We now obtain $N_{\ast}$ as
\begin{multline}
N_{\ast} \approx \ln{\left(\frac{\left(1.5637\times10^{38}\ \mbox{GeV$^{-1}$}\right)H_{\ast}}{0.002}\right)} + \frac{2}{3}\ln{\left(\frac{9\Gamma_{\varphi}}{10\pi^{2}H_{\ast}}\right)}\\+ \frac{1}{4}\ln{\left(\frac{10\pi^{2}V_{0}}{9M_{P}^{2}H_{\ast}^{2}}\right)} + \frac{2}{3}\ln{\left(\frac{9\Gamma}{10\pi^{2}H_{\textrm{TI}}}\right)}\\+ \frac{1}{4}\ln{\left(\frac{10\pi^{2}\left(9.830\times10^{-13}\ \mbox{GeV}\right)^{4}}{9M_{P}^{2}\Gamma^{2}}\right)} - N_{\textrm{TI}}
\label{Eq.: N_{*} (Gamma_{varphi} gtrsim H_{TI})}
\end{multline}
where we have used $g_{\ast}\!\approx\!10^{2}$ as the number of spin states (effective relativistic degrees of freedom) of all of the particles in the thermal bath, at the time of both primordial inflation reheating and thermal inflation reheating, as this value corresponds to the number of relativistic degrees of freedom in the Standard Model.
\subsubsection{The Case $\Gamma_{\varphi}\!\ll\!H_{\textrm{TI}}$}
\label{Subsubsection: The Case Gamma_{varphi} << H_{TI} (n_{s} and n_{s}^{'})}
We have
\begin{align}
H_{\ast}^{-1} &= \frac{a_{\ast}}{a_{\textrm{end,inf}}}\frac{a_{\textrm{end,inf}}}{a_{\textrm{start,TI}}}\frac{a_{\textrm{start,TI}}}{a_{\textrm{end,TI}}}\frac{a_{\textrm{end,TI}}}{a_{\textrm{reh,TI}}}\frac{a_{\textrm{reh,TI}}}{a_{\textrm{eq}}}\frac{a_{\textrm{eq}}}{a_{\textrm{piv}}}H_{\textrm{piv}}^{-1}
\\
\nonumber
\\
&= e^{-N_{\ast}}\frac{a_{\textrm{end,inf}}}{a_{\textrm{start,TI}}}e^{-N_{\textrm{TI}}}\frac{a_{\textrm{end,TI}}}{a_{\textrm{reh,TI}}}\frac{a_{\textrm{reh,TI}}}{a_{\textrm{eq}}}\frac{a_{\textrm{eq}}}{a_{\textrm{piv}}}H_{\textrm{piv}}^{-1}
\label{Eq.: H_{*}^{-1} (Gamma_{varphi} << H_{TI})}
\end{align}
Using $a\!\propto\!T^{-\frac{8}{3}}$ for the period between the end of primordial inflation and the start of thermal inflation, as well as for the period between the end of thermal inflation and thermal inflation reheating and using $a\!\propto\!T^{-1}$ for all other times, we have
\begin{equation}
e^{N_{\ast}} = \frac{H_{\ast}}{k}\left(\frac{T_{\textrm{start,TI}}}{T_{\textrm{end,inf}}}\right)^{\frac{8}{3}}\left(\frac{T_{\textrm{reh,TI}}}{T_{\textrm{end,TI}}}\right)^{\frac{8}{3}}\frac{T_{\textrm{piv}}}{T_{\textrm{reh,TI}}}e^{-N_{\textrm{TI}}}
\label{Eq.: e^{N_{*}} (Gamma_{varphi} << H_{TI})}
\end{equation}
Using $T_{\textrm{piv}}\!=\!9.830\times10^{-13}\ \mbox{GeV}$ (\cref{Eq.: T pivot Value}), we obtain $N_{\ast}$ as
\begin{multline}
N_{\ast} \approx \ln{\left(\frac{\left(1.5637\times10^{38}\ \mbox{GeV$^{-1}$}\right)H_{\ast}}{0.002}\right)} + \frac{2}{3}\ln{\left(\frac{9\Gamma}{10\pi^{2}H_{\ast}}\right)}\\+ \frac{1}{4}\ln{\left(\frac{10\pi^{2}\left(9.830\times10^{-13}\ \mbox{GeV}\right)^{4}}{9M_{P}^{2}\Gamma^{2}}\right)} - N_{\textrm{TI}}
\label{Eq.: N_{*} (Gamma_{varphi} << H_{TI})}
\end{multline}
where we have used $g_{\ast}\!\approx\!10^{2}$ as the number of spin states (effective relativistic degrees of freedom) of all of the particles in the thermal bath at the time of thermal inflation reheating, as this value corresponds to the number of relativistic degrees of freedom in the Standard Model.
\subsection{Tensor Fraction $r$}
\label{Subsection: Tensor Fraction r}
The definition of the Tensor Fraction, $r$ \cite{Lyth_&_Liddle:Prim._Den._Pert.}, is
\begin{equation}
r \equiv \frac{\mathcal{P}_{h}}{\mathcal{P}_{\zeta}}
\label{Eq.: r Definition}
\end{equation}
where $\mathcal{P}_{h}$ and $\mathcal{P}_{\zeta}$ are the spectrums of the primordial tensor and curvature perturbations respectively. The spectrum $\mathcal{P}_{h}$ is given by
\begin{equation}
\mathcal{P}_{h}(k) = \frac{8}{M_{P}^{2}}\left(\frac{H_{k}}{2\pi}\right)^{2}
\label{Eq.: P_{h}(k)}
\end{equation}
for a given wavenumber $k$. Using this, together with $\rho_{\ast}\!=\!3M_{P}^{2}H_{\ast}^{2}$, given that we are saying $H_{k}\!=\!H_{\ast}$ for our current case, as well as the observed value $\mathcal{P}_{\zeta}(k_{0})\!=\!2.142\times10^{-9}$, we obtain
\begin{equation}
r = \left(\frac{\rho_{\ast}^{\frac{1}{4}}}{3.25\times10^{16}\ \mbox{GeV}}\right)^{4}
\label{Eq.: r}
\end{equation}
\section{``End of Inflation'' Mechanism}
\label{Section: ``End of Inflation'' Mechanism}
In this section, we investigate the ``end of inflation'' mechanism. We aim to obtain a number of constraints on the model parameters and the initial conditions for the fields. Considering these constraints, we intend to determine the available parameter space (if any). In this parameter space we will calculate distinct observational signatures (such as non-Gaussianity) that may test this scenario in the near future.
\subsection{Generating $\zeta$}
\label{Subsection: Generating zeta}
As $\phi$ is coupled to $\psi$, the ``end of inflation'' mechanism will generate a contribution to the primordial curvature perturbation $\zeta$ \cite{Lyth:Generating_Curvature_Perturbation_E._of_I.}. We will use the $\delta N$ formalism to calculate this contribution. In this formalism, the final coordinate slice is the transition slice, going from thermal inflation to field oscillation. The $\delta N$ formalism allows us to calculate the primordial curvature perturbation as
\begin{equation}
\zeta = \delta N_{\textrm{TI}} = \frac{\mathrm{d}N_{\textrm{TI}}}{\mathrm{d}m}\delta m + \frac{1}{2!}\frac{\mathrm{d^{2}}N_{\textrm{TI}}}{\mathrm{d}m^{2}}\delta m^{2} + \frac{1}{3!}\frac{\mathrm{d^{3}}N_{\textrm{TI}}}{\mathrm{d}m^{3}}\delta m^{3} + ...
\label{Eq.: zeta Definition - Our Work}
\end{equation}
The number of e-folds between the start and end of thermal inflation is given by
\begin{equation}
N_{\textrm{TI}} = \ln{\left(\frac{a_2}{a_1}\right)} = \ln{\left(\frac{T_1}{T_2}\right)}
\label{Eq.: N}
\end{equation}
where $a_{1}\!=\!a_{\textrm{start,TI}}$ and $a_{2}\!=\!a_{\textrm{end,TI}}$.
\subsubsection{The Case $\Gamma_{\varphi}\!\gtrsim\!H_{\textrm{TI}}$}
\label{Subsubsection: The Case Gamma_{varphi} gtrsim H_{TI} (Generating zeta)}
Substituting $T_{1}$ and $T_{2}$, \cref{Eq.: T_{1} (Gamma_{varphi} gtrsim H_{TI}),Eq.: T_{2}} respectively, into \cref{Eq.: N} gives
\begin{equation}
N_{\textrm{TI}} \simeq \ln{\left(\frac{\sqrt{2}\,gV_{0}^{\frac{1}{4}}}{m}\right)}
\label{Eq.: N (Gamma_{varphi} gtrsim H_{TI})}
\end{equation}
Therefore the $\delta N$ formalism to third order gives
\begin{equation}
\zeta = \delta N_{\textrm{TI}} = -\frac{\delta m}{m} + \frac{1}{2}\frac{\delta m^{2}}{m^{2}} - \frac{1}{3}\frac{\delta m^{3}}{m^{3}}
\label{Eq.: Zeta (m) (Gamma_{varphi} gtrsim H_{TI})}
\end{equation}
By substituting our mass definition and its differential, \cref{Eq.: Mass Definition,Eq.: Mass Variation}, into \cref{Eq.: Zeta (m) (Gamma_{varphi} gtrsim H_{TI})} we obtain the power spectrum of the primordial curvature perturbation, which to first order is
\begin{equation}
\mathcal{P}_{\zeta}^{\frac{1}{2}} = \frac{\alpha h^{2}H_{\ast}\psi^{2\alpha-1}}{\pi m^{2}M_{P}^{2\alpha-2}}
\label{Eq.: Power Spectrum (Gamma_{varphi} gtrsim H_{TI})}
\end{equation}
It must be noted that although there will be perturbations in $\psi$ that are generated during thermal inflation that will become classical due to the inflation, the scales to which these correspond are much smaller than cosmological scales, as thermal inflation lasts for only about 10 e-folds. Therefore we do not consider them here.\\
\indent A required condition for the perturbative expansion in \cref{Eq.: Zeta (m) (Gamma_{varphi} gtrsim H_{TI})} to be suitable is that each term is much smaller than the preceding one. This requirement gives
\begin{equation}
\frac{h^{2}H_{\ast}\psi^{2\alpha-1}}{m^{2}M_{P}^{2\alpha-2}} \ll 1
\label{Eq.: Suitable Perturbative Expansion}
\end{equation}
\subsubsection{The Case $\Gamma_{\varphi}\!\ll\!H_{\textrm{TI}}$}
\label{Subsubsection: The Case Gamma_{varphi} << H_{TI} (Generating zeta)}
Substituting $T_{1}$ and $T_{2}$, \cref{Eq.: T_{1} (Gamma_{varphi} << H_{TI}),Eq.: T_{2}} respectively, into \cref{Eq.: N} gives
\begin{equation}
N_{\textrm{TI}} \simeq \frac{8}{3}\ln{\left(\frac{\sqrt{2}\,g\left(M_{P}^{2}\,H_{\textrm{TI}}\Gamma_{\varphi}\right)^{\frac{1}{4}}}{m}\right)}
\label{Eq.: N (Gamma_{varphi} << H_{TI})}
\end{equation}
Therefore the $\delta N$ formalism to third order gives
\begin{equation}
\zeta = \delta N_{\textrm{TI}} = -\frac{8}{3}\frac{\delta m}{m} + \frac{4}{3}\frac{\delta m^{2}}{m^{2}} - \frac{8}{9}\frac{\delta m^{3}}{m^{3}}
\label{Eq.: Zeta (m) (Gamma_{varphi} << H_{TI})}
\end{equation}
By substituting our mass definition and its differential, \cref{Eq.: Mass Definition,Eq.: Mass Variation}, into \cref{Eq.: Zeta (m) (Gamma_{varphi} << H_{TI})} we obtain the power spectrum of the primordial curvature perturbation, which to first order is
\begin{equation}
\mathcal{P}_{\zeta}^{\frac{1}{2}} = \frac{8}{3}\frac{\alpha h^{2}H_{\ast}\psi^{2\alpha-1}}{\pi m^{2}M_{P}^{2\alpha-2}}
\label{Eq.: Power Spectrum (Gamma_{varphi} << H_{TI})}
\end{equation}
\indent The condition for the perturbative expansion in \cref{Eq.: Zeta (m) (Gamma_{varphi} << H_{TI})} to be suitable, i.e. that each term is much smaller than the preceding one, yields the same constraint as in \cref{Eq.: Suitable Perturbative Expansion}.
\subsection{Non-Gaussianity}
\label{Subsection: End of Inflation Non-Gaussianity}
One of the distinct observational signatures that we hope to generate through this model is the production of characteristic and observable non-Gaussianity in the curvature perturbation. Non-Gaussianity refers to the departure that the distribution of, in this particular case, the curvature perturbation is from purely Gaussian, i.e. of the familiar bell-shaped distribution. For a purely Gaussian distribution, there is no correlation between different modes of the perturbation. For a non-Gaussian distribution however, there is correlation, with the 3-point correlator for the curvature perturbation being \cite{Lyth_&_Liddle:Prim._Den._Pert.}
\begin{equation}
\left\langle\zeta_{\textbf{k}_{1}}\zeta_{\textbf{k}_{2}}\zeta_{\textbf{k}_{3}}\right\rangle = (2\pi)^{3}\delta^{3}_{\textbf{k}_{1}+\textbf{k}_{2}+\textbf{k}_{3}}B_{\zeta}\left(k_{1},k_{2},k_{3}\right)
\label{Eq.: 3-point Correlator of Curvature Perturbation}
\end{equation}
where $B_{\zeta}\left(k_{1},k_{2},k_{3}\right)$ is a function called the \textbf{Bispectrum}, being given by
\begin{equation}
B_{\zeta}\left(k_{1},k_{2},k_{3}\right) = \frac{6}{5}f_{\textrm{NL}}\left(k_{1},k_{2},k_{3}\right)\left[P_{\zeta}\!\left(k_{1}\right)\!P_{\zeta}\!\left(k_{2}\right) + P_{\zeta}\!\left(k_{1}\right)\!P_{\zeta}\!\left(k_{3}\right) + P_{\zeta}\!\left(k_{2}\right)\!P_{\zeta}\!\left(k_{3}\right)\right]
\label{Eq.: Bispectrum of Curvature Perturbation}
\end{equation}
where $f_{\textrm{NL}}\left(k_{1},k_{2},k_{3}\right)$ effectively parameterises the bispectrum (it is $\frac{5}{6}$ the value of the reduced bispectrum) and $P_{\zeta}(k)$ is the spectrum of the curvature perturbation (the spectrum that is being used in this thesis is that defined by $\mathcal{P}_{\zeta}(k)\!\equiv\!\frac{k^{3}}{2\pi^{2}}P_{\zeta}(k)$).\\
\indent The 4-point (connected) correlator for the curvature perturbation is
\begin{equation}
\left\langle\zeta_{\textbf{k}_{1}}\zeta_{\textbf{k}_{2}}\zeta_{\textbf{k}_{3}}\zeta_{\textbf{k}_{4}}\right\rangle = (2\pi)^{3}\delta^{3}_{\textbf{k}_{1}+\textbf{k}_{2}+\textbf{k}_{3}+\textbf{k}_{4}}T_{\zeta}
\label{Eq.: 4-point Correlator of Curvature Perturbation}
\end{equation}
where $T_{\zeta}$ is a function called the \textbf{Trispectrum}, being given by \cite{Byrnes_et_al.:Prim._Trispectrum_Inf.}
\begin{multline}
T_{\zeta}\left(\textbf{k}_{1},\textbf{k}_{2},\textbf{k}_{3},\textbf{k}_{4}\right) = \tau_{\textrm{NL}}\left[P_{\zeta}\!\left(k_{13}\right)\!P_{\zeta}\!\left(k_{3}\right)\!P_{\zeta}\!\left(k_{4}\right) + 11\ \mbox{perms.}\right]\\+ \frac{54}{25}g_{\textrm{NL}}\left[P_{\zeta}\!\left(k_{2}\right)\!P_{\zeta}\!\left(k_{3}\right)\!P_{\zeta}\!\left(k_{4}\right) + 3\ \mbox{perms.}\right]
\label{Eq.: Trispectrum of Curvature Perturbation}
\end{multline}
where $k_{13}\!\equiv\!|\textbf{k}_{1}+\textbf{k}_{3}|$ and $\tau_{\textrm{NL}}$ and $g_{\textrm{NL}}$ effectively parameterise the trispectrum.\\
\indent We will consider what is termed local non-Gaussianity, which for the bispectrum corresponds to the ``squeezed'' configuration of the momenta triangle, in that the magnitude of one of the momentum vectors is much smaller than the other two, which are of similar magnitude to each other, e.g. $k_{3}\!\ll\!k_{1},k_{2}$ and $k_{1}\!\approx\!k_{2}$. Within the framework of the $\delta N$ formalism, the non-Gaussianity parameter $f_{\textrm{NL}}$ is obtained as \cite{Byrnes_et_al.:Prim._Trispectrum_Inf.}
\begin{equation}
f_{\textrm{NL}} = \frac{5}{6}\frac{N^{\prime\prime}}{N^{\prime^{2}}}
\label{Eq.: f_{NL} Expression - End of Inflation}
\end{equation}
where the prime denotes the derivative with respect to $\psi$. By substituting $N_{\textrm{TI}}$ from \cref{Eq.: N (Gamma_{varphi} gtrsim H_{TI})} or \cref{Eq.: N (Gamma_{varphi} << H_{TI})} into \cref{Eq.: f_{NL} Expression - End of Inflation} we obtain
\begin{equation}
f_{\textrm{NL}} \simeq A\left(1 - \frac{mm^{\prime\prime}}{m^{\prime^{2}}}\right)
\label{Eq.: f_{NL} (m) - End of Inflation}
\end{equation}
where $A\!=\!\frac{5}{6}$ for $\Gamma_{\varphi}\!\gtrsim\!H_{\textrm{TI}}$ (\cref{Eq.: N (Gamma_{varphi} gtrsim H_{TI})}) or $A\!=\!\frac{5}{16}$ for $\Gamma_{\varphi}\!\ll\!H_{\textrm{TI}}$ (\cref{Eq.: N (Gamma_{varphi} << H_{TI})}). Then, from our mass definition $m$, \cref{Eq.: Mass Definition}, we obtain
\begin{equation}
f_{\textrm{NL}} \simeq A\left[1 + \frac{2\alpha -1}{\alpha}\left(\frac{m_{0}^{2}M_{P}^{2\alpha-2}}{2h^{2}\psi^{2\alpha}} - 1\right)\right]
\label{Eq.: f_{NL} - End of Inflation}
\end{equation}
\indent The non-Gaussianity parameter $g_{\textrm{NL}}$ is obtained as \cite{Byrnes_et_al.:Prim._Trispectrum_Inf.}
\begin{equation}
g_{\textrm{NL}} = \frac{25}{54}\frac{N^{\prime\prime\prime}}{N^{\prime^{3}}}
\label{Eq.: g_{NL} Expression - End of Inflation}
\end{equation}
By substituting $N_{\textrm{TI}}$ from \cref{Eq.: N (Gamma_{varphi} gtrsim H_{TI})} or \cref{Eq.: N (Gamma_{varphi} << H_{TI})} into \cref{Eq.: g_{NL} Expression - End of Inflation} we obtain
\begin{equation}
g_{\textrm{NL}} \simeq B\left(2 - \frac{3mm^{\prime\prime}}{m^{\prime^{2}}} + \frac{m^{2}m^{\prime\prime\prime}}{m^{\prime^{3}}}\right)
\label{Eq.: g_{NL} (m) - End of Inflation}
\end{equation}
where $B\!=\!\frac{25}{54}$ for $\Gamma_{\varphi}\!\gtrsim\!H_{\textrm{TI}}$ (\cref{Eq.: N (Gamma_{varphi} gtrsim H_{TI})}) or $B\!=\!\frac{25}{384}$ for $\Gamma_{\varphi}\!\ll\!H_{\textrm{TI}}$ (\cref{Eq.: N (Gamma_{varphi} << H_{TI})}). Then, our $m$ from \cref{Eq.: Mass Definition} gives
\begin{multline}
g_{\textrm{NL}} \simeq B\left[2 + \frac{6\alpha-3}{\alpha}\left(\frac{m_{0}^{2}M_{P}^{2\alpha-2}}{2h^{2}\psi^{2\alpha}} -1\right)\right.\\\left.+ \frac{\left(2\alpha^{2}-\alpha\right)\left(\alpha-1\right)}{\alpha^{3}}\left(\frac{m_{0}^{4}M_{P}^{4\alpha-4}}{4h^{4}\psi^{4\alpha}} -\frac{m_{0}^{2}M_{P}^{2\alpha-2}}{h^{2}\psi^{2\alpha}} +1\right)\right]
\label{Eq.: g_{NL} - End of Inflation}
\end{multline}
In the parameter space available (if any), we will investigate the range of values for $f_{\textrm{NL}}$ and $g_{\textrm{NL}}$.
\subsection{Constraining the Free Parameters}
\label{Subsection: Constraining the Free Parameters - End of Inflation Section}
\subsubsection{Primordial Inflation Energy Scale}
\label{Subsubsection: Primordial Inflation Energy Scale}
We want the energy scale of primordial inflation to be
\begin{equation}
V^{\frac{1}{4}} \lesssim 10^{14}\ \mbox{GeV}
\label{Eq.: Primordial Inflation Scale}
\end{equation}
so that the inflaton contribution to the curvature perturbation is negligible. Therefore, from the Friedmann equation $3M_{P}^{2}H_{\ast}^{2}\!=\!V$ we require
\begin{equation}
H_{\ast} \lesssim 10^{10}\ \mbox{GeV}
\label{Eq.: H_{*} Constraint a}
\end{equation}
\subsubsection{Thermal Inflation Dynamics}
\label{Subsubsection: Thermal Inflation Dynamics}
We will consider only the case in which the inflationary trajectory is 1-dimensional, in that only the $\phi$ field is involved in determining the trajectory of thermal inflation in field space. We do this only to work with the simplest scenario for the trajectory. It is not a requirement on the model itself. In order that the $\psi$ field does not affect the inflationary trajectory during thermal inflation, we require from our $m$ mass definition, \cref{Eq.: Mass Definition},
\begin{equation}
m_{0} \gg h\frac{\psi^{\alpha}}{M_{P}^{\alpha-1}}
\label{Eq.: m_{0} >> Coupling Term}
\end{equation}
Therefore we have
\begin{equation}
m \simeq m_{0}
\label{Eq.: m approx m_{0}}
\end{equation}
\indent From our potential, \cref{Eq.: Full Potential}, \cref{Eq.: m_{0} >> Coupling Term} gives
\begin{equation}
m_{0}^{2} < 2g^{2}T_{1}^{2}
\label{Eq.: m_{0}^{2}<2g^{2}T_{1}^{2}}
\end{equation}
For $\Gamma_{\varphi}\!\gtrsim\!H_{\textrm{TI}}$, substituting $T_{1}$ from \cref{Eq.: T_{1} (Gamma_{varphi} gtrsim H_{TI})} into \cref{Eq.: m_{0}^{2}<2g^{2}T_{1}^{2}} gives
\begin{equation}
m_{0} < \left(\frac{g^{2n+2}}{\sqrt{\lambda}}\right)^{\frac{1}{n}}M_{P}
\label{Eq.: m_{0} Constraint c (Gamma_{varphi} gtrsim H_{TI})}
\end{equation}
and for $\Gamma_{\varphi}\!\ll\!H_{\textrm{TI}}$, substituting $T_{1}$ from \cref{Eq.: T_{1} (Gamma_{varphi} << H_{TI})} into \cref{Eq.: m_{0}^{2}<2g^{2}T_{1}^{2}} gives
\begin{equation}
m_{0} < \left[\left(g^{4}\Gamma_{\varphi}\right)^{n+1}\frac{M_{P}^{2n+1}}{\sqrt{\lambda}}\right]^{\frac{1}{3n+2}}
\label{Eq.: m_{0} Constraint c (Gamma_{varphi} << H_{TI})}
\end{equation}
\subsubsection{Lack of Observation of $\phi$ Particles}
\label{Subsubsection: Lack of Observation of phi Particles}
Given that we have not observed any $\phi$ particles, the most liberal constraint on the present value of the effective mass of $\phi$ is
\begin{equation}
m_{\phi,\textrm{now}} \gtrsim 1\ \mbox{TeV}
\label{Eq.: m_{phi,now} gtrsim 1 TeV}
\end{equation}
From our potential, \cref{Eq.: Full Potential}, we have
\begin{equation}
m_{\phi,\textrm{now}}^{2} \sim -m_{0}^{2} + (2n+4)(2n+3)\lambda\frac{\left\langle\phi\right\rangle^{2n+2}}{M_{P}^{2n}}
\label{Eq.: m_{phi,now}^{2}}
\end{equation}
Substituting the VEV of $\phi$, \cref{Eq.: Phi VEV}, into here gives
\begin{equation}
m_{\phi,\textrm{now}} \sim m_{0}
\label{Eq.: m_{phi,now} sim m_{0}}
\end{equation}
for all reasonable values of $n$. Therefore, we require
\begin{equation}
m_{0} \gtrsim 1\ \mbox{TeV}
\label{Eq.: m_{0} gtrsim 1 TeV}
\end{equation}
\subsubsection{Light $\psi$}
\label{Subsubsection: Light psi - End of Inflation Subsubsection}
In order that $\psi$ acquires classical perturbations during primordial inflation, we require the effective mass of $\psi$ to be light during this time, i.e.
\begin{equation}
|m_{\psi,\textrm{eff}}| \ll H_{\ast}
\label{Eq.: Psi Effective Mass << H_{*}}
\end{equation}
where we are using notation such that $|m_{\psi,\textrm{eff}}|\equiv\sqrt{\left|m_{\psi,\textrm{eff}}^{2}\right|}$. We have
\begin{equation}
m_{\psi,\textrm{eff}}^{2} = m_{\psi}^{2} + \left(4\alpha^{2}-2\alpha\right)h^{2}\phi^{2}\left(\frac{\psi}{M_{P}}\right)^{2\alpha-2}
\label{Eq.: Psi Effective Mass Squared - 1st appearance}
\end{equation}
Therefore we require
\begin{equation}
m_{\psi} \ll H_{\ast}
\label{Eq.: Psi Mass << H_{*}}
\end{equation}
and
\begin{equation}
h\phi_{\ast}\left(\frac{\psi_{\ast}}{M_{P}}\right)^{\alpha-1} \ll H_{\ast}
\label{Eq.: 2nd Term Psi Effective Mass << H_{*}}
\end{equation}
where $\phi_{\ast}$ and $\psi_{\ast}$ are the values of $\phi$ and $\psi$ during primordial inflation respectively.\\
\indent We require that $\psi$ remains at $\psi_{\ast}$, the value during primordial inflation, all the way up to the end of thermal inflation. The reason for this is that if $\psi$ starts to move, then its perturbation will decrease. This is because $\psi$ unfreezes when the Hubble parameter becomes less than $\psi$'s mass, i.e. $H\!<\!m_{\psi}$. In this case, the perturbation of $\psi$ also unfreezes, because it has the same mass as $\psi$. The density of the oscillating $\psi$ field decreases as matter, so $m_{\psi}^{2}\psi^{2}\!\propto\!a^{-3}$. Therefore, $\psi\!\propto\!a^{-\frac{3}{2}}$. The same is true for the perturbation, i.e. $\delta\psi\!\propto\!a^{-\frac{3}{2}}$. This means that the perturbation decreases exponentially (as $a\!\propto\!e^{Ht}$) and so the whole effect of perturbing the end of thermal inflation is diminished. Requiring that $\psi$ is light at all times up until the end of thermal inflation is sufficient to ensure that the field and its perturbation remain at $\psi_{\ast}$ and $\delta\psi_{\ast}$ respectively. Therefore we require
\begin{equation}
m_{\psi} \ll H_{\textrm{TI}}
\label{Eq.: Psi Mass << H_{TI}}
\end{equation}
which is of course stronger than requiring just $m_{\psi}\!\ll\!H_{\ast}$, \cref{Eq.: Psi Mass << H_{*}}.\\
\indent Given that we have not observed any $\psi$ particles, the most liberal constraint on the present value of the effective mass of $\psi$ is
\begin{equation}
m_{\psi,\textrm{now}} \gtrsim 1\ \mbox{TeV}
\label{Eq.: m_{psi,now} gtrsim 1 TeV}
\end{equation}
\subsubsection{The Field Value $\psi_{\ast}$}
\label{Subsubsection: The Field Value psi_{*}}
Substituting the observed spectrum value $\mathcal{P}_{\zeta}(k_{0})\!=\!2.142\times10^{-9}$ into both \cref{Eq.: Power Spectrum (Gamma_{varphi} gtrsim H_{TI})} and \cref{Eq.: Power Spectrum (Gamma_{varphi} << H_{TI})}, with $\alpha \sim 1$, gives the same constraint, which is
\begin{equation}
\psi_{\ast} \sim \left(\frac{10^{-4}m_{0}^{2}M_{P}^{2\alpha-2}}{h^{2}H_{\ast}}\right)^{\frac{1}{2\alpha-1}}
\label{Eq.: Psi_{*}}
\end{equation}
This constraint automatically satisfies the requirement of a suitable perturbative expansion, \cref{Eq.: Suitable Perturbative Expansion}. Substituting \cref{Eq.: Psi_{*}} into \cref{Eq.: m_{0} >> Coupling Term}, regarding the dynamics of thermal inflation, gives
\begin{equation}
h \gg \left(\frac{10^{-4}}{H_{\ast}}\right)^{\alpha}m_{0}\,M_{P}^{\alpha-1}
\label{Eq.: h Constraint h}
\end{equation}
Rearranging this for $m_{0}$ gives the constraint
\begin{equation}
m_{0} \ll \left(10^{4}H_{\ast}\right)^{\alpha}\frac{h}{M_{P}^{\alpha-1}}
\label{Eq.: m_{0} Constraint h}
\end{equation}
\indent We require the field value of $\psi$ to be much larger than its perturbation, i.e. $\psi_{\ast} \gg \delta\psi_{\ast}$, so that the perturbative approach is valid. Therefore we obtain, with $\delta\psi_{\ast} \sim H_{\ast}$,
\begin{equation}
\psi_{\ast} \gg H_{\ast}
\label{Eq.: Psi_{*} >> H_{*}}
\end{equation}
and
\begin{equation}
\frac{\delta\psi_{\ast}}{\psi_{\ast}} \ll 1
\label{Eq.: Contrast Constraint a}
\end{equation}
Combining the frozen value $\psi_{\ast}$, \cref{Eq.: Psi_{*}}, with $\psi_{\ast}\!\gg\!\delta\psi_{\ast}$ and $\delta\psi_{\ast}^{2\alpha-1}\!\sim\!H_{\ast}^{2\alpha-1}$ gives
\begin{equation}
m_{0} \gg 100h\frac{H_{\ast}^{\alpha}}{M_{P}^{\alpha-1}}
\label{Eq.: m_{0} Constraint e}
\end{equation}
\subsubsection{Thermal Fluctuation of $\phi$}
\label{Subsubsection: Thermal Fluctuation of phi}
The effective mass of $\phi$ at the end of primordial inflation is
\begin{equation}
m_{\phi,\textrm{end,inf}}^{2} \sim g^{2}T_{\textrm{end,inf}}^{2} - m_{0}^{2}
\label{Eq.: phi Effective Mass End Prim. Inf. Squared}
\end{equation}
We have $gT_{\textrm{end,inf}}\!\gg\!m_{0}$, which therefore gives
\begin{equation}
m_{\phi,\textrm{end,inf}} \sim g\,T_{\textrm{end,inf}}
\label{Eq.: phi Effective Mass End Prim. Inf.}
\end{equation}
The proof is as follows. Instead of considering the time of the end of primordial inflation, we consider the time at the start of thermal inflation, which is of course a time at a lower temperature. For the case $\Gamma_{\varphi}\!\gtrsim\!H_{\textrm{TI}}$, we would be saying that
\begin{align}
gT_{1} &\gg m_{0}
\\
gV_{0}^{\frac{1}{4}} &\gg m_{0}
\label{Eq.: gV_{0}^{1/4} >> m_{0}}
\end{align}
Substituting $V_{0}$, \cref{Eq.: V_{0}}, into \cref{Eq.: gV_{0}^{1/4} >> m_{0}} gives
\begin{equation}
m_{0} \ll \left(\frac{g^{2n+2}}{\sqrt{\lambda}}\right)^{\frac{1}{n}}M_{P}
\label{Eq.: m_{0} Constraint q (Gamma_{varphi} gtrsim H_{TI})}
\end{equation}
which is identical to \cref{Eq.: m_{0} Constraint c (Gamma_{varphi} gtrsim H_{TI})} except for the difference in the limit. For the case $\Gamma_{\varphi}\!\ll\!H_{\textrm{TI}}$, we would be saying that
\begin{align}
gT_{1} &\gg m_{0}
\\
g\left(M_{P}^{2}\,H_{\textrm{TI}}\Gamma_{\varphi}\right)^{\frac{1}{4}} &\gg m_{0}
\label{Eq.: g(M_{P}^{2}H_{TI}Gamma_{varphi})^{1/4} >> m_{0}}
\end{align}
Substituting $H_{\textrm{TI}}$, \cref{Eq.: H_{TI}}, into \cref{Eq.: g(M_{P}^{2}H_{TI}Gamma_{varphi})^{1/4} >> m_{0}} gives
\begin{equation}
m_{0} \ll \left[\left(g^{4}\Gamma_{\varphi}\right)^{n+1}\frac{M_{P}^{2n+1}}{\sqrt{\lambda}}\right]^{\frac{1}{3n+2}}
\label{Eq.: m_{0} Constraint q (Gamma_{varphi} << H_{TI})}
\end{equation}
which is identical to \cref{Eq.: m_{0} Constraint c (Gamma_{varphi} << H_{TI})} except for the difference in the limit.\\
\indent As we are dealing with the thermal fluctuation of $\phi$ about $\phi\!=\!0$, we have $\left\langle\delta\phi\right\rangle_{T}\!=\!\left\langle\phi\right\rangle_{T}$. The thermal fluctuation of $\phi$ is
\begin{equation}
\sqrt{\left\langle\phi^{2}\right\rangle_{T}} \sim T
\label{Eq.: phi Thermal Fluctuation}
\end{equation}
and we require
\begin{equation}
g \ll 1
\label{Eq.: g << 1 (Therm. Fluc. phi Subsubsection)}
\end{equation}
A detailed derivation of this is given in \cref{Appendix: Derivation of Thermal Fluctuation of phi}.\\
\indent In order to keep $m_{\psi,\textrm{eff}}$ light, we require, from \cref{Subsubsection: Light psi - End of Inflation Subsubsection},
\begin{equation}
hT_{1}\left(\frac{\psi_{\ast}}{M_{P}}\right)^{\alpha-1} \ll H_{\textrm{TI}}
\label{Eq.: Thermal Fluctuation Constraint a}
\end{equation}
During the time between the end of primordial inflation and primordial inflation reheating, $T\!\propto\!a^{-\frac{3}{8}}$ and $H\!\propto\!a^{-\frac{3}{2}}$ and during radiation domination, $T\!\propto\!a^{-1}$ and $H\!\propto\!a^{-2}$. Therefore, if \cref{Eq.: Thermal Fluctuation Constraint a} is satisfied, then equivalent constraints for higher $T$ and $H$ are guaranteed to be satisfied as well. In the case of $\Gamma_{\varphi}\!\gtrsim\!H_{\textrm{TI}}$, by substituting \cref{Eq.: V_{0},Eq.: H_{TI},Eq.: Psi_{*},Eq.: T_{1} (Gamma_{varphi} gtrsim H_{TI})} into \cref{Eq.: Thermal Fluctuation Constraint a} we obtain the constraint
\begin{equation}
h \ll \frac{1}{\lambda^{\frac{2\alpha-1}{4n+4}}}\left(\frac{m_{0}}{M_{P}}\right)^{\frac{\left(2\alpha-1\right)\left(n+2\right)}{2n+2}}\left(\frac{100\sqrt{H_{\ast}M_{P}}}{m_{0}}\right)^{2\alpha-2}
\label{Eq.: h Constraint e (Gamma_{varphi} gtrsim H_{TI})}
\end{equation}
Rearranging this for $m_{0}$ gives the following. For $\alpha\!=\!1$ and any value of $n$, or for $\alpha\!=\!2$ and $n\!=\!1$ we have
\begin{equation}
m_{0} \gg \left(\frac{\left(10^{4}H_{\ast}\right)^{\left(2\alpha-2\right)\left(n+1\right)}}{h^{2n+2}\lambda^{\frac{2\alpha-1}{2}}M_{P}^{2\alpha+n}}\right)^{\frac{1}{2\alpha n -3n-2}}
\label{Eq.: m_{0} Constraint i (Gamma_{varphi} gtrsim H_{TI})}
\end{equation}
For all other $\alpha$ and $n$ combinations, the above inequality is reversed, giving
\begin{equation}
m_{0} \ll \left(\frac{\left(10^{4}H_{\ast}\right)^{\left(2\alpha-2\right)\left(n+1\right)}}{h^{2n+2}\lambda^{\frac{2\alpha-1}{2}}M_{P}^{2\alpha+n}}\right)^{\frac{1}{2\alpha n -3n-2}}
\label{Eq.: m_{0} Constraint v (Gamma_{varphi} gtrsim H_{TI})}
\end{equation}
In the case of $\Gamma_{\varphi}\!\ll\!H_{\textrm{TI}}$, by substituting \cref{Eq.: H_{TI},Eq.: Psi_{*},Eq.: T_{1} (Gamma_{varphi} << H_{TI})} into \cref{Eq.: Thermal Fluctuation Constraint a} we obtain the constraint
\begin{equation}
h \ll \left(\frac{m_{0}^{n+2}}{\sqrt{\lambda}\,M_{P}}\right)^{\frac{6\alpha-3}{4n+4}}\frac{1}{\left(M_{P}\sqrt{\Gamma_{\varphi}}\right)^{\frac{2\alpha-1}{2}}}\left(\frac{100\sqrt{H_{\ast}M_{P}}}{m_{0}}\right)^{2\alpha-2}
\label{Eq.: h Constraint e (Gamma_{varphi} << H_{TI})}
\end{equation}
Rearranging this for $m_{0}$ gives the following. For the values of $n$ and $\alpha$ given in \cref{Table: Values for "Eq.: m_{0} Constraint i (Gamma_{varphi} << H_{TI})"} we have
\begin{equation}
m_{0} \gg \left(\frac{\left(10^{4}H_{\ast}M_{P}\right)^{\left(4\alpha-4\right)\left(n+1\right)}}{h^{4n+4}\left(\sqrt{\lambda}\,M_{P}\right)^{6\alpha-3}\left(M_{P}^{2}\,\Gamma_{\varphi}\right)^{\left(2\alpha-1\right)\left(n+1\right)}}\right)^{\frac{1}{2\alpha n -4\alpha-5n-2}}
\label{Eq.: m_{0} Constraint i (Gamma_{varphi} << H_{TI})}
\end{equation}
\begin{table}[h!]
\centering
\begin{tabular}{c @{\hskip 30pt} c}
$n$ & $\alpha$\\
\hline
\\
[-8pt]
$1$ & $3\text{--}\infty$\\
[2pt]
$2$ & All\\
[2pt]
$3$ & $1\text{--}8$\\
[2pt]
$4$ & $1\text{--}5$\\
[2pt]
$5$ & $1\text{--}4$\\
[2pt]
$6\text{--}13$ & $1\text{--}3$\\
[2pt]
$14\text{--}\infty$ & $1,2$
\end{tabular}
\caption{Values for which \cref{Eq.: m_{0} Constraint i (Gamma_{varphi} << H_{TI})} applies.}
\label{Table: Values for "Eq.: m_{0} Constraint i (Gamma_{varphi} << H_{TI})"}
\end{table}
while for all other $n$ and $\alpha$ values, the above inequality is reversed, giving
\begin{equation}
m_{0} \ll \left(\frac{\left(10^{4}H_{\ast}M_{P}\right)^{\left(4\alpha-4\right)\left(n+1\right)}}{h^{4n+4}\left(\sqrt{\lambda}\,M_{P}\right)^{6\alpha-3}\left(M_{P}^{2}\,\Gamma_{\varphi}\right)^{\left(2\alpha-1\right)\left(n+1\right)}}\right)^{\frac{1}{2\alpha n -4\alpha-5n-2}}
\label{Eq.: m_{0} Constraint v (Gamma_{varphi} << H_{TI})}
\end{equation}
\subsubsection{Thermalization of $\phi$}
\label{Subsubsection: Thermalization of phi}
In order that $\phi$ interacts with the thermal bath and therefore that we actually have the $g^{2}T^{2}\phi^{2}$ term in our potential, \cref{Eq.: Full Potential}, we require
\begin{equation}
\Gamma_{\textrm{therm}} > H
\label{Eq.: Gamma_{therm} > H}
\end{equation}
where $\Gamma_{\textrm{therm}}$ is the thermalization rate of $\phi$, which is given by
\begin{align}
\Gamma_{\textrm{therm}} &= n\left\langle\sigma v\right\rangle
\\
&\sim \sigma\,T^{3}
\label{Eq.: Thermalization Rate Definition}
\end{align}
where $n\!\sim\!T^{3}$ is the number density of particles in the thermal bath, $\sigma$ is the scattering cross-section for the interaction of $\phi$ and the particles in the thermal bath, $v$ is the relative velocity between a $\phi$ particle and a thermal bath particle (which in our case is $\approx\!c\!=\!1$) and $\left\langle\,\right\rangle$ denotes a thermal average. The scattering cross-section $\sigma$ is given by
\begin{equation}
\sigma \sim \frac{g^{4}}{E_{\textrm{c.m.}}^{2}}
\label{Eq.: Cross-section -- Thermalization}
\end{equation}
where $E_{\textrm{c.m.}}$ is the centre-of-mass energy, which is
\begin{equation}
E_{\textrm{c.m.}} \sim T
\label{Eq.: Centre-of-mass Energy}
\end{equation}
Substituting \cref{Eq.: Centre-of-mass Energy} into \cref{Eq.: Cross-section -- Thermalization} gives
\begin{equation}
\sigma \sim \frac{g^{4}}{T^{2}}
\label{Eq.: Cross-section Final Result -- Thermalization}
\end{equation}
This scattering cross-section is the total cross-section for all types of scattering (e.g. elastic) that can take place between $\phi$ and the particles in the thermal bath. For a complete Field Theory derivation of the elastic scattering cross-section between $\phi$ and the thermal bath, see \cref{Appendix: Field Theory Derivation Elastic Scattering Cross-section}. The thermalization rate now becomes
\begin{equation}
\Gamma_{\textrm{therm}} \sim g^{4}T
\label{Eq.: Thermalization Rate}
\end{equation}
During the time between the end of primordial inflation and primordial inflation reheating, $T\!\propto\!a^{-\frac{3}{8}}$ and $H\!\propto\!a^{-\frac{3}{2}}$ and during radiation domination $T\!\propto\!a^{-1}$ and $H\!\propto\!a^{-2}$. Therefore, if the constraint $\Gamma_{\textrm{therm}}\!>\!H$ is satisfied at the time of the end of primordial inflation, then it is satisfied all the way up to the start of thermal inflation. Therefore we have the constraint
\begin{equation}
\Gamma_{\textrm{therm}} > H_{\ast}
\label{Eq.: Gamma_{therm} > H_{*}}
\end{equation}
Taking \cref{Eq.: Thermalization Rate} with $T\!\sim\!\left(M_{P}^{2}\,H_{\ast}\,\Gamma_{\varphi}\right)^{\frac{1}{4}}$ gives
\begin{equation}
\Gamma_{\varphi} > \frac{H_{\ast}^{3}}{g^{16}M_{P}^{2}}
\label{Eq.: Gamma_{varphi} Constraint a}
\end{equation}
We also require $\Gamma_{\textrm{therm}}\!>\!H$ to be satisfied throughout the whole of thermal inflation. Therefore, we have the constraint
\begin{equation}
g^{4}\,T_{2} > H_{\textrm{TI}}
\label{Eq.: g^{4}T_{2} > H_{TI}}
\end{equation}
Substituting $H_{\textrm{TI}}$ and $T_{2}$, \cref{Eq.: T_{2},Eq.: H_{TI}} respectively, into the above gives
\begin{equation}
m_{0} < \left(\frac{g^{3}}{\sqrt{2}}\right)^{n+1}\sqrt{\lambda}\,M_{P}
\label{Eq.: m_{0} Constraint u}
\end{equation}
\subsubsection{The Field Value $\phi_{\ast}$}
\label{Subsubsection: The Field Value phi_{*}}
We consider three possible cases for the value of the thermal waterfall field $\phi$ during primordial inflation, with $m_{\phi,\textrm{inf}}$ being the effective mass of $\phi$ during primordial inflation:
\begin{enumerate}[label=\Alph*)]
\item $\phi$ heavy, i.e. $|m_{\phi,\textrm{inf}}| \gg H_{\ast}$, in which $\phi$ rolls down to its VEV.
\item $\phi$ light, i.e. $|m_{\phi,\textrm{inf}}| \ll H_{\ast}$, in which $\phi$ is at the Bunch-Davies value (to be explained below).
\item $\phi$ light, in which a SUGRA correction to the potential is appreciable, with $\phi$ rolling down to $\phi=0$.
\end{enumerate}
\underline{\textbf{Case A}}\vspace{0.2em}
\\
Substituting $\left\langle\phi\right\rangle$ and $\psi_{\ast}$, \cref{Eq.: Phi VEV,Eq.: Psi_{*}} respectively, into \cref{Eq.: 2nd Term Psi Effective Mass << H_{*}} gives
\begin{equation}
h \ll H_{\ast}^{3\alpha-2}\left(\frac{\sqrt{\left(2n+4\right)\lambda}}{m_{0}M_{P}^{n}}\right)^{\frac{2\alpha-1}{n+1}}\left(\frac{10^{4}M_{P}}{m_{0}^{2}}\right)^{\alpha-1}
\label{Eq.: h Constraint f}
\end{equation}
Rearranging this for $m_{0}$ gives
\begin{equation}
m_{0} \ll \left[\left(\frac{\sqrt{\left(2n+4\right)\lambda}}{M_{P}^{n}}\right)^{2\alpha-1}\left(\frac{\left(10^{4}M_{P}\right)^{\alpha-1}H_{\ast}^{3\alpha-2}}{h}\right)^{n+1}\right]^{\frac{1}{2\alpha n +4\alpha-2n-3}}
\label{Eq.: m_{0} Constraint m}
\end{equation}
\\
\underline{\textbf{Case B}}\vspace{0.2em}
\\
We consider $\phi$ to be at the Bunch-Davies value
\begin{equation}
\phi_{\textrm{BD}} \sim \left(\frac{M_{P}^{n}H_{\ast}^{2}}{\sqrt{\lambda}}\right)^{\frac{1}{n+2}}
\label{Eq.: Phi_* B.-D. Value}
\end{equation}
corresponding to the Bunch-Davies vacuum \cite{Bunch_&_Davies:Quan._Field_Theory_de_Sit._Space:Renorm._Point_Splitting}, which is the unique quantum state that corresponds to the vacuum, i.e. no particle quanta, in the infinite past in conformal time in a de Sitter spacetime. $\phi_{\textrm{BD}}$ is of this form as $\lambda\frac{\phi^{2n+4}}{M_{P}^{2n}} \sim H_{\ast}^{4}$, this being because the probability of this Bunch-Davies state is proportional to the factor $e^{-\frac{V}{H^{4}}}$. Substituting $\psi_{\ast}$ and $\phi_{\textrm{BD}}$, \cref{Eq.: Phi_* B.-D. Value,Eq.: Psi_{*}} respectively, into \cref{Eq.: 2nd Term Psi Effective Mass << H_{*}} gives
\begin{equation}
h \ll H_{\ast}^{3\alpha-2}\left(\frac{100\sqrt{M_{P}}}{m_{0}}\right)^{2\alpha-2}\left(\frac{\sqrt{\lambda}}{M_{P}^{n}H_{\ast}^{2}}\right)^{\frac{2\alpha-1}{n+2}}
\label{Eq.: h Constraint g}
\end{equation}
Rearranging this for $m_{0}$ gives
\begin{equation}
m_{0} \ll 100\sqrt{M_{P}}\left[\frac{H_{\ast}^{3\alpha-2}}{h}\left(\frac{\sqrt{\lambda}}{M_{P}^{n}H_{\ast}^{2}}\right)^{\frac{2\alpha-1}{n+2}}\right]^{\frac{1}{2\alpha-2}}
\label{Eq.: m_{0} Constraint n}
\end{equation}
\\
\underline{\textbf{Case C}}\vspace{0.2em}
\\
If $\phi$ is light during primordial inflation, i.e. $|m_{\phi,\textrm{inf}}|\!\ll\!H_{\ast}$, our potential, \cref{Eq.: Full Potential}, can receive an appreciable SUGRA correction during primordial inflation of \cite{Dine_et_al.:SUSY_Break._Early_Uni. , Dine_et_al.:Baryo._Flat_Dirs._Supersym._SM , Lyth_&_Moroi:Masses_Weakly_Coupled_Scalar_Fields_Early_Uni.}
\begin{equation}
\Delta V \sim c_{\ast}H_{\ast}^{2}\phi^{2}
\label{Eq.: SUGRA Correction}
\end{equation}
where $c_{\ast}$ is a coupling constant. The SUGRA correction is appreciable only from the time of primordial inflation up until primordial inflation reheating, as it is suppressed at all times after this \cite{Lyth_&_Moroi:Masses_Weakly_Coupled_Scalar_Fields_Early_Uni.}. As the scale factor $a$ grows (almost) exponentially during inflation, $\phi$ is driven rapidly to $0$, i.e. we have $\phi_{\ast}=0$. Therefore the effective mass of $\phi$ during primordial inflation is
\begin{equation}
m_{\phi,\textrm{inf}}^{2} \sim -m_{0}^{2} + c_{\ast}H_{\ast}^{2}
\label{Eq.: Phi Effective Mass Prim. Inf. Squared}
\end{equation}
In order to keep this light we therefore need
\begin{equation}
m_{0} \ll H_{\ast}
\label{Eq.: m_{0} Constraint l}
\end{equation}
and
\begin{equation}
c_{\ast} < 1
\label{Eq.: SUGRA c < 1}
\end{equation}
\subsubsection{Energy Density of the Thermal Waterfall Field}
\label{Subsubsection: Energy Density of the Thermal Waterfall Field}
We require the energy density of $\phi$ to be subdominant at all times, in order that it does not cause any inflation by itself. During the period between the end of primordial inflation and the start of thermal inflation, the energy density of $\phi$ is
\begin{align}
\rho_{\phi} &\sim g^{2}T^{2}\phi^{2}
\\
&\sim g^{2}T^{4}
\label{Eq.: phi Energy Density between End Primordial Inflation and Start Thermal Inflation}
\end{align}
the second line coming from the thermal fluctuation of $\phi$, which is $\sim\!T$. Therefore, considering the Friedmann equation, we require
\begin{equation}
g\,T_{1}^{2} \ll M_{P}H_{\textrm{TI}}
\label{Eq.: gT_{1}^{2} << M_{P}H_{TI}}
\end{equation}
During the time between the end of primordial inflation and primordial inflation reheating, $T\!\propto\!a^{-\frac{3}{8}}$ and $H\!\propto\!a^{-\frac{3}{2}}$ and during radiation domination $T\!\propto\!a^{-1}$ and $H\!\propto\!a^{-2}$. Therefore, if \cref{Eq.: gT_{1}^{2} << M_{P}H_{TI}} is satisfied, then equivalent constraints for higher $T$ and $H$ are guaranteed to be satisfied as well. For the case $\Gamma_{\varphi}\!\gtrsim\!H_{\textrm{TI}}$, by substituting $H_{\textrm{TI}}$ and $T_{1}$, \cref{Eq.: T_{1} (Gamma_{varphi} gtrsim H_{TI}),Eq.: H_{TI}} respectively, into \cref{Eq.: gT_{1}^{2} << M_{P}H_{TI}} we obtain
\begin{equation}
g \ll 1
\label{Eq.: g << 1 (phi Ener. Den.)}
\end{equation}
which is the same constraint as \cref{Eq.: g << 1 (Therm. Fluc. phi Subsubsection)}. For the case $\Gamma_{\varphi}\!\ll\!H_{\textrm{TI}}$, by substituting $H_{\textrm{TI}}$ and $T_{1}$, \cref{Eq.: T_{1} (Gamma_{varphi} << H_{TI}),Eq.: H_{TI}} respectively, into \cref{Eq.: gT_{1}^{2} << M_{P}H_{TI}} we obtain
\begin{equation}
m_{0} \gg \left[\left(g^{2}\Gamma_{\varphi}\right)^{n+1}\sqrt{\lambda}\,M_{P}\right]^{\frac{1}{n+2}}
\label{Eq.: m_{0} Constraint w}
\end{equation}
\\
\underline{\textbf{$\phi_{\ast}$ Case A}}\vspace{0.2em}
\\
The energy density of $\phi$ during primordial inflation is
\begin{align}
\rho_{\phi,\textrm{inf}} &= \left(-\frac{1}{2}m_{0}^{2} + h^{2}\frac{\psi_{\ast}^{2\alpha}}{M_{P}^{2\alpha-2}}\right)\left\langle\phi\right\rangle^{2} + \lambda\frac{\left\langle\phi\right\rangle^{2n+4}}{M_{P}^{2n}}
\\
\nonumber
\\
&\sim -\frac{1}{2}m_{0}^{2}\left\langle\phi\right\rangle^{2} + \lambda\frac{\left\langle\phi\right\rangle^{2n+4}}{M_{P}^{2n}}
\label{Eq.: Phi Ener. Den. Prim. Inf. - Phi_{*} Case A}
\end{align}
with the second line coming from \cref{Eq.: m_{0} >> Coupling Term} regarding the dynamics of thermal inflation. Therefore, with the energy density of the Universe being $\sim\!M_{P}^{2}H_{\ast}^{2}$, we require
\begin{equation}
m_{0}\left\langle\phi\right\rangle \ll M_{P}H_{\ast}
\label{Eq.: Phi Ener. Den. Prim. Inf. - Phi_{*} Case A Constraint 1}
\end{equation}
and
\begin{equation}
\sqrt{\lambda}\left\langle\phi\right\rangle^{n+2} \ll M_{P}^{n+1}H_{\ast}
\label{Eq.: Phi Ener. Den. Prim. Inf. - Phi_{*} Case A Constraint 2}
\end{equation}
Substituting $\left\langle\phi\right\rangle$, \cref{Eq.: Phi VEV}, into \cref{Eq.: Phi Ener. Den. Prim. Inf. - Phi_{*} Case A Constraint 1} gives the same constraint as from substituting $\left\langle\phi\right\rangle$ into \cref{Eq.: Phi Ener. Den. Prim. Inf. - Phi_{*} Case A Constraint 2}. This constraint is
\begin{equation}
m_{0} \ll \left(\sqrt{\lambda}\,M_{P}H_{\ast}^{n+1}\right)^{\frac{1}{n+2}}
\label{Eq.: m_{0} Constraint o}
\end{equation}
However, for all viable parameter values in our model, this constraint is never the dominant constraint when we consider it alongside all of the other constraints that are detailed in this thesis for this Thermal Inflation model.\\
\\
\underline{\textbf{$\phi_{\ast}$ Case B}}\vspace{0.2em}
\\
The energy density of $\phi$ during primordial inflation is
\begin{align}
\rho_{\phi,\textrm{inf}} &= \left(-\frac{1}{2}m_{0}^{2} + h^{2}\frac{\psi_{\ast}^{2\alpha}}{M_{P}^{2\alpha-2}}\right)\phi_{\textrm{BD}}^{2} + \lambda\frac{\phi_{\textrm{BD}}^{2n+4}}{M_{P}^{2n}}
\\
\nonumber
\\
&\sim -\frac{1}{2}m_{0}^{2}\phi_{\textrm{BD}}^{2} + \lambda\frac{\phi_{\textrm{BD}}^{2n+4}}{M_{P}^{2n}}
\label{Eq.: Phi Ener. Den. Prim. Inf. - Phi_{*} Case B}
\end{align}
with the second line coming from \cref{Eq.: m_{0} >> Coupling Term} regarding the dynamics of thermal inflation. Therefore, with the energy density of the Universe being $\sim\!M_{P}^{2}H_{\ast}^{2}$, we require
\begin{equation}
m_{0}\,\phi_{\textrm{BD}} \ll M_{P}H_{\ast}
\label{Eq.: Phi Ener. Den. Prim. Inf. - Phi_{*} Case B Constraint 1}
\end{equation}
and
\begin{equation}
\sqrt{\lambda}\,\phi_{\textrm{BD}}^{n+2} \ll M_{P}^{n+1}H_{\ast}
\label{Eq.: Phi Ener. Den. Prim. Inf. - Phi_{*} Case B Constraint 2}
\end{equation}
Substituting $\phi_{\textrm{BD}}$, \cref{Eq.: Phi_* B.-D. Value}, into \cref{Eq.: Phi Ener. Den. Prim. Inf. - Phi_{*} Case B Constraint 1} gives
\begin{equation}
m_{0} \ll \left(\sqrt{\lambda}\,M_{P}^{2}H_{\ast}^{n}\right)^{\frac{1}{n+2}}
\label{Eq.: m_{0} Constraint p}
\end{equation}
and substituting $\phi_{\textrm{BD}}$ into \cref{Eq.: Phi Ener. Den. Prim. Inf. - Phi_{*} Case B Constraint 2} gives just
\begin{equation}
H_{\ast} \ll M_{P}
\label{Eq.: H_{*} << M_{P}}
\end{equation}
However, for all viable parameter values in our model, \cref{Eq.: m_{0} Constraint p,Eq.: H_{*} << M_{P}} are both individually never the dominant constraint when we consider them alongside all of the other constraints that are detailed in this thesis for this Thermal Inflation model.\\
\\
\underline{\textbf{$\phi_{\ast}$ Case C}}\vspace{0.2em}
\\
We do not have any additional constraints here for this case, as $\phi=0$.
\subsubsection{Time of Transition from Thermal Inflation to Thermal Waterfall Field Oscillation}
\label{Subsubsection: Time of Transition from Thermal Inflation to Thermal Waterfall Field Oscillation}
In order for the equations of the $\delta N$ formalism that are derived within the context of the ``end of inflation'' mechanism to be valid, we require the transition from thermal inflation to thermal waterfall field oscillation to be sufficiently fast \cite{Lyth:Hybrid_Waterfall_and_Curvature_Perturbation}. More specifically, we require
\begin{equation}
\Delta t \ll \delta t_{1\rightarrow2}
\label{Eq.: Delta t << delta t_{1->2}}
\end{equation}
where $\Delta t\!\equiv\!t_{2}-t_{1}$ is the time taken for the transition to occur and $\delta t_{1\rightarrow2}$ is the proper time between a uniform energy density spacetime slice just before the transition at $t_{1}$ and one just after the transition at $t_{2}$ when $\phi$ starts to oscillate around its VEV. Qualitatively, we require the thickness of the transition slice to be much smaller than its warping. The primordial curvature perturbation that is generated by the ``end of inflation'' mechanism is given by
\begin{equation}
\zeta = H_{\textrm{TI}}\, \delta t_{1\rightarrow2}
\label{Eq.: zeta = H delta t_{1->2}}
\end{equation}
Therefore, from \cref{Eq.: Delta t << delta t_{1->2}} we require
\begin{equation}
\zeta \gg H_{\textrm{TI}} \Delta t
\label{Eq.: zeta >> H Delta t}
\end{equation}
\indent To calculate $\phi_{1}$ and $\phi_{2}$, the value of $\phi$ at times $t_{1}$ and $t_{2}$ respectively, we use the fact that the process is so rapid that it takes place in less than a Hubble time, so that the Universe expansion can be ignored. Then the equation of motion is
\begin{equation}
\ddot{\phi} + \frac{\partial V}{\partial \phi} = 0
\label{Eq.: Fast-roll EoM}
\end{equation}
At the end of thermal inflation, $\phi$ is not centered on the origin, but has started to roll down the potential slightly. At this time, $g^{2}T^{2}$ is much smaller than $m_{0}^{2}$. Therefore we have
\begin{equation}
\frac{\partial V}{\partial \phi} \sim -m_{0}^{2}\phi
\label{Eq.: V' at end of inflation}
\end{equation}
So we have the equation of motion
\begin{equation}
\ddot{\phi} \sim m_{0}^{2}\phi
\label{Eq.: Fast-roll EoM -- Phi}
\end{equation}
The solution is
\begin{equation}
\phi \sim Ae^{m_{0}t}
\label{Eq.: Fast-roll Phi Solution}
\end{equation}
where $A$ is a constant and we are keeping only the growing mode. Therefore we have
\begin{align}
\ln{\left(\frac{\phi_{2}}{\phi_{1}}\right)} &\sim m_{0}\left(t_{2}-t_{1}\right)
\\
\nonumber
\\
&\sim m_{0} \Delta t
\label{Eq.: ln(Phi_{2}/Phi_{1})}
\end{align}
We know that
\begin{equation}
\phi_{1} \sim T \sim m_{0}
\label{Eq.: Phi_{1}}
\end{equation}
and
\begin{equation}
\phi_{2} \sim \left\langle\phi\right\rangle
\label{Eq.: Phi_{2}}
\end{equation}
Therefore we have
\begin{equation}
\ln{\left(\left[\frac{1}{\sqrt{\left(2n+4\right)\lambda}}\left(\frac{M_{P}}{m_{0}}\right)^{n}\right]^{\frac{1}{n+1}}\right)} \sim m_{0} \Delta t
\label{Eq.: ln() sim m_{0} Delta T Final}
\end{equation}
For all values of $n$, $\lambda$ and $m_{0}$, we have $\Delta t\!\geq\!m_{0}^{-1}$. Therefore, from \cref{Eq.: zeta >> H Delta t} we have
\begin{equation}
\zeta \gg \frac{H_{\textrm{TI}}}{m_{0}}
\label{Eq.: zeta >> H/m_{0}}
\end{equation}
Given that $\zeta\sim10^{-5}$, we require
\begin{equation}
H_{\textrm{TI}} \ll 10^{-5}m_{0}
\label{Eq.: H_{TI} Constraint e}
\end{equation}
\indent We obtain an additional constraint by substituting \cref{Eq.: H_{TI} Constraint e} into the requirement of $m_{\psi}\!\ll\!H_{\textrm{TI}}$, \cref{Eq.: Psi Mass << H_{TI}}. This gives
\begin{equation}
m_{\psi} \ll 10^{-5}m_{0}
\label{Eq.: Psi Mass Constraint a}
\end{equation}
\indent A further constraint is obtained by substituting $H_{TI}$, \cref{Eq.: H_{TI}}, into \cref{Eq.: H_{TI} Constraint e}. We obtain
\begin{equation}
m_{0} \ll 10^{-5n-5}\sqrt{\lambda}\,M_{P}
\label{Eq.: m_{0} Constraint f}
\end{equation}
\subsubsection{Energy Density of the Oscillating Spectator Field}
\label{Subsubsection: Energy Density of the Oscillating Spectator Field}
As $\psi$ has acquired perturbations from primordial inflation, we require it not to dominant the energy density of the Universe after the end of thermal inflation when it is oscillating, at which time the effective mass of $\psi$ is increased significantly due to the coupling of $\psi$ to $\phi$. This is so as not to allow $\psi$ to act as a curvaton, i.e. not to allow $\psi$'s perturbations to generate a dominant contribution to the primordial curvature perturbation when $\psi$ decays. The reason for this is just so that we do not have a curvaton inflation scenario, as the perturbations that are generated via the modulated mass in our model that could give the dominant contribution to $\zeta$ would be negligible.\\
\indent The energy density of the oscillating $\psi$ field after the end of thermal inflation is
\begin{align}
\rho_{\psi,\textrm{osc}} &= h^{2}\frac{\overline{\psi}^{2\alpha}}{M_{P}^{2\alpha-2}}\overline{\phi}^{2} + \frac{1}{2}m_{\psi}^{2}\overline{\psi}^{2}
\\
\nonumber
\\
&\sim h^{2}\frac{\psi_{\ast}^{2\alpha}}{M_{P}^{2\alpha-2}}\left\langle\phi\right\rangle^{2} + \frac{1}{2}m_{\psi}^{2}\psi_{\ast}^{2}
\label{Eq.: Psi Energy Density Oscillating}
\end{align}
For simplicity, we assume that $\psi$ decays around the same time as $\phi$, i.e. that $H$ does not change much between the time when $\phi$ decays and the time when $\psi$ decays. Therefore, the energy density of the Universe at the time when $\psi$ decays is $\sim\!M_{P}^{2}\Gamma^{2}$. We therefore require
\begin{align}
\rho_{\psi,\textrm{osc}} &\ll M_{P}^{2}\Gamma^{2}
\\
\nonumber
\\
h^{2}\frac{\psi_{\ast}^{2\alpha}}{M_{P}^{2\alpha-2}}\left\langle\phi\right\rangle^{2} + \frac{1}{2}m_{\psi}^{2}\psi_{\ast}^{2} &\ll M_{P}^{2}\Gamma^{2}
\label{Eq.: Psi Energy Density Oscillating Constraint a}
\end{align}
Therefore we require
\begin{equation}
m_{\psi} \ll \frac{M_{P}\Gamma}{\psi_{\ast}}
\label{Eq.: Psi Mass Constraint d - Enf of Inflation Section}
\end{equation}
and
\begin{equation}
h\left\langle\phi\right\rangle\psi_{\ast}^{\alpha} \ll M_{P}^{\alpha}\Gamma
\label{Eq.: Psi Energy Density Oscillating Constraint b}
\end{equation}
Substituting $\left\langle\phi\right\rangle$, $\Gamma$ and $\psi_{\ast}$, \cref{Eq.: Phi VEV,Eq.: Psi_{*},Eq.: phi Decay Rate - Decay Rate Subsection} respectively, into \cref{Eq.: Psi Energy Density Oscillating Constraint b} gives the constraint
\begin{equation}
h \gg \frac{1}{\left(g^{2}m_{0}\right)^{2\alpha-1}}\left(\frac{m_{0}M_{P}^{n}}{\sqrt{\left(2n+4\right)\lambda}}\right)^{\frac{2\alpha-1}{n+1}}\left(\frac{m_{0}}{100\sqrt{H_{\ast}M_{P}}}\right)^{2\alpha}
\label{Eq.: h Constraint d - End of Inflation Section}
\end{equation}
\subsection{Results}
\label{Subsection: ``End of Inflation'' Mechanism Results}
In this section, we combine the above constraints to find out the allowed parameter space (if any).
\subsubsection{The Case $\alpha=1$}
\label{Subsubsection: The Case alpha=1}
\begin{figure}[h!]
\begin{center}
\includegraphics[scale=1.2]{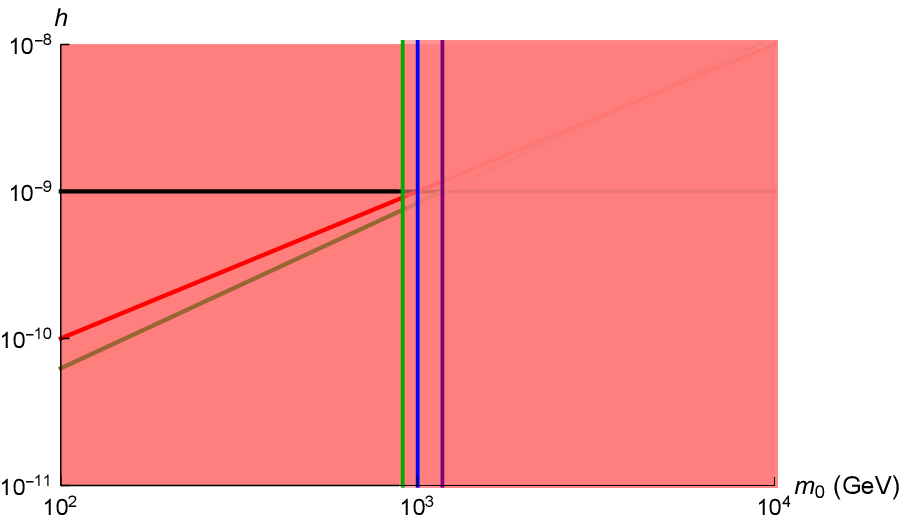}
\caption[Allowed parameter space for $h$]{The allowed parameter space for $h$, with $\alpha\!=\!1$, $\Gamma_{\varphi}\!\ll\!H_{TI}$ and the parameter values of \cref{Table: Free Parameter Values (alpha=1 and Gamma_{varphi}<<H_{TI})}. The \textcolor{black}{Black} line is $h\!=\!10^{-9}$, the \textcolor{red}{Red} line is the lower bound given by \cref{Eq.: h Constraint h} and the \textcolor{RawSienna}{Brown} line is the upper bound given by \cref{Eq.: h Constraint e (Gamma_{varphi} << H_{TI})}. The constraints on $m_{0}$ that are shown are the following: \textcolor{OliveGreen}{Green}: \cref{Eq.: m_{0} Constraint f} \ \ \textcolor{blue}{Blue}: \cref{Eq.: m_{0} Constraint h} \ \ \textcolor{violet}{Purple}: \cref{Eq.: m_{0} Constraint v (Gamma_{varphi} << H_{TI})}}
\label{Figure: h}
\end{center}
\end{figure}

Even with $\alpha$ set to the value $\alpha\!=\!1$, we still have six free parameters in the model. Therefore, the parameter space is very multi-dimensional and the number of allowed regions is potentially very vast. Given this, we show results only for one such allowed region of parameter space that we have explored. Additionally, we also only show results for the case where $\Gamma_{\varphi}\!\ll\!H_{\textrm{TI}}$, in that reheating from primordial inflation occurs at some time after the end of thermal inflation, as this scenario was found to yield more parameter space than the case where $\Gamma_{\varphi}\!\gtrsim\!H_{\textrm{TI}}$.\\
\indent From \cref{Eq.: g << 1 (Therm. Fluc. phi Subsubsection)} we require $g\!\ll\!1$. We also require the constraint given by \cref{Eq.: Gamma_{varphi} Constraint a} to be satisfied, where $g$ is present as $g^{-16}$. Therefore, this latter constraint will start to become very strong very quickly as we decrease $g$. We find that a value of $g\!=\!0.4$ yields allowed parameter space, for reasonable values of $H_{\ast}$ and $\Gamma_{\varphi}$. The parameter space that we find here however, when all constraints are considered together and regardless of the $\phi_{\ast}$ case, is actually a sharp prediction of single values for all but one of the free parameters and the other quantities in the model, to within an order of magnitude, rather than a range of parameter space. The values of the free parameters are displayed in \cref{Table: Free Parameter Values (alpha=1 and Gamma_{varphi}<<H_{TI})}.\\
\begin{table}[h!]
\centering
\begin{tabular}{c @{\hskip 30pt} c}
Parameter & Value\\
\hline
\\
[-8pt]
$n$ & $1$\\
[5pt]
$g$ & $0.4$\\
[5pt]
$H_{\ast}$ & $10^{8}\ \mbox{GeV}$\\
[5pt]
$\Gamma_{\varphi}$ & $4\times10^{-7}\ \mbox{GeV}$\\
[5pt]
$\lambda$ & $\frac{10^{-8}}{6!}$\\
[5pt]
$h$ & $10^{-9}$
\end{tabular}
\caption{Values of the free parameters for which parameter space exists, for $\alpha\!=\!1$ and $\Gamma_{\varphi}\!\ll\!H_{\textrm{TI}}$.}
\label{Table: Free Parameter Values (alpha=1 and Gamma_{varphi}<<H_{TI})}
\end{table}
\indent The allowed range of values for $\psi_{\ast}$, the contrast $\frac{\delta\psi_{\ast}}{\psi_{\ast}}$ and $H_{TI}$ are displayed in \cref{Figure: psi_{*},Figure: delta psi_{*} / psi_{*},Figure: H_{TI}} respectively. \Cref{Figure: m_{psi}} shows the allowed parameter space for $m_{\psi}$. Within the range $m_{0}\!\sim\!10^{2}\,\text{--}\,10^{3}\ \mbox{GeV}$, the mass $m_{\psi}$ can span many orders of magnitude, with only an upper limit of $\sim\!10^{-4}\,\text{--}\,10^{-2}\ \mbox{GeV}$. Within the model, there is no effective lower bound on $m_{\psi}$. Finally, \cref{Figure: h} shows the allowed parameter space for $h$, for a value of $h\!=\!10^{-9}$. From looking at this plot, it may initially seem as if there is no allowed parameter space available for our given value of $h$. However, as we are working within an order of magnitude for each value, we can see that all of the visible constraints do allow for our given value of $h$, for a thermal waterfall field mass value of $m_{0}\!\sim\!10^{3}\ \mbox{GeV}$.\\
\begin{figure}[h!]
\begin{center}
\includegraphics[scale=1.2]{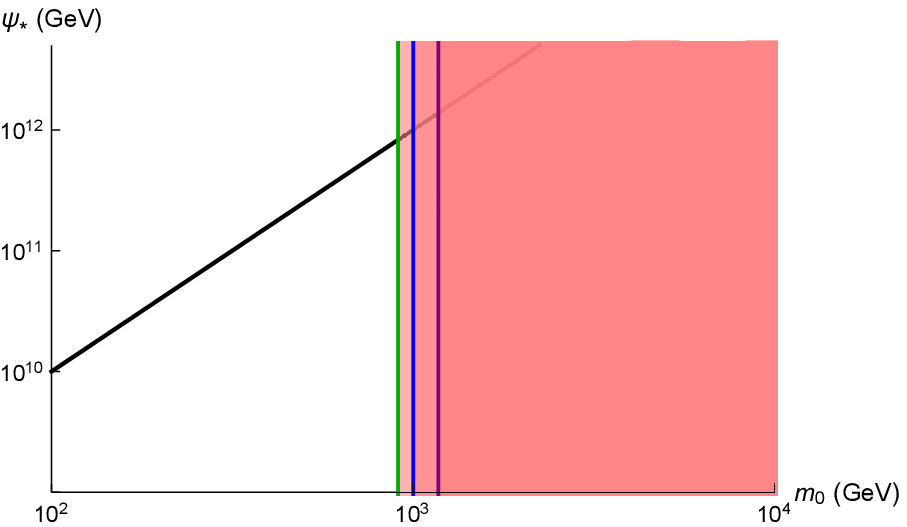}
\caption[Allowed parameter space for $\psi_{\ast}$]{The allowed parameter space for $\psi_{\ast}$, with $\alpha\!=\!1$, $\Gamma_{\varphi}\!\ll\!H_{TI}$ and the parameter values of \cref{Table: Free Parameter Values (alpha=1 and Gamma_{varphi}<<H_{TI})}. The constraints on $m_{0}$ that are shown are the following: \textcolor{OliveGreen}{Green}: \cref{Eq.: m_{0} Constraint f} \ \ \textcolor{blue}{Blue}: \cref{Eq.: m_{0} Constraint h} \ \ \textcolor{violet}{Purple}: \cref{Eq.: m_{0} Constraint v (Gamma_{varphi} << H_{TI})}}
\label{Figure: psi_{*}}
\end{center}
\end{figure}
\begin{figure}[h!]
\begin{center}
\includegraphics[scale=1.2]{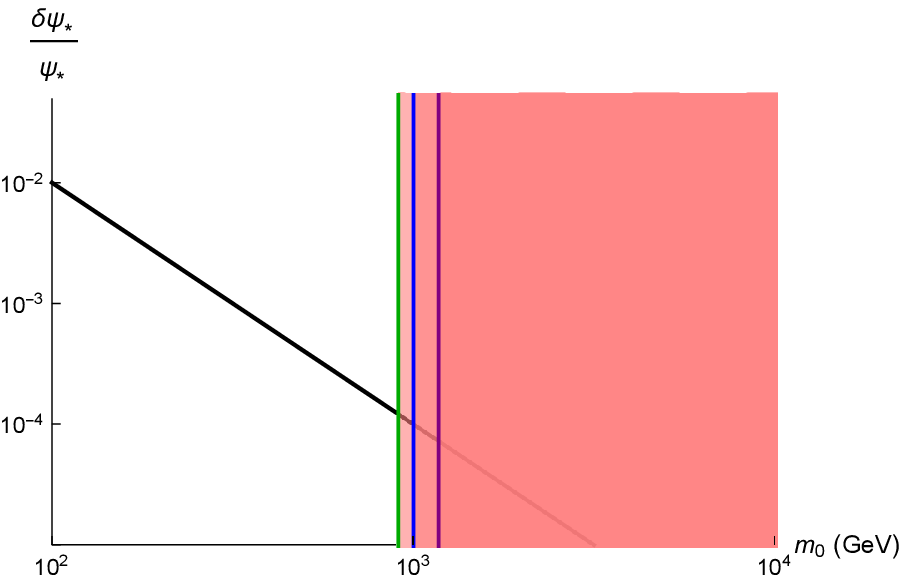}
\caption[Allowed parameter space for $\frac{\delta\psi_{\ast}}{\psi_{\ast}}$]{The allowed parameter space for $\frac{\delta\psi_{\ast}}{\psi_{\ast}}$, with $\alpha\!=\!1$, $\Gamma_{\varphi}\!\ll\!H_{TI}$ and the parameter values of \cref{Table: Free Parameter Values (alpha=1 and Gamma_{varphi}<<H_{TI})}. The constraints on $m_{0}$ that are shown are the following: \textcolor{OliveGreen}{Green}: \cref{Eq.: m_{0} Constraint f} \ \ \textcolor{blue}{Blue}: \cref{Eq.: m_{0} Constraint h} \ \ \textcolor{violet}{Purple}: \cref{Eq.: m_{0} Constraint v (Gamma_{varphi} << H_{TI})}}
\label{Figure: delta psi_{*} / psi_{*}}
\end{center}
\end{figure}
\begin{figure}[h!]
\begin{center}
\includegraphics[scale=1.2]{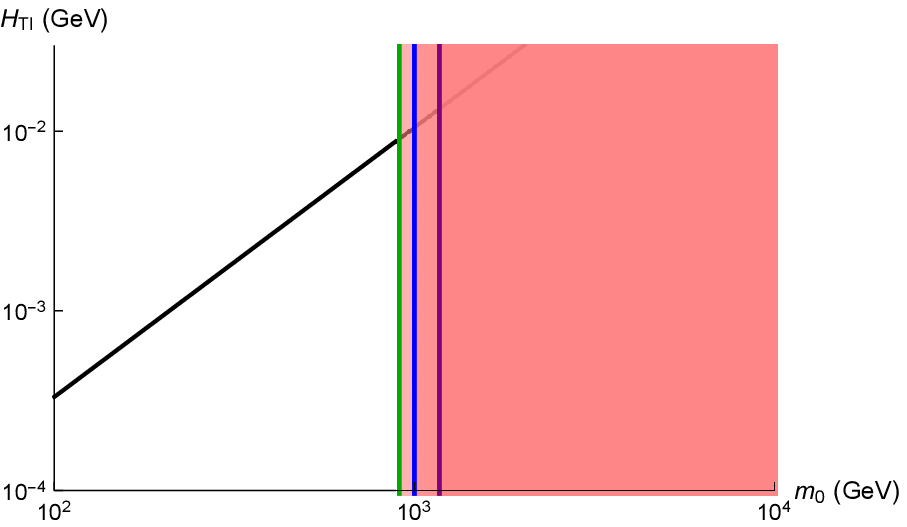}
\caption[Allowed parameter space for $H_{TI}$]{The allowed parameter space for $H_{TI}$, with $\alpha\!=\!1$, $\Gamma_{\varphi}\!\ll\!H_{TI}$ and the parameter values of \cref{Table: Free Parameter Values (alpha=1 and Gamma_{varphi}<<H_{TI})}. The constraints on $m_{0}$ that are shown are the following: \textcolor{OliveGreen}{Green}: \cref{Eq.: m_{0} Constraint f} \ \ \textcolor{blue}{Blue}: \cref{Eq.: m_{0} Constraint h} \ \ \textcolor{violet}{Purple}: \cref{Eq.: m_{0} Constraint v (Gamma_{varphi} << H_{TI})}}
\label{Figure: H_{TI}}
\end{center}
\end{figure}
\begin{figure}[h!]
\begin{center}
\includegraphics[scale=1.2]{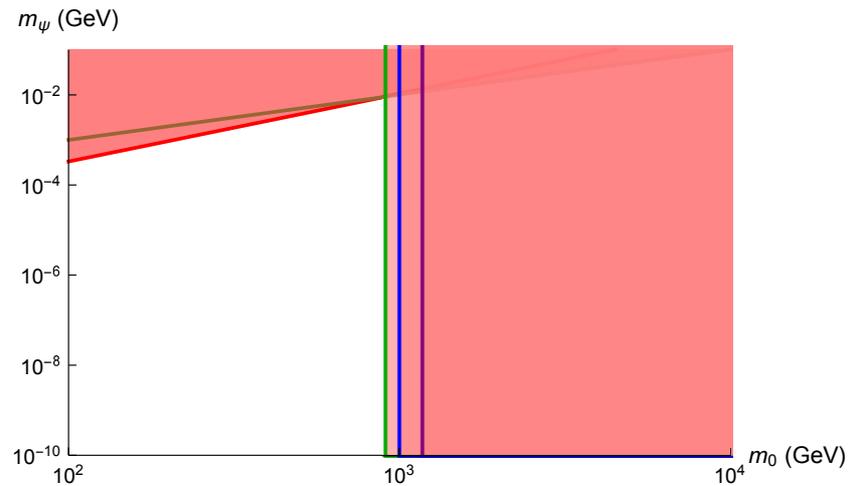}
\caption[Allowed parameter space for $m_{\psi}$]{The allowed parameter space for $m_{\psi}$, with $\alpha\!=\!1$, $\Gamma_{\varphi}\!\ll\!H_{TI}$ and the parameter values of \cref{Table: Free Parameter Values (alpha=1 and Gamma_{varphi}<<H_{TI})}. The upper bound given by the \textcolor{red}{Red} line is that of \cref{Eq.: Psi Mass << H_{TI}} and the upper bound given by the \textcolor{RawSienna}{Brown} line is that of \cref{Eq.: Psi Mass Constraint a}. (We choose to display only down to $10^{-10}\ \mbox{GeV}$, i.e. this value is not a lower bound.) The constraints on $m_{0}$ that are shown are the following: \textcolor{OliveGreen}{Green}: \cref{Eq.: m_{0} Constraint f} \ \ \textcolor{blue}{Blue}: \cref{Eq.: m_{0} Constraint h} \ \ \textcolor{violet}{Purple}: \cref{Eq.: m_{0} Constraint v (Gamma_{varphi} << H_{TI})}}
\label{Figure: m_{psi}}
\end{center}
\end{figure}
\indent Values of other quantities in the model for a mass value of $m_{0}\!\sim\!10^{3}\ \mbox{GeV}$ and the parameter values of \cref{Table: Free Parameter Values (alpha=1 and Gamma_{varphi}<<H_{TI})} are shown in \cref{Table: Quantity Values for m_{0}sim10^{3} (alpha=1 and Gamma_{varphi}<<H_{TI})}. In this table we include the tensor fraction, which for a value of $H_{\ast}\!\sim\!10^{8}\ \mbox{GeV}$ yields the negligible value $r\!\sim\!10^{-13}$.\\
\begin{table}[h!]
\centering
\begin{tabular}{c @{\hskip 30pt} c}
Quantity & Order of Magnitude Value\\
\hline
\\
[-8pt]
$\psi_{\ast}$ & $10^{12}\ \mbox{GeV}$\\
[5pt]
$\frac{\delta\psi_{\ast}}{\psi_{\ast}}$ & $10^{-4}$\\
[5pt]
$H_{TI}$ & $10^{-2}\ \mbox{GeV}$\\
[5pt]
$\left\langle\phi\right\rangle$ & $10^{13}\ \mbox{GeV}$\\
[5pt]
$V_{0}^{\frac{1}{4}}$ & $10^{8}\ \mbox{GeV}$\\
[5pt]
$T_{1}$ & $10^{7}\ \mbox{GeV}$\\
[5pt]
$T_{2}$ & $10^{3}\ \mbox{GeV}$\\
[5pt]
$\Gamma$ & $10^{2}\ \mbox{GeV}$\\
[5pt]
$r$ & $10^{-13}$
\end{tabular}
\caption{Values of quantities in the model for $\alpha\!=\!1$, $\Gamma_{\varphi}\!\ll\!H_{\textrm{TI}}$, $m_{0}\!\sim\!10^{3}\ \mbox{GeV}$ and the parameter values of \cref{Table: Free Parameter Values (alpha=1 and Gamma_{varphi}<<H_{TI})}.}
\label{Table: Quantity Values for m_{0}sim10^{3} (alpha=1 and Gamma_{varphi}<<H_{TI})}
\end{table}
\indent \Cref{Figure: f_{NL},Figure: g_{NL}} show the prediction of the model for the non-Gaussianity parameters $f_{\textrm{NL}}$ and $g_{\textrm{NL}}$ respectively, with $h$ and $H_{\ast}$ values from \cref{Table: Free Parameter Values (alpha=1 and Gamma_{varphi}<<H_{TI})}, together with the central value and range for the parameters as obtained by the Planck spacecraft \cite{Planck_2015_Non-Gaussianity_Results}. From looking at \cref{Figure: g_{NL}}, it may initially seem as if our predicted value for $g_{\textrm{NL}}$ for a thermal waterfall field mass value of $m_{0}\!\sim\!10^{3}\ \mbox{GeV}$ is ruled-out. However, as we are working within an order of magnitude for each value, we can see that our predicted value, i.e. the Black curve, is allowed within an order of magnitude below $m_{0}\!\sim\!10^{3}\ \mbox{GeV}$. The values of $f_{\textrm{NL}}$ and $g_{\textrm{NL}}$ for a thermal waterfall field mass of $m_{0}\!\sim\!10^{3}\ \mbox{GeV}$ are shown in \cref{Table: Non-Gaussianity Parameters for m_{0}sim10^{3} (alpha=1 and Gamma_{varphi}<<H_{TI})}, with them both being within current observational bounds \cite{Planck_2015_Non-Gaussianity_Results}.\\
\begin{figure}[h!]
\begin{center}
\includegraphics[scale=1.2]{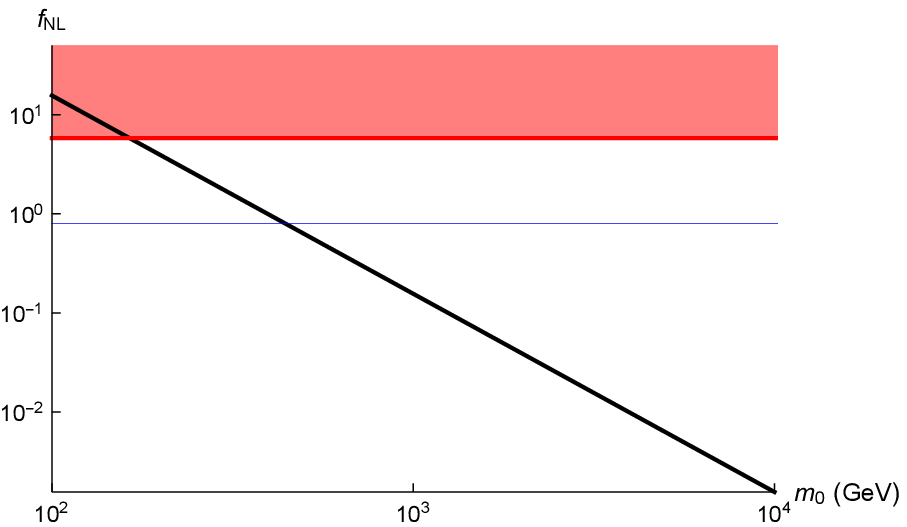}
\caption[Prediction for $f_{\textrm{NL}}$]{Prediction of the model for the non-Gaussianity parameter $f_{\textrm{NL}}$, with $\alpha\!=\!1$, $\Gamma_{\varphi}\!\ll\!H_{TI}$ and $h$ and $H_{\ast}$ values from \cref{Table: Free Parameter Values (alpha=1 and Gamma_{varphi}<<H_{TI})}. (A plot of \cref{Eq.: f_{NL} - End of Inflation}, with $\psi\!=\!\psi_{\ast}$.) The \textcolor{blue}{Blue} and \textcolor{red}{Red} lines are the central value and upper bound of $f_{\textrm{NL}}$ respectively as obtained by the Planck spacecraft \cite{Planck_2015_Non-Gaussianity_Results}, with the lower bound being outside the displayed range of $f_{\textrm{NL}}$.}
\label{Figure: f_{NL}}
\end{center}
\end{figure}
\begin{figure}[h!]
\begin{center}
\includegraphics[scale=1.2]{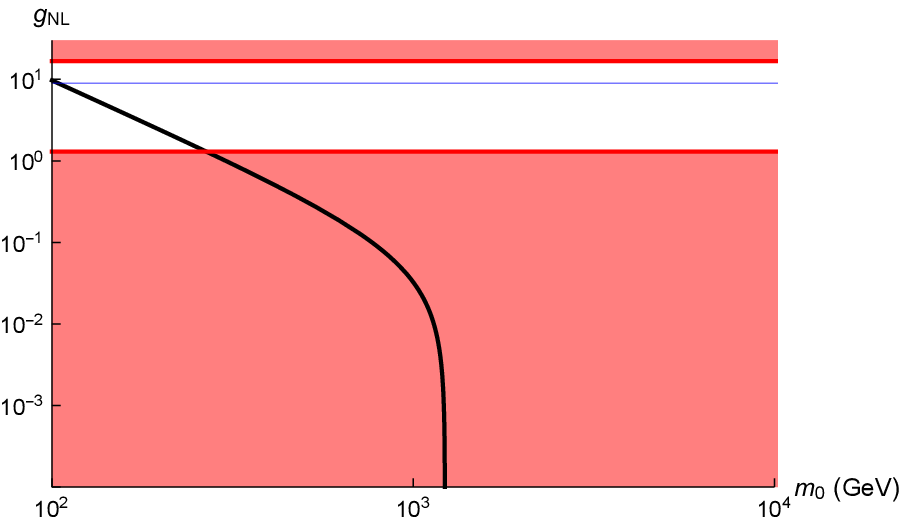}
\caption[Prediction for $g_{\textrm{NL}}$]{Prediction of the model for the non-Gaussianity parameter $g_{\textrm{NL}}$, with $\alpha\!=\!1$, $\Gamma_{\varphi}\!\ll\!H_{TI}$ and $h$ and $H_{\ast}$ values from \cref{Table: Free Parameter Values (alpha=1 and Gamma_{varphi}<<H_{TI})}. (A plot of \cref{Eq.: g_{NL} - End of Inflation}, with $\psi\!=\!\psi_{\ast}$.) The \textcolor{blue}{Blue} and \textcolor{red}{Red} lines are the central value and lower/upper bounds of $g_{\textrm{NL}}$ respectively as obtained by the Planck spacecraft \cite{Planck_2015_Non-Gaussianity_Results}.}
\label{Figure: g_{NL}}
\end{center}
\end{figure}
\begin{table}[h!]
\centering
\begin{tabular}{c @{\hskip 30pt} c}
Parameter & Order of Magnitude Value\\
\hline
\\
[-8pt]
$f_{\textrm{NL}}$ & $10^{-1}$\\
[5pt]
$g_{\textrm{NL}}$ & $10^{-2}$
\end{tabular}
\caption{Prediction for non-Gaussianity parameters of the model, with $\alpha\!=\!1$, $\Gamma_{\varphi}\!\ll\!H_{\textrm{TI}}$, $m_{0}\!\sim\!10^{3}\ \mbox{GeV}$ and $h$ and $H_{\ast}$ values from \cref{Table: Free Parameter Values (alpha=1 and Gamma_{varphi}<<H_{TI})}.}
\label{Table: Non-Gaussianity Parameters for m_{0}sim10^{3} (alpha=1 and Gamma_{varphi}<<H_{TI})}
\end{table}
\\
\underline{\textbf{$n_{s}$ and $n_{s}^{\prime}$: Chaotic Inflation}}\vspace{0.2em}
\\
We provide results for the spectral index and its running when the period of primordial inflation is that of slow-roll Chaotic Inflation, with the potential
\begin{equation}
V(\varphi) = \frac{1}{2}m_{\varphi}^{2}\varphi^{2}
\label{Eq.: Chaotic Inflation Potential}
\end{equation}
From \cref{Subsection: Spectral Index --- n_{s} and n_{s}^{'}}, the spectral index $n_{s}$ is given by
\begin{equation}
n_{s} \simeq 1 - 2\epsilon + 2\eta_{\psi\psi}
\label{Eq.: Spectral Index - End of Inflation Results Section}
\end{equation}
with $\epsilon$ being given by \cref{Eq.: epsilon Definition} and $\eta_{\psi\psi}$ being given by \cref{Eq.: eta_{psi psi} Definition} and where both are to be evaluated at the point where cosmological scales exit the horizon during primordial inflation. The potential of \cref{Eq.: Chaotic Inflation Potential} gives
\begin{equation}
\epsilon = \frac{2M_{P}^{2}}{\varphi_{\ast}^{2}}
\label{Eq.: epsilon - varphi_{*}}
\end{equation}
We obtain an expression for $\varphi_{\ast}$ in terms of $N_{\ast}$ by using the equation
\begin{equation}
N_{\ast} \approx \frac{1}{M_{P}^{2}}\int_{\varphi_{end}}^{\varphi_{\ast}}\frac{V(\varphi)}{V^{\prime}(\varphi)}\,\mathrm{d}\varphi
\label{Eq.: N_{*}}
\end{equation}
We define the end of primordial inflation to be when $\epsilon\!=\!1$. This gives
\begin{equation}
\varphi_{end} = \sqrt{2}M_{P}
\label{Eq.: varphi_{end}}
\end{equation}
Therefore we have
\begin{align}
N_{\ast} &\approx \frac{1}{M_{P}^{2}}\int_{\sqrt{2}M_{P}}^{\varphi_{\ast}}\frac{V(\varphi)}{V^{\prime}(\varphi)}\,\mathrm{d}\varphi
\\
\nonumber
\\
&\approx \frac{1}{2M_{P}^{2}}\int_{\sqrt{2}M_{P}}^{\varphi_{\ast}}\varphi\,\mathrm{d}\varphi
\\
\nonumber
\\
\varphi_{\ast} &\approx \sqrt{4N_{\ast}+2}\,M_{P}
\label{Eq.: varphi_{*}}
\end{align}
Substituting \cref{Eq.: varphi_{*}} into \cref{Eq.: epsilon - varphi_{*}} gives
\begin{equation}
\epsilon \approx \frac{1}{2N_{\ast}+1}
\label{Eq.: epsilon}
\end{equation}
We also need to calculate $\eta_{\psi\psi}$. Using our potential, \cref{Eq.: Full Potential}, we obtain $V_{\psi\psi}$ at the time cosmological scales exit the horizon as
\begin{equation}
V_{\psi\psi}|_{\ast} = m_{\psi}^{2} + \left(4\alpha^{2}-2\alpha\right)h^{2}\phi_{\ast}^{2}\left(\frac{\psi_{\ast}}{M_{P}}\right)^{2\alpha-2}
\label{Eq.: V_{psi psi}|_{*}}
\end{equation}
Therefore we obtain $\eta_{\psi\psi}$ as
\begin{equation}
\eta_{\psi\psi} = \frac{1}{3H_{\ast}^{2}}\left[m_{\psi}^{2} + \left(4\alpha^{2}-2\alpha\right)h^{2}\phi_{\ast}^{2}\left(\frac{\psi_{\ast}}{M_{P}}\right)^{2\alpha-2}\right]
\label{Eq.: eta_{psi psi}}
\end{equation}
Our final result for the spectral index is therefore
\begin{equation}
n_{s} \approx 1 - \frac{2}{2N_{\ast}+1} + \frac{2}{3H_{\ast}^{2}}\left[m_{\psi}^{2} + \left(4\alpha^{2}-2\alpha\right)h^{2}\phi_{\ast}^{2}\left(\frac{\psi_{\ast}}{M_{P}}\right)^{2\alpha-2}\right]
\label{Eq.: n_{s}}
\end{equation}
\indent From \cref{Subsection: Spectral Index --- n_{s} and n_{s}^{'}}, the running of the spectral index $n_{s}^{\prime}$ is given by
\begin{equation}
n_{s}^{\prime} \simeq -8\epsilon^{2} + 4\epsilon\eta + 4\epsilon\eta_{\psi\psi}
\label{Eq.: Running of Spectral Index - End of Inflation Results Section}
\end{equation}
with $\eta$ being given by \cref{Eq.: eta Definition}, which is to be evaluated at the point where cosmological scales exit the horizon during primordial inflation. The potential of \cref{Eq.: Chaotic Inflation Potential} gives
\begin{equation}
\eta = \frac{2M_{P}^{2}}{\varphi_{\ast}^{2}}
\label{Eq.: eta - varphi_{*}}
\end{equation}
which is identical to the value of $\epsilon$, \cref{Eq.: epsilon - varphi_{*}}. Substituting \cref{Eq.: varphi_{*}} into \cref{Eq.: eta - varphi_{*}} gives
\begin{equation}
\eta \approx \frac{1}{2N_{\ast}+1}
\label{Eq.: eta}
\end{equation}
Our final result for the running of the spectral index is therefore
\begin{equation}
n_{s}^{\prime} \approx -\frac{4}{\left(2N_{\ast}+1\right)^{2}} + \frac{4}{\left(6N_{\ast}+3\right)H_{\ast}^{2}}\left[m_{\psi}^{2} + \left(4\alpha^{2}-2\alpha\right)h^{2}\phi_{\ast}^{2}\left(\frac{\psi_{\ast}}{M_{P}}\right)^{2\alpha-2}\right]
\label{Eq.: n_{s}'}
\end{equation}
\indent In order to obtain $n_{s}$ and $n_{s}^{\prime}$, we first need to obtain $N_{\ast}$. The prediction of the model for $N_{\textrm{TI}}$ and $N_{\ast}$ are shown in \cref{Figure: N_{TI},Figure: N_{*}} respectively, with $n$, $g$, $H_{\ast}$, $\Gamma_{\varphi}$ and $\lambda$ values from \cref{Table: Free Parameter Values (alpha=1 and Gamma_{varphi}<<H_{TI})}. The kink that is visible in the plot of $N_{\ast}$ at around $m_{0}\!\approx\!10^{9}\ \mbox{GeV}$ is a result of the fact that for $m_{0}$ values larger than this, we do not have any period of thermal inflation, as can be seen in the plot of $N_{\textrm{TI}}$. The values of $N_{\textrm{TI}}$ and $N_{\ast}$ for a thermal waterfall field mass of $m_{0}\!\sim\!10^{3}\ \mbox{GeV}$ are shown in \cref{Table: N_{TI} and N_{*} for m_{0}sim10^{3} (Gamma_{varphi}<<H_{TI})}.
\begin{figure}[h!]
\begin{center}
\includegraphics[scale=1.2]{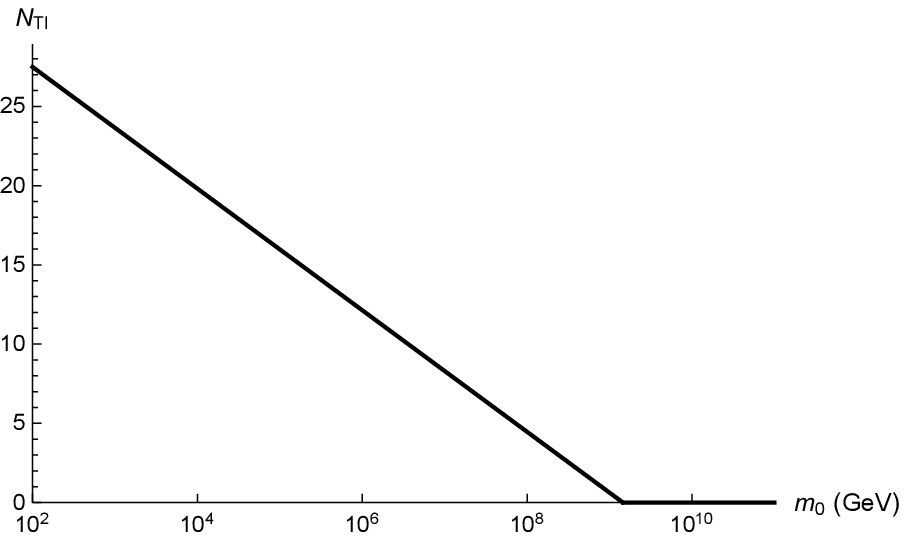}
\caption[Prediction for $N_{\textrm{TI}}$]{Prediction of the model for $N_{\textrm{TI}}$, with $\Gamma_{\varphi}\!\ll\!H_{TI}$ and $n$, $g$, $\Gamma_{\varphi}$ and $\lambda$ values from \cref{Table: Free Parameter Values (alpha=1 and Gamma_{varphi}<<H_{TI})}. (A plot of \cref{Eq.: N (Gamma_{varphi} << H_{TI})}, with $m\!=\!m_{0}$.)}
\label{Figure: N_{TI}}
\end{center}
\end{figure}
\begin{figure}[h!]
\begin{center}
\includegraphics[scale=1.2]{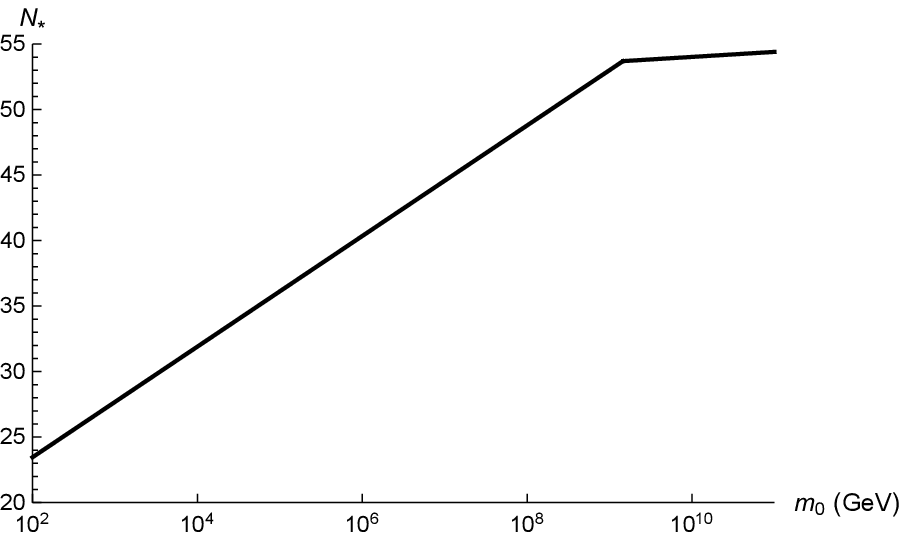}
\caption[Prediction for $N_{\ast}$]{Prediction of the model for $N_{\ast}$, with $\Gamma_{\varphi}\!\ll\!H_{TI}$ and $n$, $g$, $H_{\ast}$, $\Gamma_{\varphi}$ and $\lambda$ values from \cref{Table: Free Parameter Values (alpha=1 and Gamma_{varphi}<<H_{TI})}. (A plot of \cref{Eq.: N_{*} (Gamma_{varphi} << H_{TI})}, with $m\!=\!m_{0}$ and $\Gamma\!=\!g^{2}m_{0}$.)}
\label{Figure: N_{*}}
\end{center}
\end{figure}
\begin{table}[h!]
\centering
\begin{tabular}{c @{\hskip 30pt} c}
Parameter & Value\\
\hline
\\
[-8pt]
$N_{\textrm{TI}}$ & $\approx\!24$\\
[5pt]
$N_{\ast}$ & $\approx\!28$
\end{tabular}
\caption{Prediction for $N_{\textrm{TI}}$ and $N_{\ast}$ of the model, with $\Gamma_{\varphi}\!\ll\!H_{\textrm{TI}}$, $m_{0}\!\sim\!10^{3}\ \mbox{GeV}$ and $n$, $g$, $H_{\ast}$, $\Gamma_{\varphi}$ and $\lambda$ values from \cref{Table: Free Parameter Values (alpha=1 and Gamma_{varphi}<<H_{TI})}.}
\label{Table: N_{TI} and N_{*} for m_{0}sim10^{3} (Gamma_{varphi}<<H_{TI})}
\end{table}
\\
\indent The prediction of the model for $n_{s}$ and $n_{s}^{\prime}$ for each $\phi_{\ast}$ case and for a spectator field mass at the upper bound of $m_{\psi}\!=\!10^{-2}\ \mbox{GeV}$ are shown in \cref{Figure: n_{s} for phi_{*} Case A,Figure: n_{s}' for phi_{*} Case A,Figure: n_{s} for phi_{*} Case B,Figure: n_{s}' for phi_{*} Case B,Figure: n_{s} for phi_{*} Case C,Figure: n_{s}' for phi_{*} Case C}, with the parameter values of \cref{Table: Free Parameter Values (alpha=1 and Gamma_{varphi}<<H_{TI})}. The predicted values of $n_{s}$ and $n_{s}^{\prime}$ of the model for a thermal waterfall field mass of $m_{0}\!\sim\!10^{3}\ \mbox{GeV}$ for all three $\phi_{\ast}$ Cases are the same to within at least four significant figures. They are also both insensitive to the value of $m_{\psi}$ within its allowed range. $n_{s}$ and $n_{s}^{\prime}$ are shown in \cref{Table: n_{s} and n_{s}' for m_{0}sim10^{3} (alpha=1 and Gamma_{varphi}<<H_{TI})}, with them both being within current observational bounds \cite{Planck_2015_Cosmo._Results}.
\begin{figure}[h!]
\begin{center}
\includegraphics[scale=1.2]{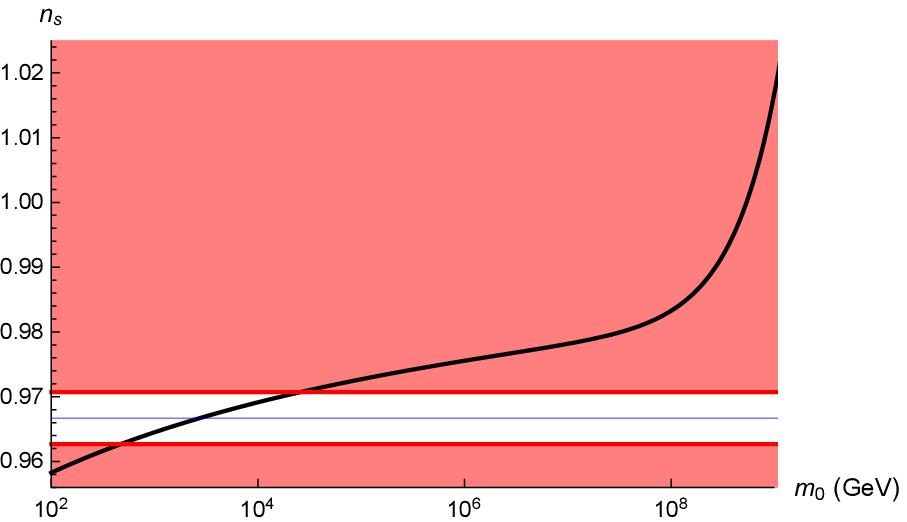}
\caption[Prediction for $n_{s}$ for $\phi_{\ast}$ Case A: Chaotic Inflation]{Prediction of the model for $n_{s}$ for $\phi_{\ast}$ Case A with primordial inflation being Chaotic Inflation, with $\alpha\!=\!1$, $\Gamma_{\varphi}\!\ll\!H_{TI}$, $m_{\psi}\!=\!10^{-2}\ \mbox{GeV}$ and the parameter values from \cref{Table: Free Parameter Values (alpha=1 and Gamma_{varphi}<<H_{TI})}. (A plot of \cref{Eq.: n_{s}}, with $\phi_{\ast}\!=\!\left\langle\phi\right\rangle$, $m\!=\!m_{0}$ and $\Gamma\!=\!g^{2}m_{0}$.) The \textcolor{blue}{Blue} and \textcolor{red}{Red} lines are the central value and lower/upper bounds of $n_{s}$ respectively as obtained by the Planck spacecraft \cite{Planck_2015_Cosmo._Results}.}
\label{Figure: n_{s} for phi_{*} Case A}
\end{center}
\end{figure}
\begin{figure}[h!]
\begin{center}
\includegraphics[scale=1.2]{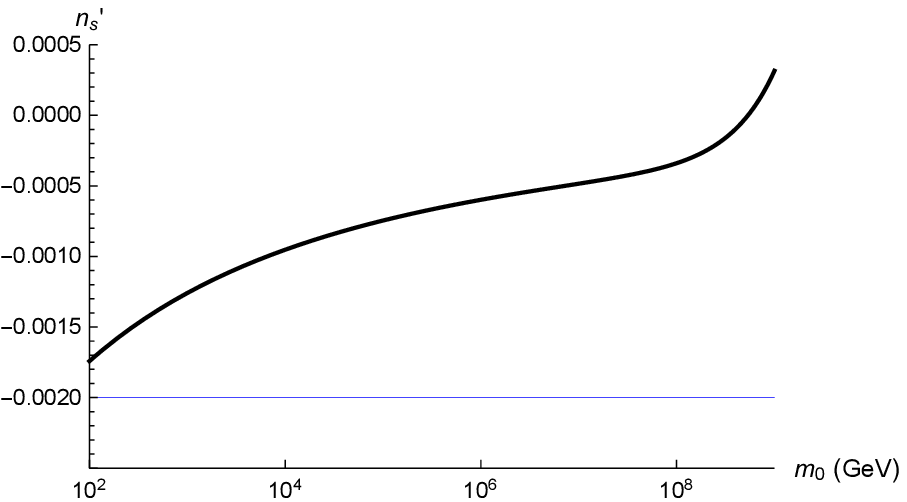}
\caption[Prediction for $n_{s}^{\prime}$ for $\phi_{\ast}$ Case A: Chaotic Inflation]{Prediction of the model for $n_{s}^{\prime}$ for $\phi_{\ast}$ Case A with primordial inflation being Chaotic Inflation, with $\alpha\!=\!1$, $\Gamma_{\varphi}\!\ll\!H_{TI}$, $m_{\psi}\!=\!10^{-2}\ \mbox{GeV}$ and the parameter values from \cref{Table: Free Parameter Values (alpha=1 and Gamma_{varphi}<<H_{TI})}. (A plot of \cref{Eq.: n_{s}'}, with $\phi_{\ast}\!=\!\left\langle\phi\right\rangle$, $m\!=\!m_{0}$ and $\Gamma\!=\!g^{2}m_{0}$.) The \textcolor{blue}{Blue} line is the central value of $n_{s}^{\prime}$ as obtained by the Planck spacecraft \cite{Planck_2015_Cosmo._Results}, with the lower and upper bounds being outside the displayed range of $n_{s}^{\prime}$.}
\label{Figure: n_{s}' for phi_{*} Case A}
\end{center}
\end{figure}
\begin{figure}[h!]
\begin{center}
\includegraphics[scale=1.2]{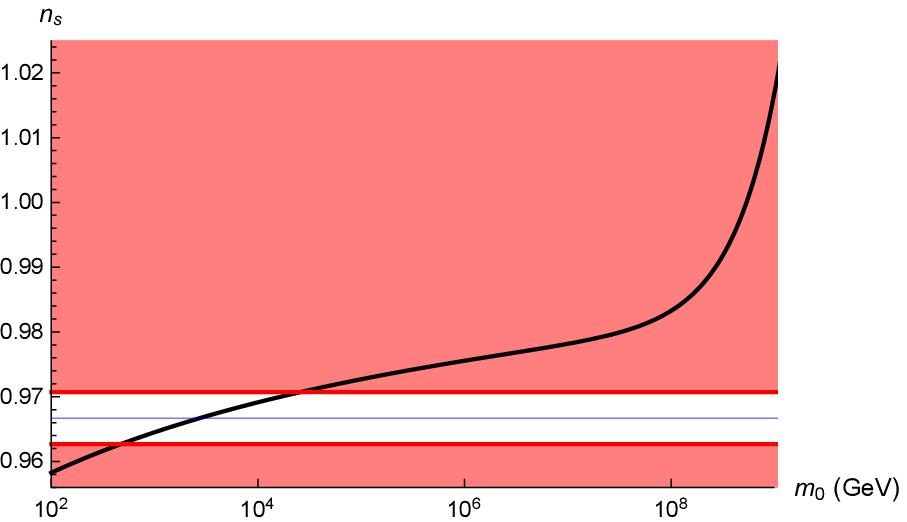}
\caption[Prediction for $n_{s}$ for $\phi_{\ast}$ Case B: Chaotic Inflation]{Prediction of the model for $n_{s}$ for $\phi_{\ast}$ Case B with primordial inflation being Chaotic Inflation, with $\alpha\!=\!1$, $\Gamma_{\varphi}\!\ll\!H_{TI}$, $m_{\psi}\!=\!10^{-2}\ \mbox{GeV}$ and the parameter values from \cref{Table: Free Parameter Values (alpha=1 and Gamma_{varphi}<<H_{TI})}. (A plot of \cref{Eq.: n_{s}}, with $\phi_{\ast}\!=\!\phi_{\textrm{BD}}$, $m\!=\!m_{0}$ and $\Gamma\!=\!g^{2}m_{0}$.) The \textcolor{blue}{Blue} and \textcolor{red}{Red} lines are the central value and lower/upper bounds of $n_{s}$ respectively as obtained by the Planck spacecraft \cite{Planck_2015_Cosmo._Results}.}
\label{Figure: n_{s} for phi_{*} Case B}
\end{center}
\end{figure}
\begin{figure}[h!]
\begin{center}
\includegraphics[scale=1.2]{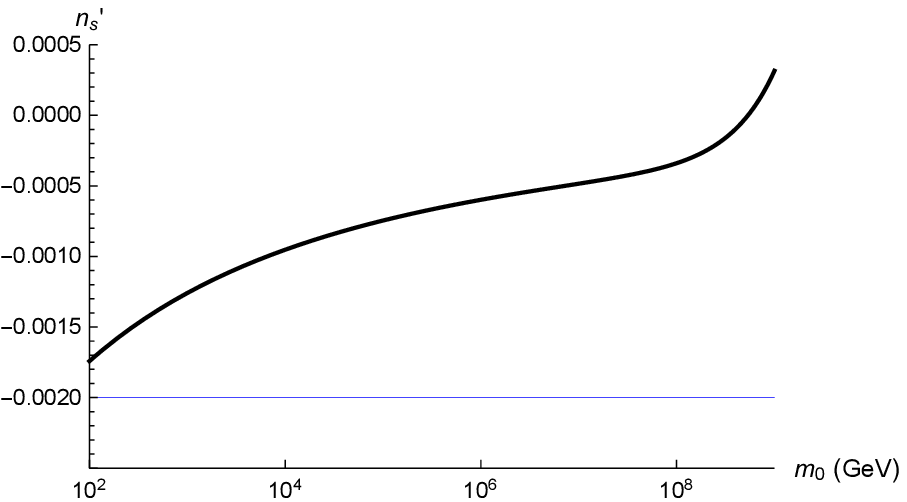}
\caption[Prediction for $n_{s}^{\prime}$ for $\phi_{\ast}$ Case B: Chaotic Inflation]{Prediction of the model for $n_{s}^{\prime}$ for $\phi_{\ast}$ Case B with primordial inflation being Chaotic Inflation, with $\alpha\!=\!1$, $\Gamma_{\varphi}\!\ll\!H_{TI}$, $m_{\psi}\!=\!10^{-2}\ \mbox{GeV}$ and the parameter values from \cref{Table: Free Parameter Values (alpha=1 and Gamma_{varphi}<<H_{TI})}. (A plot of \cref{Eq.: n_{s}'}, with $\phi_{\ast}\!=\!\phi_{\textrm{BD}}$, $m\!=\!m_{0}$ and $\Gamma\!=\!g^{2}m_{0}$.) The \textcolor{blue}{Blue} line is the central value of $n_{s}^{\prime}$ as obtained by the Planck spacecraft \cite{Planck_2015_Cosmo._Results}, with the lower and upper bounds being outside the displayed range of $n_{s}^{\prime}$.}
\label{Figure: n_{s}' for phi_{*} Case B}
\end{center}
\end{figure}
\begin{figure}[h!]
\begin{center}
\includegraphics[scale=1.2]{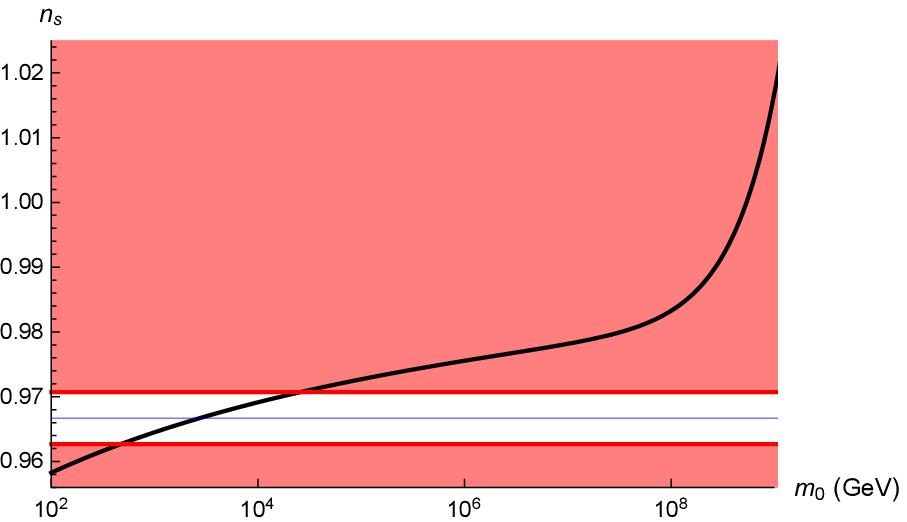}
\caption[Prediction for $n_{s}$ for $\phi_{\ast}$ Case C: Chaotic Inflation]{Prediction of the model for $n_{s}$ for $\phi_{\ast}$ Case C with primordial inflation being Chaotic Inflation, with $\alpha\!=\!1$, $\Gamma_{\varphi}\!\ll\!H_{TI}$, $m_{\psi}\!=\!10^{-2}\ \mbox{GeV}$ and the parameter values from \cref{Table: Free Parameter Values (alpha=1 and Gamma_{varphi}<<H_{TI})}. (A plot of \cref{Eq.: n_{s}}, with $\phi_{\ast}\!=\!0$, $m\!=\!m_{0}$ and $\Gamma\!=\!g^{2}m_{0}$.) The \textcolor{blue}{Blue} and \textcolor{red}{Red} lines are the central value and lower/upper bounds of $n_{s}$ respectively as obtained by the Planck spacecraft \cite{Planck_2015_Cosmo._Results}.}
\label{Figure: n_{s} for phi_{*} Case C}
\end{center}
\end{figure}
\begin{figure}[h!]
\begin{center}
\includegraphics[scale=1.2]{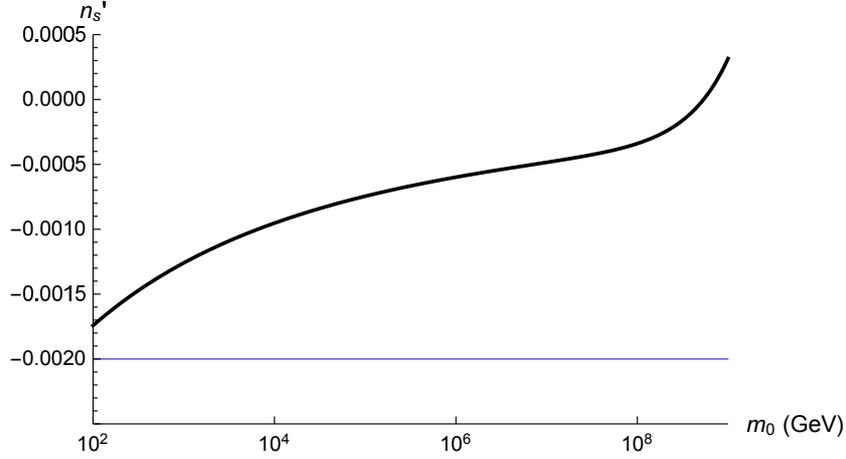}
\caption[Prediction for $n_{s}^{\prime}$ for $\phi_{\ast}$ Case C: Chaotic Inflation]{Prediction of the model for $n_{s}^{\prime}$ for $\phi_{\ast}$ Case C with primordial inflation being Chaotic Inflation, with $\alpha\!=\!1$, $\Gamma_{\varphi}\!\ll\!H_{TI}$, $m_{\psi}\!=\!10^{-2}\ \mbox{GeV}$ and the parameter values from \cref{Table: Free Parameter Values (alpha=1 and Gamma_{varphi}<<H_{TI})}. (A plot of \cref{Eq.: n_{s}'}, with $\phi_{\ast}\!=\!0$, $m\!=\!m_{0}$ and $\Gamma\!=\!g^{2}m_{0}$.) The \textcolor{blue}{Blue} line is the central value of $n_{s}^{\prime}$ as obtained by the Planck spacecraft \cite{Planck_2015_Cosmo._Results}, with the lower and upper bounds being outside the displayed range of $n_{s}^{\prime}$.}
\label{Figure: n_{s}' for phi_{*} Case C}
\end{center}
\end{figure}
\begin{table}[h!]
\centering
\begin{tabular}{c @{\hskip 30pt} c}
Quantity & Value\\
\hline
\\
[-8pt]
$n_{s}$ & $\approx\!0.9645$\\
[5pt]
$n_{s}^{\prime}$ & $\approx\!-0.001259$
\end{tabular}
\caption{Prediction for $n_{s}$ and $n_{s}^{\prime}$ of the model with primordial inflation being Chaotic Inflation, with $\alpha\!=\!1$, $\Gamma_{\varphi}\!\ll\!H_{\textrm{TI}}$, $m_{\psi}\!=\!10^{-2}\ \mbox{GeV}$, $m_{0}\!\sim\!10^{3}\ \mbox{GeV}$ and the parameter values from \cref{Table: Free Parameter Values (alpha=1 and Gamma_{varphi}<<H_{TI})}.}
\label{Table: n_{s} and n_{s}' for m_{0}sim10^{3} (alpha=1 and Gamma_{varphi}<<H_{TI})}
\end{table}
\subsubsection{The Case $\alpha\neq1$}
\label{Subsubsection: The Case alpha!=1}
The present value of the effective mass of $\psi$ is
\begin{align}
m_{\psi,\textrm{now}}^{2} &= \left(4\alpha^{2}-2\alpha\right)h^{2}\left(\frac{\left\langle\psi\right\rangle}{M_{P}}\right)^{2\alpha-2}\left\langle\phi\right\rangle^{2} + m_{\psi}^{2}
\\
\nonumber
\\
m_{\psi,\textrm{now}} &= m_{\psi}
\label{Eq.: m_{psi,now} for alpha!=1}
\end{align}
the second line coming from the fact that $\left\langle\psi\right\rangle\!=\!0$ today. Therefore, from \cref{Eq.: m_{psi,now} gtrsim 1 TeV} we have
\begin{equation}
m_{\psi} \gtrsim 1\ \mbox{TeV}
\label{Eq.: m_{psi} gtrsim 1 TeV}
\end{equation}
In order for this and \cref{Eq.: Psi Mass Constraint a} to both be satisfied, we require the lower bound in \cref{Eq.: m_{psi} gtrsim 1 TeV} to be much smaller than the upper bound in \cref{Eq.: Psi Mass Constraint a}. This gives
\begin{equation}
m_{0} \gg 10^{8}\ \mbox{GeV}
\label{Eq.: m_{0} >> 10^{8} GeV}
\end{equation}
We now require the lower bound here to be much smaller than the upper bound in \cref{Eq.: m_{0} Constraint f}. This gives
\begin{equation}
\lambda \gg (2n+4)!\ \frac{10^{10n+26}\ \mbox{GeV$^{2}$}}{M_{P}^{2}}
\label{Eq.: lambda Constraint}
\end{equation}
where we have explicitly factored out the $\frac{1}{(2n+4)!}$ term from our definition of $\lambda$. Given that $n\!\geq\!1$, this constraint is in conflict with the requirement that $\lambda\!\lesssim\!1$. Therefore, for $\alpha\!\neq\!1$, we find that the ``end of inflation'' mechanism cannot produce the dominant contribution to the observed primordial curvature perturbation within this Thermal Inflation model as it currently stands.
\section{Modulated Decay Rate}
\label{Section: Modulated Decay Rate}
Now we investigate the modulated decay scenario to see if it can produce the dominant contribution to the primordial curvature perturbation $\zeta$. As in \cref{Section: ``End of Inflation'' Mechanism}, we aim to obtain a number of constraints on the model parameters and the initial conditions for the fields. Considering these constraints, we intend to determine the available parameter space (if any). In this parameter space we will calculate distinct observational signatures (such as non-Gaussianity) that may test this scenario in the near future.\\
\indent From \cref{Subsection: phi Decay Rate}, the decay rate of the $\phi$ field is given by
\begin{equation}
\Gamma \sim g^{2}m
\label{Eq.: phi Decay Rate - Modulated Decay Rate Section}
\end{equation}
The primordial curvature perturbation that is produced by a varying decay rate \cite{Dvali_et_al.:New_Mech._Generating_Den._Perts._from_Inflation} is given to first order by
\begin{equation}
\zeta = \delta N = -\frac{1}{6}\frac{\delta\Gamma}{\Gamma}
\label{Eq.: Zeta -- Gamma (Decay Rate)}
\end{equation}
Differentiating $\Gamma\!\sim\!g^{2}m$ with respect to $m$ gives
\begin{equation}
\delta\Gamma \sim g^{2}\delta m
\label{Eq.: Phi Thermal Decay Rate Variation}
\end{equation}
Therefore we obtain the primordial curvature perturbation as
\begin{equation}
\zeta = \delta N \sim -\frac{\delta m}{m}
\label{Eq.: Zeta (Decay Rate)}
\end{equation}
This is of the same order of magnitude as the primordial curvature perturbation that is produced by the ``end of inflation'' mechanism, \cref{Eq.: Zeta (m) (Gamma_{varphi} gtrsim H_{TI}),Eq.: Zeta (m) (Gamma_{varphi} << H_{TI})}.
\subsection{Non-Gaussianity}
\label{Subsection: Modulated Decay Rate Non-Gaussianity}
As with the ``end of inflation'' mechanism scenario, we will consider what is termed local non-Gaussianity, which for the bispectrum corresponds to the ``squeezed'' configuration of the momenta triangle, in that the magnitude of one of the momentum vectors is much smaller than the other two, which are of similar magnitude to each other, e.g. $k_{3}\!\ll\!k_{1},k_{2}$ and $k_{1}\!\approx\!k_{2}$. Within the framework of the $\delta N$ formalism, the non-Gaussianity parameter $f_{\textrm{NL}}$ is obtained as \cite{Lyth_&_Liddle:Prim._Den._Pert.}
\begin{equation}
f_{\textrm{NL}} = 5\left(\frac{\Gamma\Gamma^{\prime\prime}}{\Gamma^{\prime^{2}}} -1\right)
\label{Eq.: f_{NL} Expression - Modulated Decay Rate}
\end{equation}
where the prime denotes the derivative with respect to $\psi$. By substituting $\Gamma$ from \cref{Eq.: phi Decay Rate - Modulated Decay Rate Section} into \cref{Eq.: f_{NL} Expression - Modulated Decay Rate} we obtain
\begin{equation}
f_{\textrm{NL}} \sim 5\left(\frac{mm^{\prime\prime}}{m^{\prime^{2}}} -1\right)
\label{Eq.: f_{NL} (m) - Modulated Decay Rate}
\end{equation}
Then, from our mass definition $m$, \cref{Eq.: Mass Definition}, we obtain
\begin{equation}
f_{\textrm{NL}} \sim -5\left[1 + \frac{2\alpha -1}{\alpha}\left(\frac{m_{0}^{2}M_{P}^{2\alpha-2}}{2h^{2}\psi_{\ast}^{2\alpha}} - 1\right)\right]
\label{Eq.: f_{NL} - Modulated Decay Rate}
\end{equation}
\subsection{Constraining the Free Parameters}
\label{Subsection: Constraining the Free Parameters - Modulated Decay Rate Section}
Regarding the constraints that appear in \cref{Subsection: Constraining the Free Parameters - End of Inflation Section}, all are relevant to this modulated decay scenario \textit{except} for those that appear in \cref{Subsubsection: Time of Transition from Thermal Inflation to Thermal Waterfall Field Oscillation}, as we do not require that the transition from thermal inflation to thermal waterfall field oscillation be sufficiently fast here, as well as those appearing in \cref{Subsubsection: Energy Density of the Oscillating Spectator Field}, as the spectator field will not be oscillating in this scenario (see \cref{Subsubsection: Light psi - Modulated Decay Rate Section}).
\subsubsection{Time for $\phi$ Decay}
\label{Subsubsection: Time for phi Decay}
After the end of thermal inflation, we require there to exist an amount of time, i.e. an amount of Universe expansion, prior to the decay of the $\phi$ field. This is so that the modulated decay rate mechanism can have an effect. If the $\phi$ field decayed \textit{immediately} after the end of thermal inflation, the decay rate would not be effectively modulated. Therefore, we require
\begin{equation}
H_{TI} > \Gamma
\label{Eq.: H_{TI} > Gamma}
\end{equation}
Substituting \cref{Eq.: H_{TI},Eq.: phi Decay Rate - Modulated Decay Rate Section} into here gives
\begin{equation}
m_{0} > \sqrt{\lambda}\ g^{2n+2}M_{P}
\label{Eq.: m_{0} Constraint y}
\end{equation}
\subsubsection{Light $\psi$}
\label{Subsubsection: Light psi - Modulated Decay Rate Section}
For the ``end of inflation'' mechanism scenario, we required the effective mass of $\psi$ to be light all the way up until the end of thermal inflation. However, in this modulated decay rate scenario, we require the effective mass to be light for longer, all the way up until $\phi$ decays, in order that $\psi$ does not start oscillating, so that the perturbations in $\psi$ remain at decay time and thus have the effect of perturbing the decay rate. Therefore we require
\begin{equation}
|m_{\psi,\textrm{eff}}| \ll H
\label{Eq.: Psi Effective Mass << H}
\end{equation}
The $\phi$ field decays when $H$ falls to $\sim\!\Gamma$. Therefore we require
\begin{equation}
|m_{\psi,\textrm{eff}}| \ll \Gamma
\label{Eq.: Psi Effective Mass << Gamma}
\end{equation}
We have
\begin{equation}
m_{\psi,\textrm{eff}}^{2} = m_{\psi}^{2} + \left(4\alpha^{2}-2\alpha\right)h^{2}\phi^{2}\left(\frac{\psi}{M_{P}}\right)^{2\alpha-2}
\label{Eq.: Psi Effective Mass Squared - 2nd appearance}
\end{equation}
Therefore we require
\begin{equation}
m_{\psi} \ll \Gamma
\label{Eq.: Psi Mass << Gamma}
\end{equation}
and
\begin{align}
h\overline{\phi}\left(\frac{\psi_{\ast}}{M_{P}}\right)^{\alpha-1} &\ll \Gamma
\\
\nonumber
\\
h\left\langle\phi\right\rangle\left(\frac{\psi_{\ast}}{M_{P}}\right)^{\alpha-1} &\ll \Gamma
\label{Eq.: 2nd Term Psi Effective Mass << Gamma}
\end{align}
Substituting $\left\langle\phi\right\rangle$, $\psi_{\ast}$ and $\Gamma$, \cref{Eq.: Phi VEV,Eq.: Psi_{*},Eq.: phi Decay Rate - Modulated Decay Rate Section} respectively, into \cref{Eq.: 2nd Term Psi Effective Mass << Gamma} gives the constraint
\begin{equation}
h \ll \left(g^{2}m_{0}M_{P}^{\alpha-1}\right)^{2\alpha-1}\left(\frac{\sqrt{\left(2n+4\right)\lambda}}{m_{0}M_{P}^{n}}\right)^{\frac{2\alpha-1}{n+1}}\left(\frac{100\sqrt{H_{\ast}}}{m_{0}M_{P}^{\alpha-1}}\right)^{2\alpha-2}
\label{Eq.: h Constraint i}
\end{equation}
\subsubsection{Energy Density of the Spectator Field}
\label{Subsubsection: Energy Density of the Spectator Field}
We require the energy density of $\psi$ to be subdominant after thermal inflation up until it decays, in order that it does not cause any inflation by itself. The energy density of $\psi$ after thermal inflation is
\begin{align}
\rho_{\psi} &= h^{2}\frac{\psi_{\ast}^{2\alpha}}{M_{P}^{2\alpha-2}}\overline{\phi}^{2} + \frac{1}{2}m_{\psi}^{2}\psi_{\ast}^{2}
\\
\nonumber
\\
&\sim h^{2}\frac{\psi_{\ast}^{2\alpha}}{M_{P}^{2\alpha-2}}\left\langle\phi\right\rangle^{2} + \frac{1}{2}m_{\psi}^{2}\psi_{\ast}^{2}
\label{Eq.: Psi Energy Density after TI}
\end{align}
For simplicity, we assume that $\psi$ decays around the same time as $\phi$, i.e. that $H$ does not change much between the time when $\phi$ decays and the time when $\psi$ decays. Therefore, the energy density of the Universe at the time when $\psi$ decays is $\sim\!M_{P}^{2}\Gamma^{2}$. We therefore require
\begin{align}
\rho_{\psi} &\ll M_{P}^{2}\Gamma^{2}
\\
\nonumber
\\
h^{2}\frac{\psi_{\ast}^{2\alpha}}{M_{P}^{2\alpha-2}}\left\langle\phi\right\rangle^{2} + \frac{1}{2}m_{\psi}^{2}\psi_{\ast}^{2} &\ll M_{P}^{2}\Gamma^{2}
\label{Eq.: Psi Energy Density after TI Constraint a}
\end{align}
Therefore we require
\begin{equation}
m_{\psi} \ll \frac{M_{P}\Gamma}{\psi_{\ast}}
\label{Eq.: Psi Mass Constraint d - Modulated Decay Rate Section}
\end{equation}
and
\begin{equation}
h\left\langle\phi\right\rangle\psi_{\ast}^{\alpha} \ll M_{P}^{\alpha}\Gamma
\label{Eq.: Psi Energy Density after TI Constraint b}
\end{equation}
These two constraints are identical to \cref{Eq.: Psi Mass Constraint d - Enf of Inflation Section,Eq.: Psi Energy Density Oscillating Constraint b} that appear in \cref{Subsubsection: Energy Density of the Oscillating Spectator Field}. This is easy to understand, in that the only difference between the derivation of the two constraints in \cref{Subsubsection: Energy Density of the Oscillating Spectator Field} and those that appear here, is that, in the latter, we start with $\psi$ taking its frozen value $\psi_{\ast}$, whereas in the former, we start with $\psi$ taking its average oscillation value. However, the average oscillation value is just $\psi_{\ast}$. Substituting $\left\langle\phi\right\rangle$, $\psi_{\ast}$ and $\Gamma$, \cref{Eq.: Phi VEV,Eq.: Psi_{*},Eq.: phi Decay Rate - Modulated Decay Rate Section} respectively, into \cref{Eq.: Psi Energy Density after TI Constraint b} gives the constraint
\begin{equation}
h \gg \frac{1}{\left(g^{2}m_{0}\right)^{2\alpha-1}}\left(\frac{m_{0}M_{P}^{n}}{\sqrt{\left(2n+4\right)\lambda}}\right)^{\frac{2\alpha-1}{n+1}}\left(\frac{m_{0}}{100\sqrt{H_{\ast}M_{P}}}\right)^{2\alpha}
\label{Eq.: h Constraint d - Modulated Decay Rate Section}
\end{equation}
which is identical to \cref{Eq.: h Constraint d - End of Inflation Section}.
\subsection{Results}
\label{Subsection: Modulated Decay Rate Results}
In order for \cref{Eq.: h Constraint i,Eq.: h Constraint d - Modulated Decay Rate Section} to both be satisfied, we require the upper bound in \cref{Eq.: h Constraint i} to be much larger than the lower bound in \cref{Eq.: h Constraint d - Modulated Decay Rate Section}. This gives
\begin{equation}
m_{0} \ll \frac{\sqrt{(2n+4)\lambda}\left(100g^{2}\sqrt{H_{\ast}}\right)^{n+1}}{\sqrt{M_{P}}^{n-1}}
\label{Eq.: m_{0} Constraint z}
\end{equation}
We now require this upper bound to be much larger than the lower bound in \cref{Eq.: m_{0} Constraint y}. This gives
\begin{equation}
H_{\ast} \gg \frac{M_{P}}{10^{4}(2n+4)^{\frac{1}{n+1}}}
\label{Eq.: H_{*} Constraint b}
\end{equation}
We now require this lower bound to be much smaller than the upper bound in \cref{Eq.: H_{*} Constraint a}. This gives
\begin{equation}
(2n+4)^{\frac{1}{n+1}} \gg \frac{M_{P}}{10^{14}\ \mbox{GeV}}
\label{Eq.: n Constraint}
\end{equation}
For all values of $n$, this constraint is grossly violated. Therefore, we find that this modulated decay rate scenario cannot produce the dominant contribution to the observed primordial curvature perturbation within this Thermal Inflation model as it currently stands.
\chapter{Summary and Conclusions}
\label{Chapter: Summary and Conclusions}
The research activity and success that has occurred within the field of Cosmology over the last several decades, both observationally and theoretically/computationally, is vast. We have reached a point in which the amount of data being obtained observationally, as well as its precision, is of a sufficient level to strongly guide the direction that theoretical topics within Cosmology should take. This era of ``Precision Cosmology'' is, for example, having an ever-greater input into the topic of Inflation. It has the power to rule-out with high confidence certain models of Inflation, whilst giving further (strong) support to others.\\
\indent The momentum that Cosmology research is experiencing at the moment is not just down to the field of Cosmology however. Several other fields are providing significant input into Cosmology, the main one being, at least for our work, particle physics. The topics of QFT and, in general, the Standard Model of particle physics play key roles in the development of modern Particle Cosmology.\\
\indent This thesis explores one such contribution to the current state of Particle Cosmology. We have developed a model of Thermal Inflation in which a thermal waterfall scalar field, $\phi$, is coupled to a light spectator scalar field, $\psi$. If this spectator field remains light from the time of primordial inflation up until the time when the thermal waterfall field decays, then a contribution to the primordial curvature perturbation $\zeta$ will be generated by two mechanisms: ``end of inflation'' and modulated decay. The motivation for the creation of this new model was to determine whether it can produce the dominant contribution to the primordial curvature perturbation and then to scrutinise its predictions for various quantities that can be observationally (and theoretically) tested.\\
\indent Our model explores two different cases for the decay of the inflaton $\varphi$: Inflaton decay is complete before the start of thermal inflation and inflaton decay occurs after the end of thermal inflation. We have also explored several different cases for the initial value of $\phi$ during primordial inflation, which we label $\phi_{\ast}$ Case A, B and C. With regard to the decay of the thermal waterfall field $\phi$, we have only considered the case that the decay is via direct interactions with the particles in the thermal bath that exist due to the (partial) reheating of the inflaton $\varphi$. Regarding the decay of the spectator field, we assume that this field, $\psi$, decays around the same time as the thermal waterfall field $\phi$.\\
\indent We have used the $\delta N$ formalism to study the perturbations that are generated from both the ``end of inflation'' and modulated decay mechanisms within this model. We find that $\zeta$ is of the same magnitude to first order for both cases.\\
\indent We have constrained the two mechanisms within the model by using a large array of constraints, coming from both observational and theoretical considerations. Some of these are commonplace constraints that appear in many inflation models, such as the requirement that the model does not spoil the high-degree of success of BBN. However, some of the constraints that we employ are rarely seen elsewhere in inflation model-building. Therefore, we believe that, in general, our model is significantly more comprehensive than most others and that some of our constraints should be applied to other inflation models.\\
\indent We first discuss the results for the modulated decay scenario within our model. After considering all constraints together, we found strong tension between several constraints relating to $h$, $m_{0}$ and $H_{\ast}$, as these yield the constraint given by \cref{Eq.: n Constraint}, which can never be satisfied. In conclusion therefore, we find that the modulated decay rate mechanism scenario within our model cannot produce the dominant contribution to the observed primordial curvature perturbation as it currently stands.\\
\indent Now we discuss the results for the ``end of inflation'' scenario within our model. We first start with the case $\alpha\neq1$. Similarly with the modulated decay scenario, after considering all constraints together, we found tension between several constraints, relating to $m_{\psi}$ and $m_{0}$. These combined constraints yield the constraint given by \cref{Eq.: lambda Constraint}, which can never be satisfied, given that we require $\lambda\!\lesssim\!1$. In conclusion therefore, we find that the ``end of inflation'' mechanism scenario for $\alpha\neq1$ within our model cannot produce the dominant contribution to the observed primordial curvature perturbation as it currently stands.\\
\indent We now turn to the case $\alpha\!=\!1$. We have given results only for the case $\Gamma_{\varphi}\!\ll\!H_{\textrm{TI}}$. The reason for this is that this case yielded more liberal constraints than the case $\Gamma_{\varphi}\!\gtrsim\!H_{\textrm{TI}}$. For reasonable values of $H_{\ast}$ and $\Gamma_{\varphi}$, a value of $g\!=\!0.4$ yields a sharp prediction to within an order of magnitude for all but one of the quantities in the model, regardless of the $\phi_{\ast}$ case. The quantity $m_{\psi}$ can span many orders of magnitude when looking within the range $m_{0}\!\sim\!10^{2}\,\text{--}\,10^{3}\ \mbox{GeV}$ and we report only an upper bound of $m_{\psi}\!\sim\!10^{-4}\,\text{--}\,10^{-2}\ \mbox{GeV}$.\\
\indent Regarding the tensor fraction, we obtain a prediction of $r\!\sim\!10^{-13}$ for a value of $H_{\ast}\!\sim\!10^{8}\ \mbox{GeV}$.\\
\indent We also report values for the non-Gaussianity parameters $f_{\textrm{NL}}$ and $g_{\textrm{NL}}$ for local-type non-Gaussianity, i.e. for when the momenta triangle is in the squeezed configuration. We predict values of $f_{\textrm{NL}}\!\sim\!10^{-1}$ and $g_{\textrm{NL}}\!\sim\!10^{-2}$, for a value of $m_{0}\!\sim\!10^{3}\ \mbox{GeV}$, which are within current observational bounds.\\
\indent In order to obtain predictions from our model for the scalar spectral index and its running, we needed to choose an inflation model for the period of primordial inflation. We chose to use a simple slow-roll Chaotic Inflation model, with the potential given by \cref{Eq.: Chaotic Inflation Potential}. We obtain values of $n_{s}\!\approx\!0.9645$ and $n_{s}^{\prime}\!\approx\!-0.001259$, for values of $m_{0}\!\sim\!10^{3}\ \mbox{GeV}$, $N_{\textrm{TI}}\!\approx\!24$ and $N_{\ast}\!\approx\!28$ and independent of the $\phi_{\ast}$ Case and the value of $m_{\psi}$ within its allowed range. These values for $n_{s}$ and $n_{s}^{\prime}$ are within current observational bounds.\\
\indent As already mentioned, the analysis that we have performed on our model has been limited to a relatively small region of the vast parameter space. Therefore, it is possible that there exist other allowed regions, for different values of some or all of the free parameters in the model. This is a potential area for future research. In addition, as our analysis has been purely analytical, where we have analysed most quantities to within an order of magnitude, a future area of work could be to perform numerical analysis. In particular, the constraint that was imposed in order that the inflationary trajectory was 1-dimensional, in that only the $\phi$ field was involved in determining the trajectory of thermal inflation in field space, could be removed, as a numerical analysis lends itself better to such multi-dimensional inflationary trajectories than does an analytic one.\\
\indent Cosmology is still experiencing a golden era at present, with ever more data and theoretical insight progressing the field and giving rise to breakthroughs. We can hope that this situation and momentum grows, with it continuing to lead us on an incredibly exciting journey further down the avenue of the most fundamental questions that humankind can ask.
\appendix
\begin{appendices}
\chapter{Derivation of Thermal Fluctuation of $\phi$}
\label{Appendix: Derivation of Thermal Fluctuation of phi}
Our aim is to calculate the thermal fluctuation of $\phi$ that exists when the field is in thermal equilibrium with a thermal bath. This derivation has been worked through following Ref. \cite{Mukhanov:Physical_Founds._Cosmo.}. As we are dealing with the thermal fluctuation of $\phi$ about $\phi=0$, we will simply call $\delta\phi$ by $\phi$.\\
\indent The solution to the perturbation equation
\begin{equation}
\ddot{\phi} - \nabla^{2}\phi + V^{\prime\prime}\phi = 0
\label{Eq.: Perturbation Field Equation - Appendix}
\end{equation}
is
\begin{equation}
\phi = \frac{1}{\sqrt{16\pi^{3}}}\int\frac{1}{\sqrt{\omega_{k}}}\left(e^{-i\omega_{k}t+i\textbf{k}\cdot\textbf{x}}a_{\textbf{k}}^{-} + e^{i\omega_{k}t-i\textbf{k}\cdot\textbf{x}}a_{\textbf{k}}^{+}\right)\,\mathrm{d^{3}}\!k
\label{Eq.: Perturbation Field Equation Solution}
\end{equation}
where
\begin{equation}
\omega_{k} = \sqrt{k^{2}+\left|m_{\phi,\textrm{eff}}^{2}\right|}
\label{Eq.: omega_{k}}
\end{equation}
\indent Using
\begin{equation}
\left\langle\hat{a}_{\textbf{k}}^{+}\hat{a}_{\textbf{k}^{\prime}}^{-}\right\rangle \equiv \frac{\left\langle n_{\textbf{k}}\left|\hat{a}_{\textbf{k}}^{+}\hat{a}_{\textbf{k}^{\prime}}^{-}\right|n_{\textbf{k}}\right\rangle}{\left\langle n_{\textbf{k}}|n_{\textbf{k}}\right\rangle} = n_{\textbf{k}}\delta\!\left(\textbf{k}-\textbf{k}^{\prime}\right)
\label{Eq.: Creat. Annih. 2-point Corr.}
\end{equation}
and
\begin{equation}
\left\langle\hat{a}_{\textbf{k}}^{+}\hat{a}_{\textbf{k}^{\prime}}^{+}\right\rangle = \left\langle\hat{a}_{\textbf{k}}^{-}\hat{a}_{\textbf{k}^{\prime}}^{-}\right\rangle = 0
\label{Eq.: Creat. 2-point Corr. & Annih. 2-point Corr.}
\end{equation}
where $n_{\textbf{k}}$ is the occupation number, as well as the commutation relation
\begin{equation}
\left[\hat{a}_{\textbf{k}}^{-},\hat{a}_{\textbf{k}^{\prime}}^{+}\right] = \delta\!\left(\textbf{k}-\textbf{k}^{\prime}\right)
\label{Eq.: Annih. Creat. Comm. - Appendix}
\end{equation}
we obtain the 2-point correlator as
\begin{multline}
\left\langle\phi^{2}\right\rangle = \frac{1}{16\pi^{3}}\iint\frac{1}{\sqrt{\omega_{k}\omega_{k^{\prime}}}}\left(e^{-i\omega_{k}t+i\textbf{k}\cdot\textbf{x}+i\omega_{k^{\prime}}t-i\textbf{k}^{\prime}\cdot\textbf{x}}\left(\delta\!\left(\textbf{k}-\textbf{k}^{\prime}\right) + \left\langle\hat{a}_{\textbf{k}^{\prime}}^{+}\hat{a}_{\textbf{k}}^{-}\right\rangle\right)\right.\\+ \left.e^{i\omega_{k}t-i\textbf{k}\cdot\textbf{x}-i\omega_{k^{\prime}}t+i\textbf{k}^{\prime}\cdot\textbf{x}}\left\langle\hat{a}_{\textbf{k}}^{+}\hat{a}_{\textbf{k}^{\prime}}^{-}\right\rangle\right)\,\mathrm{d^{3}}\!k\ \mathrm{d^{3}}\!k^{\prime}
\label{Eq.: <phi^2> 2-point correlator}
\end{multline}
Working through the integral, we obtain
\begin{align}
\left\langle\phi^{2}\right\rangle &= \frac{1}{8\pi^{3}}\int\frac{1}{\omega_{k}}\left(\frac{1}{2} + n_{\textbf{k}}\right)\,\mathrm{d^{3}}\!k
\\
\nonumber
\\
&= \frac{1}{8\pi^{3}}\int_{0}^{\infty}\frac{k^{2}}{\omega_{k}}\left(\frac{1}{2} + n_{\textbf{k}}\right)\,\mathrm{d}k\,\int_{0}^{\pi}\mathrm{\sin}\!\left(\theta\right)\,\mathrm{d}\theta\,\int_{0}^{2\pi}\mathrm{d}\varphi
\\
\nonumber
\\
&= \frac{1}{2\pi^{2}}\int_{0}^{\infty}\frac{k^{2}}{\omega_{k}}\left(\frac{1}{2}+n_{\textbf{k}}\right)\,\mathrm{d}k
\label{Eq.: <phi^2> -- n_{k}}
\end{align}
\indent The occupation number $n_{\textbf{k}}$ is given by the Bose-Einstein distribution
\begin{equation}
n_{\omega_{k}} = \frac{1}{e^{\frac{\omega_{k}}{T}}-1}
\label{Eq.: n_{omega_{k}}}
\end{equation}
where we are neglecting the chemical potential. Substituting this into \cref{Eq.: <phi^2> -- n_{k}} gives
\begin{equation}
\left\langle\phi^{2}\right\rangle_{T} = \frac{1}{2\pi^{2}}\int_{0}^{\infty}\frac{k^{2}}{\omega_{k}\left(e^{\frac{\omega_{k}}{T}}-1\right)}\,\mathrm{d}k
\label{Eq.: <phi^2>_{T} -- k}
\end{equation}
The $\frac{1}{2}$ term that appears in \cref{Eq.: <phi^2> -- n_{k}} is not present in \cref{Eq.: <phi^2>_{T} -- k} as it can be removed by one of the following:
\begin{itemize}
\item Having $n_{\omega_{k}} \gg \frac{1}{2}$
\item Subtracting the vacuum
\item Performing normal ordering
\end{itemize}
\Cref{Eq.: <phi^2>_{T} -- k} can be expressed as
\begin{equation}
\left\langle\phi^{2}\right\rangle_{T} = \frac{T^{2}}{4\pi^{2}}J_{-}^{\left(1\right)}\left(\frac{|m_{\phi,\textrm{eff}}|}{T},0\right)
\label{Eq.: <phi^{2}>_{T} -- J}
\end{equation}
The proof is as follows. The definition of the $J$ term is
\begin{equation}
J_{\mp}^{\left(\nu\right)}(\kappa,\tau) \equiv \int_{\kappa}^{\infty}\frac{\left(x^{2}-\kappa^{2}\right)^{\frac{\nu}{2}}}{e^{x-\tau}\mp1}\,\mathrm{d}x + \int_{\kappa}^{\infty}\frac{\left(x^{2}-\kappa^{2}\right)^{\frac{\nu}{2}}}{e^{x+\tau}\mp1}\,\mathrm{d}x
\label{Eq.: J Def.}
\end{equation}
where for our derivation we define
\begin{equation}
\kappa \equiv \frac{|m_{\phi,\textrm{eff}}|}{T} \hspace{5em} \tau \equiv \frac{\mu}{T}
\label{Eq.: kappa & tau Def.}
\end{equation}
where $\mu$ is the chemical potential. Therefore we obtain (with $\tau\!=\!0$ as $\mu\!=\!0$)
\begin{equation}
J_{-}^{\left(1\right)}\left(\frac{|m_{\phi,\textrm{eff}}|}{T},0\right) \equiv 2\int_{\frac{\left|m_{\phi,\textrm{eff}}\right|}{T}}^{\infty}\frac{\sqrt{x^{2}-\left(\frac{|m_{\phi,\textrm{eff}}|}{T}\right)^{2}}}{e^{x}-1}\,\mathrm{d}x
\label{Eq.: J_{-}^{1}}
\end{equation}
Substituting this into \cref{Eq.: <phi^{2}>_{T} -- J} gives
\begin{equation}
\left\langle\phi^{2}\right\rangle_{T} = \frac{T^{2}}{2\pi^{2}}\int_{\frac{\left|m_{\phi,\textrm{eff}}\right|}{T}}^{\infty}\frac{\sqrt{x^{2}-\left(\frac{|m_{\phi,\textrm{eff}}|}{T}\right)^{2}}}{e^{x}-1}\,\mathrm{d}x
\label{Eq.: <phi^{2}>_{T} -- x}
\end{equation}
Now, we can obtain \cref{Eq.: <phi^{2}>_{T} -- x} from \cref{Eq.: <phi^2>_{T} -- k} by simply changing the integration variable as
\begin{align}
k \rightarrow x &\equiv \frac{\omega_{k}}{T}
\\
\nonumber
\\
&= \frac{\sqrt{k^{2}+\left|m_{\phi,\textrm{eff}}^{2}\right|}}{T}
\\
\nonumber
\\
\mathrm{d}x &= \frac{k}{\omega_{k}T}\,\mathrm{d}k
\label{Eq.: k->x}
\end{align}
\indent We can make progress in trying to solve the integral of $J_{-}^{\left(1\right)}$ as follows. $J_{-}^{\left(-1\right)}$ can be written as
\begin{equation}
J_{-}^{\left(-1\right)} = 2\sum_{n=1}^{\infty}K_{0}(n\kappa)
\label{Eq.: J_{-}^{-1}}
\end{equation}
where $K$ is the modified Bessel function of the second kind. There exists a recurrence relation
\begin{equation}
\frac{\partial J_{\mp}^{\left(\nu\right)}}{\partial\kappa} = -\nu\kappa J_{\mp}^{\left(\nu-2\right)}
\label{Eq.: J Recur. Rel.}
\end{equation}
Therefore we have
\begin{align}
J_{-}^{\left(1\right)} &= -\int\kappa J_{-}^{\left(-1\right)}\,\mathrm{d}\kappa
\\
\nonumber
\\
&= -2\int\kappa \sum_{n=1}^{\infty}K_{0}(n\kappa)\,\mathrm{d}\kappa
\label{Eq.: J_{-}^{1} from Recur. Rel.}
\end{align}
The summation of the modified Bessel functions can be expressed as
\begin{equation}
\sum_{n=1}^{\infty}K_{0}(n\kappa) = \frac{1}{2}\left[\underline{C}+\ln{\left(\frac{\kappa}{4\pi}\right)}\right] + \frac{\pi}{2\kappa} + \pi\sum_{l=1}^{\infty}\left(\frac{1}{\sqrt{\kappa^{2}+4l^{2}\pi^{2}}}-\frac{1}{2l\pi}\right)
\label{Eq.: Summation of K}
\end{equation}
as given in Ref. \cite{Gradshteyn_&_Ryzhik:Table_Integrals_Series_Products}, where $\underline{C}$ is Euler's constant. Therefore we obtain
\begin{equation}
J_{-}^{\left(1\right)} = -\int\left[\underline{C}\kappa+\kappa\ln{\left(\frac{\kappa}{4\pi}\right)}+\pi+2\pi\kappa\sum_{l=1}^{\infty}\left(\frac{1}{\sqrt{\kappa^{2}+4l^{2}\pi^{2}}}-\frac{1}{2l\pi}\right)\right]\,\mathrm{d}\kappa
\label{Eq.: J_{-}^{1} after summation of K}
\end{equation}
In general, this integral doesn't converge, due to the summation term. However, for $\kappa\!\ll\!1$, the summation term vanishes and the integral does converge. We adopt this case, with only a mild constraint on $g$ being introduced. From the definition of $\kappa$, \cref{Eq.: kappa & tau Def.}, $\kappa\!\ll\!1$ requires
\begin{equation}
T \gg |m_{\phi,\textrm{eff}}|
\label{Eq.: T >> |m_{phi,eff}|}
\end{equation}
As we are considering the time of the end of primordial inflation, $|m_{\phi,\textrm{eff}}|\!\sim\!g\,T_{\textrm{end,inf}}$, as given by \cref{Eq.: phi Effective Mass End Prim. Inf.}. Therefore we have
\begin{align}
T_{\textrm{end,inf}} &\gg gT_{\textrm{end,inf}}
\\
g &\ll 1
\label{Eq.: g << 1 (Appendix)}
\end{align}
So, we now have
\begin{equation}
J_{-}^{\left(1\right)} \simeq -\frac{1}{2}\underline{C}\kappa^{2} + \frac{1}{4}\kappa^{2} - \frac{1}{2}\kappa^{2}\ln{\left(\frac{\kappa}{4\pi}\right)} - \pi\kappa + A
\label{Eq.: J_{-}^{1} -- Integral Solved}
\end{equation}
where $A$ is a constant of integration. We can obtain $A$ by equating \cref{Eq.: J_{-}^{1} -- Integral Solved} with \cref{Eq.: J Def.} and setting $\kappa\!=\!\tau\!=\!0$. We obtain
\begin{align}
A &= 2\int_{0}^{\infty}\frac{x}{e^{x}-1}\,\mathrm{d}x
\\
\nonumber
\\
&= \frac{\pi^{2}}{3}
\label{Eq.: Constant of Integration A}
\end{align}
Given that $|m_{\phi,\textrm{eff}}|\!\sim\!g\,T_{\textrm{end,inf}}$, we have $\kappa\!=\!g$. Therefore \cref{Eq.: J_{-}^{1} -- Integral Solved} becomes
\begin{equation}
J_{-}^{\left(1\right)} \simeq \frac{\pi^{2}}{3} - \pi g - \frac{1}{2}g^{2}\ln{\left(\frac{g}{4\pi}\right)} - \frac{1}{2}\underline{C}\,g^{2} + \frac{1}{4}g^{2}
\label{Eq.: J_{-}^{1} -- g}
\end{equation}
For all values of $g$, i.e. $g\!\leq\!1$, we have
\begin{equation}
J_{-}^{\left(1\right)} \sim \frac{\pi^{2}}{3}
\label{Eq.: J_{-}^{1} Final Result}
\end{equation}
Substituting this into \cref{Eq.: <phi^{2}>_{T} -- J} gives
\begin{align}
\left\langle\phi^{2}\right\rangle_{T} &\sim \frac{T^{2}}{12}
\\
\nonumber
\\
\sqrt{\left\langle\phi^{2}\right\rangle_{T}} &\sim T
\label{Eq.: phi Thermal Fluctuation (Appendix)}
\end{align}
\chapter{Field Theory Derivation of Elastic Scattering Cross-section between $\phi$ and a Thermal Bath}
\label{Appendix: Field Theory Derivation Elastic Scattering Cross-section}
We assume the thermal bath to only consist of one scalar field, which we call $\chi$. We define the following 4-momenta: $p_{1}$ for a $\phi$ particle before the collision, $p_{2}$ for a $\chi$ particle before the collision, $p_{3}$ for a $\phi$ particle after the collision and $p_{4}$ for a $\chi$ particle after the collision. Working within the centre-of-mass frame, the 4-momenta are:
\begin{align}
p_{1} &= (E_{1},\textbf{p})
\label{Eq.: p_{1}}
\\
p_{2} &= (p,-\textbf{p})
\label{Eq.: p_{2}}
\\
p_{3} &= \left(E_{3},\textbf{p}^{\prime}\right)
\label{Eq.: p_{3}}
\\
p_{4} &= \left(p^{\prime},-\textbf{p}^{\prime}\right)
\label{Eq.: p_{4}}
\end{align}
The differential elastic cross-section is given by
\begin{equation}
\mathrm{d}\sigma = \frac{1}{|\textbf{v}_{1}-\textbf{v}_{2}|}\frac{1}{2E_{1}}\frac{1}{2p}\left|-ig^{2}\right|^{2}\frac{\mathrm{d^{3}}\textbf{p}^{\prime}}{(2\pi)^{3}2E_{3}}\frac{\mathrm{d^{3}}(-\textbf{p}^{\prime})}{(2\pi)^{3}2p^{\prime}}(2\pi)^{4}\delta^{4}(p_{1}+p_{2}-p_{3}-p_{4})
\label{Eq.: Diff. Cross-sect. Formula}
\end{equation}
where $\textbf{v}_{1}$ and $\textbf{v}_{2}$ are the velocities of the $\phi$ and $\chi$ particles before the collision respectively, where $\textbf{v}_{2}\!=\!c$. As we are setting $c\!=\!1$, the first term in \cref{Eq.: Diff. Cross-sect. Formula} $\approx\!1$. Collecting terms together gives
\begin{equation}
\mathrm{d}\sigma \approx -\frac{g^{4}}{64\pi^{2}E_{1}E_{3}pp^{\prime}}\delta^{4}(p_{1}+p_{2}-p_{3}-p_{4})\mathrm{d^{3}}\textbf{p}^{\prime}\,\mathrm{d^{3}}(-\textbf{p}^{\prime})
\label{Eq.: Diff. Cross-sect.}
\end{equation}
Integrating this gives
\begin{align}
\sigma &\approx -\frac{g^{4}}{64\pi^{2}E_{1}E_{3}}\iint\frac{1}{pp^{\prime}}\delta^{4}(p_{1}+p_{2}-p_{3}-p_{4})\mathrm{d^{3}}\textbf{p}^{\prime}\,\mathrm{d^{3}}(-\textbf{p}^{\prime})
\\
\nonumber
\\
&\approx -\frac{g^{4}}{64\pi^{2}E_{1}E_{3}}\int\frac{1}{pp^{\prime}}\delta(E_{1}+p-E_{3}-p^{\prime})\mathrm{d^{3}}\textbf{p}^{\prime}
\\
\nonumber
\\
&\approx -\frac{g^{4}}{64\pi^{2}E_{1}E_{3}}\int\frac{1}{p}\delta(E_{1}+p-E_{3}-p^{\prime})p^{\prime}\mathrm{d}p^{\prime}\int\limits_{\Omega}\mathrm{d}\Omega
\label{Eq.: Cross-sect. (1)}
\end{align}
From the 4-momenta of $p_{1}$ and $p_{3}$, \cref{Eq.: p_{1},Eq.: p_{3}} respectively, we have
\begin{equation}
E_{1} = \sqrt{p^{2}+m_{\phi,\textrm{eff}}^{2}}
\label{Eq.: E_{1}}
\end{equation}
and
\begin{equation}
E_{3} = \sqrt{p^{\prime2}+m_{\phi,\textrm{eff}}^{2}}
\label{Eq.: E_{3}}
\end{equation}
Substituting these equations into \cref{Eq.: Cross-sect. (1)} gives
\begin{equation}
\sigma \approx -\frac{g^{4}}{64\pi^{2}E_{1}E_{3}}\int\frac{1}{p}\delta\left(\sqrt{p^{2}+m_{\phi,\textrm{eff}}^{2}}+p-\sqrt{p^{\prime2}+m_{\phi,\textrm{eff}}^{2}}-p^{\prime}\right)p^{\prime}\mathrm{d}p^{\prime}\int\limits_{\Omega}\mathrm{d}\Omega
\label{Eq.: Cross-sect. (2)}
\end{equation}
The argument of the delta function is only $=\!0$ if $p^{\prime}\!=\!p$. Therefore we have
\begin{equation}
\sigma \approx -\frac{g^{4}}{64\pi^{2}E_{1}^{2}}\int\limits_{\Omega}\mathrm{d}\Omega
\label{Eq.: Cross-sect. (3)}
\end{equation}
We have
\begin{equation}
E_{1}^{2} = p^{2} + m_{\phi,\textrm{eff}}^{2}
\label{Eq.: E_{1}^{2} (1)}
\end{equation}
From the 4-momenta of $p_{2}$, \cref{Eq.: p_{2}}, we know that $E_{2}\!=\!p$. In addition, given that the energy of a $\chi$ particle is $T$, we therefore have
\begin{equation}
E_{1}^{2} = T^{2} + m_{\phi,\textrm{eff}}^{2}
\label{Eq.: E_{1}^{2} (2)}
\end{equation}
Let us consider the time of primordial inflation reheating. We have
\begin{equation}
m_{\phi,\textrm{eff}} \sim gT
\label{Eq.: m_{phi,eff} order gT}
\end{equation}
Therefore
\begin{align}
E_{1}^{2} &\sim \left(1+g^{2}\right)T^{2}
\\
&\sim T^{2}
\label{Eq.: E_{1}^{2} (3)}
\end{align}
Substituting \cref{Eq.: E_{1}^{2} (3)} into \cref{Eq.: Cross-sect. (3)} gives
\begin{align}
\sigma &\approx -\frac{g^{4}}{64\pi^{2}T^{2}}\int\limits_{\Omega}\mathrm{d}\Omega
\\
\nonumber
\\
&\approx -\frac{g^{4}}{16\pi T^{2}}
\\
\nonumber
\\
|\sigma| &\approx \frac{g^{4}}{16\pi T^{2}}
\label{Eq.: Cross-sect.}
\end{align}
\end{appendices}
\addcontentsline{toc}{chapter}{Bibliography}
\bibliography{Thesis_References}
\end{document}